\documentclass[preprint,12pt]{elsarticle}

\usepackage{amssymb}
\usepackage{amsthm}

\journal{Applied Mathematics and Computation}

\usepackage{graphicx}
\usepackage[utf8]{inputenc}
\usepackage{amsmath}
\usepackage[labelformat=simple]{subcaption}
\usepackage[top=2.5cm,bottom=2.5cm,left=2cm,right=2cm]{geometry}
\usepackage[english]{babel}

\usepackage[toc,page]{appendix} 
\usepackage{tabularx} 
\usepackage[table,xcdraw]{xcolor}

\usepackage{pgfplots}
\pgfplotsset{compat=newest}
\usepgfplotslibrary{groupplots}

\definecolor{myblue}{rgb}{0.050980,0.34118,0.69804}
\definecolor{myred}{rgb}{1,0.21961,0.28627}
\definecolor{mygrey}{rgb}{0.43137,0.43137,0.43137}

\usepackage{enumerate}
\usepackage[mathscr]{euscript}
\usepackage[T1]{fontenc}
\usepackage{caption}
\usepackage{multirow}

\usepackage{color}
\usepackage{amsfonts}
\usepackage{amstext}
\usepackage{alltt}
\usepackage{verbatim}
\usepackage{array}
\usepackage{booktabs}
\usepackage[colorlinks=true,linkcolor=blue,citecolor=blue,urlcolor=blue]{hyperref}
\usepackage[linesnumbered,ruled,resetcount]{algorithm2e}
\usepackage{nicefrac}
\usepackage{mathtools}
\usepackage{pdfpages}
\usepackage{bbm}
\usepackage{chngcntr}  
\usepackage{etoolbox}
\usepackage{empheq}
\usepackage{enumitem}
\usepackage{float}

\usepackage[symbol]{footmisc}

\makeatletter
\let\old@algocf@pre@ruled\@algocf@pre@ruled
\renewcommand{\@algocf@pre@ruled}{%
  \Hy@raisedlink{\hyper@anchorstart{algocf.\thealgocf}\hyper@anchorend}%
  \old@algocf@pre@ruled}
\makeatother

\newcommand{\me}{m_\epsilon}
\newcommand{\eps}{\epsilon}
\newcommand{\non}{\nonumber}
\newcommand{\N}{\mathbb{N}}
\newcommand{\laggr}{\lambda_{\mathrm{a}}}
\newcommand{\lcrit}{\lambda_{\mathrm{c}}}
\newcommand{\ldiff}{\lambda_{\mathrm{d}}}
\newcommand{\lphtr}{\lambda_{\mathrm{m}}}
\newcommand{\lnucl}{\lambda_{\mathrm{n}}}
\newcommand{\lpoly}{\lambda_{\mathrm{p}}}
\newcommand{\lpolone}{\lambda_{\mathrm{pol1}}}
\newcommand{\pqtaggr}{\pi_0}
\newcommand{\NUMparam}{\tilde{N}}
\newcommand{\kaggr}{k_{\mathrm{a}}}
\newcommand{\kdiff}{k_{\mathrm{d}}}
\newcommand{\kpoly}{k_{\mathrm{p}}}
\newcommand{\ksepar}{k_{\mathrm{n}}}
\newcommand{\kphtr}{k_{\mathrm{m}}}
\newcommand{\vcrit}{v_{\mathrm{c}}}
\newcommand{\NUMparticles}{N_{\mathrm{p}}}
\newcommand{\KAPPAalpha}{\kappa_{\mathrm{a}}}
\newcommand{\KAPPAdi}{\kappa_{\mathrm{d}}}
\newcommand{\KAPPApi}{\kappa_{\mathrm{p}}}
\newcommand{\KAPPAenne}{\kappa_{\mathrm{n}}}
\newcommand{\LITREunits}{\mathrm{L}}
\newcommand{\SECONDunits}{\mathrm{s}}
\newcommand{\MOLunits}{\mathrm{mol}}
\newcommand{\AGGRRATEunits}{\tilde{\mathrm{L}}}
\newcommand{\DIFFRATEunits}{\tilde{\mathrm{L}}}
\newcommand{\TDFdiffusion}{\tilde{\varrho}_{\mathrm{d}}}
\newcommand{\TDFpolymeriz}{\tilde{\varrho}_{\mathrm{p}}}

\newif\ifproofread

\newcommand{\changegreen}[1]{%
\ifproofread
{\color{green} #1}%
\else
#1%
\fi
}

\newcommand{\changeorange}[1]{%
\ifproofread
{\color{orange} #1}
\else
#1%
\fi
}

\newif\ifrevision

\newcommand{\changered}[1]{%
\ifrevision
{\color{red} #1}
\else
#1%
\fi
}

\usepackage[automake,nonumberlist]{glossaries}
\usepackage{glossaries-extra}
\usepackage{glossary-longbooktabs}
\setglossarystyle{long-booktabs}
\renewcommand{\glossarysection}[2][]{}

\makeglossaries

\newglossaryentry{m}
{name=$m$,description={Scaled size distribution of clusters of Polymer 2 in non-equilibrium position},sort={l>m}}
\newglossaryentry{w}
{name=$w$,description={Scaled size distribution of clusters of Polymer 2 in equilibrium position},sort={l>w}}
\newglossaryentry{v}
{name=$v$,description={Scaled size, or volume, of clusters of Polymer 2},sort={l>v}}
\newglossaryentry{t}
{name=$t$,description={Scaled time},sort={l>t}}

\newglossaryentry{g}
{name=$g$,description={Scaled rate of growth in volume of clusters of Polymer 2},sort={l>g}}
\newglossaryentry{n}
{name=$n$,description={Scaled rate of nucleation of clusters of Polymer 2 in non-equilibrium position},sort={l>n}}
\newglossaryentry{mu}
{name=$\mu$,description={Scaled rate of migration of clusters of Polymer 2 from non-equilibrium to equilibrium position},sort={g>$\mu$}}
\newglossaryentry{alpha}
{name=$\alpha$,description={Scaled rate of aggregation of clusters of Polymer 2},sort={g>$\alpha$}}

\newglossaryentry{Psi}
{name=$\Psi$,description={Total amount of Monomer 2 divided by total amount of Polymer 1 and Polymer 2},sort={g<$\Psi$}}
\newglossaryentry{Vp}
{name=$V_p$,description={Scaled total volume of particles},sort={l<Vp}}
\newglossaryentry{Phi}
{name=$\Phi$,description={Volume fraction of the amount of Polymer 2 that exceeds the saturation level $\Phi_s$ in matrix phase},sort={g<$\Phi$}}

\newglossaryentry{Vmatpol2}
{name=$V^{\mathrm{mat}}_{\mathrm{pol2}}$,description={Scaled total volume of Polymer 2 in matrix phase},sort={l<Vpol2mat}}
\newglossaryentry{Vcmpol2}
{name=$V^{\mathrm{c_m}}_{\mathrm{pol2}}$,description={Scaled total volume of Polymer 2 in non-equilibrium clusters},sort={l<Vpol2cm}}
\newglossaryentry{Vcwpol2}
{name=$V^{\mathrm{c_w}}_{\mathrm{pol2}}$,description={Scaled total volume of Polymer 2 in equilibrium clusters},sort={l<Vpol2cw}}
\newglossaryentry{Sigma_m}
{name=$\Sigma_m$,description={Scaled total surface of clusters of Polymer 2 in non-equilibrium position},sort={g<$\Sigma_m$}}
\newglossaryentry{Sigma_w}
{name=$\Sigma_w$,description={Scaled total surface of clusters of Polymer 2 in equilibrium position},sort={g<$\Sigma_w$}}
\newglossaryentry{Vpol2}
{name=$V_{\mathrm{pol2}}$,description={Scaled total volume of Polymer 2},sort={l<Vpol2}}

\newglossaryentry{ka}
{name=$\kaggr$,description={Reaction rate constant of aggregation of clusters of Polymer 2},sort={l>ka}}
\newglossaryentry{kd}
{name=$\kdiff$,description={Reaction rate constant of diffusion of Polymer 2 chains from matrix to clusters phase},sort={l>kd}}
\newglossaryentry{km}
{name=$\kphtr$,description={Reaction rate constant of migration of clusters from non-equilibrium to equilibrium position},sort={l>km}}
\newglossaryentry{kn}
{name=$\ksepar$,description={Reaction rate constant of nucleation of non-equilibrium clusters},sort={l>kn}}
\newglossaryentry{kp}
{name=$\kpoly$,description={Reaction rate constant of polymerization of Monomer 2 into Polymer 2 chains},sort={l>kp}}
\newglossaryentry{vc}
{name=$v_{\mathrm{c}}$,description={Volume of Polymer 2 agglomerates nucleating into non-equilibrium clusters},sort={l>vc}}
\newglossaryentry{Np}
{name=$N_{\mathrm{p}}$,description={Total number of particles in the reactor},sort={l<Np}}
\newglossaryentry{R}
{name=$R$,description={Total amount of radicals in the particles},sort={l<R}}
\newglossaryentry{Vpol1}
{name=$V_{\mathrm{pol1}}$,description={Total volume of Polymer 1},sort={l<Vpol1}}
\newglossaryentry{barVmon2}
{name=$\bar{V}_{\mathrm{mon2}}$,description={Molar volume of Monomer 2},sort={l<Vmon2bar}}
\newglossaryentry{barVpol2}
{name=$\bar{V}_{\mathrm{pol2}}$,description={Molar volume of Polymer 2},sort={l<Vpol2bar}}

\newglossaryentry{nu0}
{name=$\nu_0$,description={Characteristic size of clusters of Polymer 2},sort={g>$\nu_0$}}
\newglossaryentry{t0}
{name=$t_0$,description={Characteristic time},sort={l>t0}}
\newglossaryentry{d0}
{name=$d_0$,description={Characteristic value of size distributions},sort={l>d0}}

\newglossaryentry{a}
{name=$a$,description={Power of clusters' size in the aggregation rate $\alpha$},sort={l>a}}
\newglossaryentry{b}
{name=$b$,description={Power of clusters' size in the growth rate $g$},sort={l>b}}
\newglossaryentry{Psi_bar}
{name=$\bar{\Psi}$,description={Initial value of the ratio $\Psi$},sort={g<$\Psi$bar}}
\newglossaryentry{Psi_r}
{name=$\Psi_r$,description={Ratio between molar volumes of Monomer 2 and Polymer 2},sort={g<$\Psi_r$}}
\newglossaryentry{Phi_s}
{name=$\Phi_s$,description={Volume fraction of the amount of Polymer 2 that saturates the matrix phase},sort={g<$\Phi_s$}}

\begin{document}
\proofreadfalse 

\revisionfalse 

\renewcommand{\sectionautorefname}{Section}
\renewcommand{\subsectionautorefname}{Section}
\renewcommand{\subsubsectionautorefname}{Section}
\renewcommand{\algorithmautorefname}{Algorithm}

\newtheorem{theorem}{Proposition}
\renewcommand{\theoremautorefname}{Proposition}
\begin{frontmatter}



\title{Reducing model complexity by means of the Optimal Scaling:\\Population Balance Model for latex particles morphology formation}

\author[bcam,cunef]{Simone Rusconi\corref{cor1}}
\author[bcam,imdea]{Christina Schenk} 
\author[bcam,iker,stoilow]{Arghir Zarnescu}
\author[bcam,iker]{Elena Akhmatskaya\corref{cor2}}

\cortext[cor1]{Contact address: \texttt{rusconis89@gmail.com}}
\cortext[cor2]{Contact address: \texttt{akhmatskaya@bcamath.org}}

\address[bcam]{BCAM - Basque Center for Applied Mathematics, Alameda de Mazarredo 14, 48009 Bilbao, Spain}
\address[cunef]{CUNEF Universidad, C/ de los Pirineos 55, 28040 Madrid, Spain}
\address[imdea]{IMDEA Materials Institute, C/ Eric Kandel 2, Tecnogetafe, 28906 Getafe (Madrid), Spain}
\address[iker]{IKERBASQUE, Basque Foundation for Science, Plaza Euskadi 5, 48009 Bilbao, Spain}
\address[stoilow]{``Simion Stoilow'' Institute of the Romanian Academy, 21 Calea Grivi\c{t}ei, 010702 Bucharest, Romania}

\date{}

\begin{abstract}
Rational computer-aided design of multiphase polymer materials is vital for rapid progress in many important applications, such as: diagnostic tests, drug delivery, coatings, additives for constructing materials, cosmetics, etc. Several property predictive models, including the prospective Population Balance Model for Latex Particles Morphology Formation (LPMF PBM), have already been developed for such materials. However, they lack computational efficiency, and the accurate prediction of materials' properties still remains a great challenge. To enhance performance of the LPMF PBM, we explore the feasibility of reducing its complexity through disregard of the aggregation terms of the model. The introduced nondimensionalization approach, which we call Optimal Scaling with Constraints, suggests a quantitative criterion for locating regions of slow and fast aggregation and helps to derive a family of dimensionless LPMF PBM of reduced complexity. The mathematical analysis of this new family is also provided. When compared with the original LPMF PBM, the resulting models demonstrate several orders of magnitude better computational efficiency.
\end{abstract}
\end{frontmatter}
\footnotetext[1]{LPMF PBM: Population Balance Model for Latex Particles Morphology Formation \label{foot:1LPMFPBM}}
\footnotetext[2]{OS: Optimal Scaling\label{foot:3OS}}
\footnotetext[3]{OSC: Optimal Scaling with Constraints\label{foot:4OSC}}
\footnotetext[4]{r-LPMF PBM: Population Balance Model for Latex Particles Morphology Formation of reduced complexity\label{foot:5rLPMFPBM}}





\section{Introduction}
\label{sec:intro}

	As a result of the performance superiority of multiphase particles over particles with uniform compositions, assembling the composite (multiphase) latex particles with well-defined morphology is of great practical interest in many important applications, such as: diagnostic tests, drug delivery, coatings, synthetic rubber, paints, leather treatments, additives for constructing materials, impact modifiers for plastic matrices, cosmetics, etc. The performance of multiphase latex particles largely depends on particle morphology, i.e. a pattern formed by the phase-separated domains comprising a multiphase particle. The synthesis of new morphologies, however, is time and resources consuming as it mainly relies on heuristic knowledge. Predictive modelling of particle morphology formation can become an important tool for saving time and resources in the synthesis of new multiphase polymer materials provided that the efficiency and performance of computational models meet standards of the technological processes. The early models for prediction of multiphase Latex Particles Morphology Formation (LPMF) \cite{GonzalezOrtiz1995, GonzalezOrtiz1996a, GonzalezOrtiz1996b, Asua2011, Akhmatskaya2012, Akhmatskaya2013} described the dynamic development of a single particle morphology and thus offered a partial view of a real system. Moreover, the detailed single particle simulations were computationally very demanding even with the use of High Performance Computers. The aforementioned deficiencies made the incorporation of such modelling approaches in the synthesis of new materials unfeasible. 

	The introduction of Population Balance models, or PBM, for calculation of the distribution of morphologies for the whole population of multiphase polymer particles in \cite{DDPM_2016, PhDThesis_Rusconi_PMCQS}, helped to significantly improve the computational efficiency of a simulation process, though the trade off between accuracy and speed remained the issue. A crucial step towards solving this problem was proposed in \cite{RUSCONI2019106944} and was based on the idea to nondimensionalize the system of governing equations. In particular, optimal and computationally tractable orders of magnitude for the terms involved in the resulting dimensionless equations were assured by the Optimal Scaling (OS) procedure, also introduced in \cite{RUSCONI2019106944}. The use of the Optimal Scaling allowed to reduce the computational complexity of the LPMF PBM\footref{foot:1LPMFPBM} and to avoid unphysical numerical oscillations present in the solutions obtained with the traditional scaling. 
	
	In this paper, we introduce an advanced variant of the OS\footref{foot:3OS}, the Optimal Scaling with Constraints (OSC), which opens a door for further improvement, namely for reducing the complexity of the LPMF PBM\footref{foot:1LPMFPBM} in a rigorous way, yet without compromising neither accuracy nor efficiency. With the help of the OSC\footref{foot:4OSC}, we derive a dimensionless model of reduced complexity and investigate its applicability as well as accuracy and performance in comparison with the previously proposed dimensionless model. We also present the mathematical analysis of the resulting model.

	The paper is structured as follows. In \autoref{sec:PBEmodel_latex} we describe the Population Balance Model for Latex Particles Morphology Formation and briefly discuss the previously proposed scaling arrangements. The new scaling regime is introduced, justified and applied along with the OSC\footref{foot:4OSC} procedure to the LPMF PBM\footref{foot:1LPMFPBM} in \autoref{sec:Optimal_Scaling_IT}. Then, \autoref{sec:Approx_Model} presents and analyses the new dimensionless model of reduced complexity, whereas \autoref{sec:Num_Test} provides numerical evidence of its accuracy and efficiency. We conclude our findings and discuss future directions in \autoref{sec:concl_disc}.
	
\section{Population Balance Model for Latex Particles Morphology Formation}
\label{sec:PBEmodel_latex}

\begin{figure}[!h]
	\centering
	\begin{subfigure}[h]{.19\linewidth}
	\centering
	\includegraphics[scale=0.65]{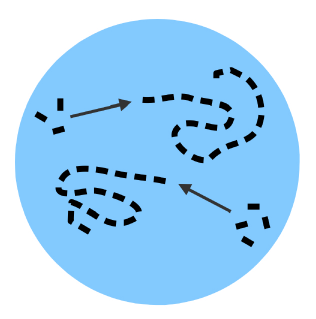}
	\caption{Polymerization}
	\label{fig1a}
	\end{subfigure}	
	\begin{subfigure}[h]{.19\linewidth}
	\centering
	\includegraphics[scale=0.65]{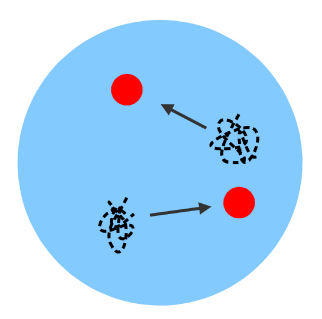}
	\caption{Nucleation}
	\label{fig1b}
	\end{subfigure}	
	\begin{subfigure}[h]{.19\linewidth}
	\centering
	\includegraphics[scale=0.65]{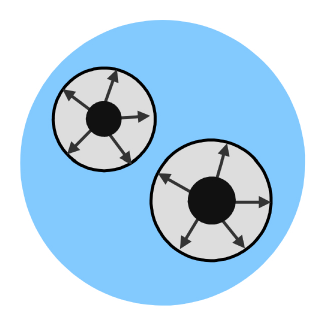}
	\caption{Growth}
	\label{fig1c}
	\end{subfigure}
	\begin{subfigure}[h]{.19\linewidth}
	\centering
	\includegraphics[scale=0.65]{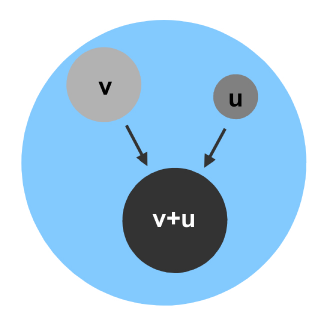}
	\caption{Aggregation}
	\label{fig1d}
	\end{subfigure}	
	\begin{subfigure}[h]{.19\linewidth}
	\centering
	\includegraphics[scale=0.65]{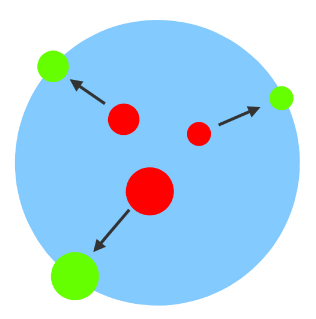}
	\caption{Migration}
	\label{fig1e}
	\end{subfigure}	
	\caption{Reaction mechanisms driving morphology development in a single polymer particle [{\color{cyan} $\bullet$}]: \subref{fig1a} \emph{polymerization} of Monomer 2 [\protect\rule[0.9mm]{0.7ex}{0.4ex}] into Polymer 2 chains [\protect\rule[0.9mm]{0.7ex}{0.4ex}$\,$\protect\rule[0.9mm]{0.7ex}{0.4ex}$\,$\protect\rule[0.9mm]{0.7ex}{0.4ex}], \subref{fig1b} \emph{nucleation} of Polymer 2 agglomerates [\protect\rule[0.9mm]{0.35ex}{0.2ex}$\,$\protect\rule[0.9mm]{0.35ex}{0.2ex}$\,$\protect\rule[0.9mm]{0.35ex}{0.2ex}] into non-equilibrium clusters [{\color{red} $\bullet$}], \subref{fig1c} \emph{growth} of equilibrium and non-equilibrium clusters in volumes, \subref{fig1d} \emph{aggregation} of equilibrium and non-equilibrium clusters with sizes $v$ and $u$ into a cluster of a size $v+u$, \subref{fig1e} \emph{migration} of non-equilibrium clusters [{\color{red} $\bullet$}] to equilibrium positions [{\color{green} $\bullet$}].}
	\label{fig:DDPM_process_description}
	\end{figure}

	We start with a brief description of the reaction mechanisms driving the morphology formation of two-phase latex particles, as proposed in \cite{DDPM_2016}. The morphology development can be summarised with help of the illustrative sketch presented in \autoref{fig:DDPM_process_description} as follows.	

	We consider a seeded emulsion or miniemulsion polymerization process and assume that the number of particles placed in a polymerization reactor is constant during a polymerization reaction. At the beginning of the process, the particles are only made by pre-formed Polymer 1, swollen with Monomer 2. The amount of Polymer 1 does not change with time and belongs to the so-called \emph{matrix} phase during the full evolution of the process. Monomer 2 gradually polymerizes into Polymer 2 chains, as the reaction evolves (\autoref{fig1a}). Polymer 2 chains form agglomerates belonging to the matrix phase, until they reach the critical size $\vcrit$ when they change their phase and nucleate into \emph{clusters} (\autoref{fig1b}). As a result, the matrix phase contains the total amount of Polymer 1, a part of Monomer 2 and Polymer 2 agglomerates with volumes smaller than $\vcrit$. On the other hand, the clusters phase holds the remaining amount of Monomer 2 and the Polymer 2 agglomerates of sizes exceeding $\vcrit$. The amount of Monomer 2 is uniformly distributed between the matrix and clusters phases\changered{, meaning that Monomer 2 swells in the same way for Polymer 1 and Polymer 2, while its total amount decreases with the polymerization process. Such an assumption of uniform monomer concentration has been proposed and supported in \cite{pola.1990.080280505} and \cite{DDPM_2016} for soft polymers, which represent the majority of the emulsion polymers.} The \emph{unswollen} volume $v$ of a given cluster is defined as the volume of Polymer 2 only, without accounting for the amount of swelling Monomer 2. The clusters can increase their unswollen volume $v$ (\autoref{fig1c}) due to $(i)$ polymerization of Monomer 2, $(ii)$ diffusion of Polymer 2 chains from the matrix to the clusters phase and $(iii)$ coagulation with other clusters. Two clusters can coagulate to build an aggregated cluster with a size equal to the sum of volumes of the aggregating clusters (\autoref{fig1d}). The clusters can be found in two positions: \emph{non-equilibrium} and \emph{equilibrium}. \changered{Polymer clusters are said in non-equilibrium position if they can move across the particle they belong to. On the other hand, the} equilibrium clusters have already reached the equilibrium position\changered{, meaning that such clusters cannot move away from their location.} The non-equilibrium clusters can migrate to the equilibrium position and become irreversibly equilibrium clusters (\autoref{fig1e}). In summary, the clusters dynamics is driven by $(i)$ the nucleation of non-equilibrium clusters from the matrix phase, $(ii)$ the growth of unswollen volumes, $(iii)$ the aggregation of clusters and $(iv)$ the migration of clusters from non-equilibrium to equilibrium positions. Different sizes and distributions of produced polymer clusters compose a particles morphology.
	
	For the reaction mechanisms discussed above, we have introduced in \cite{RUSCONI2019106944} the dimensionless Population Balance Model (PBM) for Latex Particles Morphology Formation (LPMF). From now on we shall call it LPMF PBM\footref{foot:1LPMFPBM}. The model predicts the evolution of the size distributions $m(v,t)$ and $w(v,t)$ of non-equilibrium and equilibrium polymer clusters composing the morphology of interest. The distributions $m(v,t)$ and $w(v,t)$ satisfy the system \eqref{eqn:PBE_latex}-\eqref{eqn:Sigma_m,w} of Population Balance Equations (PBE), which accounts for the \emph{polymerization}, \emph{nucleation}, \emph{growth}, \emph{aggregation} and \emph{migration} processes.
		
	\begin{equation}
	\begin{cases}	
	\frac{ \partial m(v,t) }	{ \partial t } 
	& =
	- 
	\underbrace{
	\frac{ \partial ( \, g(v,t) \, m(v,t) \, ) }
	{\partial v }
	}_{\text{Growth}}	
	\, + \, 
	\underbrace{n(v,t)}_{\text{Nucleation}}
	\, - \, 
	\underbrace{\changegreen{\mu} \, m(v,t)}_{\text{Migration}} 
	- \, 
	\underbrace{ m(v,t) \,
	\int_{0}^{\infty}
	\! \changegreen{\alpha}(v,u,t) \, m(u,t) \, du 
	}_{\text{Aggregation}} 
	\\ 
	& \quad 
	+
	\underbrace{ \frac{1}{2} \, 
	\int_0^v
	\! \changegreen{\alpha}(v-u,u,t) \, m(v-u,t) \, m(u,t) \, du
	}_{\text{Aggregation}},
	\quad \quad \quad
	\forall v,t \in \mathbb{R}^+, 
	\\	  
	\\  	
	\frac{ \partial w(v,t) }	{ \partial t } 
	& =
	- 
	\underbrace{
	\frac{ \partial ( \, g(v,t) \, w(v,t) \, ) }
	{\partial v }
	}_{\text{Growth}}	
	\, + \, 
	\underbrace{	 \changegreen{\mu} \, m(v,t) }_{\text{Migration}} 
	- \,
	\underbrace{	 w(v,t) \,
	\int_{0}^{\infty}
	\! \changegreen{\alpha}(v,u,t) \, w(u,t) \, du 
	}_{\text{Aggregation}}	
	\\ 
	& \quad 
	+
	\underbrace{ \frac{1}{2} \, 
	\int_0^v
	\! \changegreen{\alpha}(v-u,u,t) \, w(v-u,t) \, w(u,t) \, du
	}_{\text{Aggregation}},
	\quad \quad \quad
	\forall v,t \in \mathbb{R}^+, 	   
	\\
	\\     
    m(v,0) & = w(v,0) = 0,
	\quad 
	\forall v \in \mathbb{R}^+,
	\quad \quad \quad
	m(0,t) = w(0,t) = 0,
	\quad 
	\forall t \in \mathbb{R}^+,		 
	\end{cases}
	\label{eqn:PBE_latex}
	\end{equation}
	
	\begin{equation}
    \changegreen{\alpha}(v,u,t) :=
    \, \changegreen{\tilde{\alpha}_0}(t)
    \, \left[ v^a + u^a \right],
	\quad    
    \changegreen{\tilde{\alpha}_0}(t) :=
	\, \laggr
    \, (\Psi(t)+1)^{\nicefrac{14}{3}},
    \label{eqn:aggr_rate}
    \end{equation}

    \begin{equation}
	g(v,t) :=
	\, \TDFdiffusion(t) \, v^b	 
	+ 
	\, \TDFpolymeriz(t) \, v,
	\quad	
	\TDFpolymeriz(t) := 
	\, \lpoly
	\, \frac{ \Psi(t) }{ V_p(t) },
	\quad
	\TDFdiffusion(t) :=
	\, \ldiff 
	\, \Phi(t) 
	\, (\Psi(t)+1)^{\nicefrac{2}{3}},
	\label{eqn:growth_rate}
	\end{equation}
	
	\begin{equation}
    n(v,t) :=
	\, \tilde{\eta}_0(t) \, \delta(v-v_0),
	\quad
	\tilde{\eta}_0(t) := \, \lnucl \, \Phi(t),
	\quad v_0 := \lcrit,
	\quad \mbox{with } \delta(x) \mbox{ the Dirac delta},
	\label{eqn:nucl_rate}
	\end{equation}
   
	\begin{equation}
	\changegreen{\mu := \lphtr},
	\label{eqn:phase_tr_rate}
	\end{equation} 
    
	\begin{equation}
	\begin{cases}
	\frac{dV^{\mathrm{mat}}_{\mathrm{pol2}}(t)}{dt}
	& =
	\, \lpoly
	\, \frac{ \Psi(t) }{ V_p(t) }
	\, \left[ 
	\, V^{\mathrm{mat}}_{\mathrm{pol2}}(t) 
	\, + \, 
	\changegreen{\lpolone} \, 
	\right]
	- 
	\, \Phi(t) 
	\, \left[
	\, \changegreen{ \lcrit \, \lnucl }	
	+
	\, \changegreen{\ldiff} \, \Sigma_m(t) + 
	\, \changegreen{\ldiff} \, \Sigma_w(t)
	\right], \\
	V^{\mathrm{mat}}_{\mathrm{pol2}}(0) & = \, 0,
	\end{cases}
	\label{eqn:V_mat_pol2}
	\end{equation}    
    	
	\begin{equation}
	\begin{cases}
	\frac{dV^{\mathrm{c_m}}_{\mathrm{pol2}}(t)}{dt}
	& =
	\, \lpoly
	\, \frac{ \Psi(t) }{ V_p(t) }
	\, V^{\mathrm{c_m}}_{\mathrm{pol2}}(t)
	+ 
	\, \Phi(t) 
	\, \left[
	\, \changegreen{ \lcrit \, \lnucl }
	+
	\, \ldiff \, \Sigma_m(t)  
	\, \right]
	-
	\, \changegreen{\lphtr} \, V^{\mathrm{c_m}}_{\mathrm{pol2}}(t), \\
	V^{\mathrm{c_m}}_{\mathrm{pol2}}(0) & = \, 0,
	\end{cases}
	\label{eqn:V_cm_pol2}
	\end{equation}
	
	\begin{equation}
	\begin{cases}
	\frac{dV^{\mathrm{c_w}}_{\mathrm{pol2}}(t)}{dt}
	& =
	\, \lpoly
	\, \frac{ \Psi(t) }{ V_p(t) }
	\, V^{\mathrm{c_w}}_{\mathrm{pol2}}(t)
	+ 
	\, \ldiff \, \Phi(t) 
	\, \Sigma_w(t) 
	+
	\, \changegreen{\lphtr} \, V^{\mathrm{c_m}}_{\mathrm{pol2}}(t), \\
	V^{\mathrm{c_w}}_{\mathrm{pol2}}(0) & = \, 0,
	\end{cases}
	\label{eqn:V_cw_pol2}
	\end{equation}
		
	\begin{equation}
	\begin{cases}
	\frac{d\Psi(t)}{dt}
	& = 
	- \changegreen{\lpoly} 
	\, \frac{ \Psi(t) }{ \Psi(t) + 1 }
	\, \frac{ \Psi(t) + \Psi_r }
	{ V_{\mathrm{pol2}}(t) + 
	\changegreen{\lpolone} }, \\
	\Psi(0) & = \, \bar{\Psi},
	\end{cases} 
	\label{eqn:Psi}
	\end{equation}	
		
	\begin{equation}
	\begin{cases}
	\frac{dV_{\mathrm{pol2}}(t)}{dt} 
	& = 
	\, \changegreen{\lpoly} 
	\, \frac{ \Psi(t) }{ \Psi(t) + 1 }, \\
	V_{\mathrm{pol2}}(0) & = \, 0,
	\end{cases} 
	\label{eqn:V_pol2}
	\end{equation}
		
	\begin{equation}
    \Phi(t) :=
    \max \left\{
    \frac{ V^{\mathrm{mat}}_{\mathrm{pol2}}(t) }
    { ( \Psi(t) + 1 ) 
    ( V^{\mathrm{mat}}_{\mathrm{pol2}}(t) 
    + 
    \changegreen{\lpolone} ) } 
    - 
    \Phi_s,
    0 \right\},
    \label{eqn:Phi}
    \end{equation} 
    
	\begin{equation}
	V_p(t) :=
	\, \left( \Psi(t)+1 \right)
	\, \left[
	\, V^{\mathrm{mat}}_{\mathrm{pol2}}(t)
	+	
	V^{\mathrm{c_m}}_{\mathrm{pol2}}(t)
	+	
	V^{\mathrm{c_w}}_{\mathrm{pol2}}(t)
	+ 
	\changegreen{\lpolone}
	\, \right],
	\label{eqn:Vp}
	\end{equation}
		
	\begin{equation}
	\Sigma_y(t) :=
	\, (\Psi(t)+1)^{\nicefrac{2}{3}} 
	\int_0^{\infty}
	\! v^b \, y(v,t) \, dv,
	\quad
	y=m,w.
	\label{eqn:Sigma_m,w}
	\end{equation}

\noindent The coefficients appearing in 
\eqref{eqn:PBE_latex}-\eqref{eqn:Sigma_m,w} are defined in \autoref{tab:PBE_lambdas_def} and \autoref{tab:p_exp_def}\changered{, while \autoref{sec:gloss} provides a glossary for the variables in \eqref{eqn:PBE_latex}-\eqref{eqn:Sigma_m,w} and governing parameters in Tables \ref{tab:PBE_lambdas_def}-\ref{tab:p_exp_def}.} We remark that the factors $\theta$ and the experimental parameters $\tilde{p}$ in Tables \ref{tab:PBE_lambdas_def}-\ref{tab:p_exp_def} are dimensional quantities as they possess units of measure, e.g., $\nu_0$ is expressed in Litres, $t_0$ in seconds and $d_0$ in Litres$^{-1}$. Since the chosen system of units is consistent, for simplicity, in the following sections we discard the units of measure of all the dimensional quantities and treat any dimensional variable $X$ as $X/c_X$, with $c_X = 1$ to be the unit of $X$.

	Given $\tilde{p} = \tilde{p}_{\mathrm{exp}}$ (see \autoref{tab:p_exp_def}), a particular choice of scaling factors $\theta := \{ \nu_0, t_0, d_0 \}$ in \autoref{tab:PBE_lambdas_def} fully defines a scaling regime of the proposed model and provides values of corresponding coefficients $\lambda(\theta,\tilde{p})$. For example, by setting $\theta = \{ 1,\dots,1 \}$ one can restore the coefficients $\lambda$ of the unscaled LPMF PBM\footref{foot:1LPMFPBM}. Different scaling regimes of the PBM \eqref{eqn:PBE_latex}-\eqref{eqn:Sigma_m,w} were proposed and compared in \cite{RUSCONI2019106944}. The Optimal Scaling (OS\footref{foot:3OS}) was found to be best in terms of its ability to produce scaled coefficients of similar orders of magnitude. This prevented severe round-off errors, improved conditioning of the considered problem and hindered unphysical behaviour of the computed solution. In the following sections we explore yet another scaling regime for LPMF PBM\footref{foot:1LPMFPBM}.

	\begin{table}[!h]
	\centering
	\begin{tabular}[t]{lll}
	
	\hline
	\textbf{Coefficients $\lambda$} & \textbf{$\lambda(\theta,\tilde{p})$} & \\
	\hline
	
	\rowcolor[HTML]{EFEFEF}
	$\laggr$ & $ \laggr = \kappa_1 \, \nu_0^{a+1} \, t_0 \, d_0$, & $ \kappa_1 = \kaggr / \NUMparticles $ \\

	$\lcrit$ & $ \lcrit = \kappa_2 \, \nu_0^{-1}$, & $ \kappa_2 = \vcrit $ \\

	\rowcolor[HTML]{EFEFEF}
	$\ldiff$ & $ \ldiff = \kappa_3 \, \nu_0^{b-1} \, t_0$, & $ \kappa_3 = \sqrt[3]{36 \pi} \, \kdiff $ \\

	$\lphtr$ & $\lphtr = \kappa_4 \, t_0$, & $ \kappa_4 = \kphtr $ \\

	\rowcolor[HTML]{EFEFEF}
	$\lnucl$ & $ \lnucl = \kappa_5 \, \nu_0^{-1} \, t_0 \, d_0^{-1}$, & $\kappa_5 = \ksepar / \vcrit $ \\

	$\lpoly$ & $\lpoly = \kappa_6 \, \nu_0^{-2} \, t_0 \, d_0^{-1}$, & $\kappa_6 = \kpoly \, R \, \bar{V}_{\mathrm{pol2}} / \bar{V}_{\mathrm{mon2}} $ \\
		
	\rowcolor[HTML]{EFEFEF}	
	$\lpolone$ & 
	$\lpolone = \kappa_7 \, \nu_0^{-2} \, d_0^{-1}$, & $\kappa_7 = V_{\mathrm{pol1}}$ \\
	
	\hline
	\textbf{Other Coefficients} 
	& \textbf{Numerical Values} & \\ 
	\hline

	\rowcolor[HTML]{EFEFEF}		
	$a$ & $-1/3$ & \\
	
	$b$ & $2/3$ & \\
	 
	\rowcolor[HTML]{EFEFEF}		
	$\bar{\Psi}$ & $1$ & \\
	
	$\Psi_r$ & $20/19$ & \\

	\rowcolor[HTML]{EFEFEF}	
	$\Phi_s$ & $10^{-3}$ & \\
			
	\hline	
	
	\end{tabular}
    	\caption{Dimensionless coefficients $\lambda(\theta,\tilde{p}) \in (0,\infty)^{7}$, $a$, $b$, $\bar{\Psi}$, $\Psi_r$, $\Phi_s$, with dimensional scaling factors $\theta := \{ \nu_0, t_0, d_0 \}$ and physical parameters $\tilde{p} := \{ \kaggr, \kdiff, \kpoly, \ksepar, \kphtr, \vcrit, \NUMparticles, R, V_{\mathrm{pol1}}, \bar{V}_{\mathrm{mon2}}, \bar{V}_{\mathrm{pol2}} \}$ (\autoref{tab:p_exp_def}) of the equations \eqref{eqn:PBE_latex}-\eqref{eqn:Sigma_m,w} derived in \cite{RUSCONI2019106944}. The defined here coefficients $\lambda$ arise by setting the scaling factors $\{ \nu_0, t_0, m_0, w_0, M_0, P_0, \Pi_0, \delta_0 \}$ proposed in \cite{RUSCONI2019106944} as $m_0=w_0=d_0$, $M_0=P_0=\Pi_0= d_0 \, \nu_0^2$ and $\delta_0=1/\nu_0$. The numerical values of $a$, $b$, $\bar{\Psi}$, $\Psi_r$ and $\Phi_s$ correspond to the physically grounded model \cite{DDPM_2016,RUSCONI2019106944}, while the models and derivations presented in what follows are valid for any finite value of $a \le 0$, $b \in (0,1)$, $\bar{\Psi}>0$, $\Psi_r>0$ and $\Phi_s \in (0,1)$.}
	\label{tab:PBE_lambdas_def}
	\end{table}	

	\begin{table}[!h]
    \begin{subtable}[t]{0.5\textwidth}
	\centering
	\begin{tabular}[t]{ll}

	\hline
	\textbf{Parameters $\tilde{p}$} & \textbf{Values $\tilde{p}_{\mathrm{exp}}$} \\
	\hline

	\rowcolor[HTML]{EFEFEF}	
	$\kaggr$ & see \autoref{sec:Num_Test} \\	
	
	$\kdiff$ & $10^{-17} \, \LITREunits^{1-b} \, \SECONDunits^{-1}$ \\  	
	
	\rowcolor[HTML]{EFEFEF}	
	$\kpoly$ & $850 \, \LITREunits \, \MOLunits^{-1} \, \SECONDunits^{-1}$ \\ 
	
	$\ksepar$ & $2.5 \times 10^{-5} \, \LITREunits \, \SECONDunits^{-1}$ \\  
	
	\rowcolor[HTML]{EFEFEF}	
	$\kphtr$ & $10^{-5} \, \SECONDunits^{-1}$ \\

	$\vcrit$ & $2.5 \times 10^{-22} \, \LITREunits$ \\	
		
	\hline
	
	\end{tabular}
	\end{subtable}
	\hspace{\fill}
	\begin{subtable}[t]{0.5\textwidth}
	\centering
	\begin{tabular}[t]{ll}
	
	\hline
	\textbf{Parameters $\tilde{p}$} & \textbf{Values $\tilde{p}_{\mathrm{exp}}$} \\
	\hline

	\rowcolor[HTML]{EFEFEF}	
	$\NUMparticles$ & $2.8 \times 10^{17}$ \\
	
	$R$ & $2.3 \times 10^{-7} \, \MOLunits$ \\

	\rowcolor[HTML]{EFEFEF}	
	$V_{\mathrm{pol1}}$ & $0.25 \, \LITREunits$ \\  

	$\bar{V}_{\mathrm{mon2}}$ & $0.1 \, \LITREunits \, \MOLunits^{-1}$ \\  

	\rowcolor[HTML]{EFEFEF}
	$\bar{V}_{\mathrm{pol2}}$ & $0.095 \, \LITREunits \, \MOLunits^{-1}$ \\   

	& \\
	
	\hline
    
    \end{tabular}
    \end{subtable}
    	\caption{Experimental values of parameters $\tilde{p} := \{ \kaggr, \kdiff, \kpoly, \ksepar, \kphtr, \vcrit, \NUMparticles, R, V_{\mathrm{pol1}}, \bar{V}_{\mathrm{mon2}}, \bar{V}_{\mathrm{pol2}} \}$ involved in the computation of the dimensionless coefficients in \autoref{tab:PBE_lambdas_def} as proposed in \cite{DDPM_2016}. The symbols $\SECONDunits$, $\LITREunits$ and $\MOLunits$ stand for second, Litre and mole respectively. Some alternative values of $\kdiff$, $\ksepar$ and $\vcrit$ are also considered in \autoref{sec:Num_Test}.}
    	\label{tab:p_exp_def}
    	\end{table}	
	
\section{Reducing Complexity of the LPMF PBM Using Optimal Scaling}
\label{sec:Optimal_Scaling_IT}

	Here we revisit the OS\footref{foot:3OS} scaling of the model \eqref{eqn:PBE_latex}-\eqref{eqn:Sigma_m,w} and propose to apply OS\footref{foot:3OS} along with the specific constraints on $\lambda$ coefficients (\autoref{tab:PBE_lambdas_def}) with the purpose to reduce the complexity of \eqref{eqn:PBE_latex}-\eqref{eqn:Sigma_m,w}. More precisely, we wish to introduce a particular regime of coefficients 
	
	\begin{align}
	\lambda(\theta,\tilde{p}) := \{ 
	& 
	\laggr = \lambda_1(\theta,\tilde{p}), 
	\lcrit = \lambda_2(\theta,\tilde{p}), 
	\ldiff = \lambda_3(\theta,\tilde{p}), 
	\nonumber \\
	&
	\lphtr = \lambda_4(\theta,\tilde{p}), 
	\lnucl = \lambda_5(\theta,\tilde{p}), 
	\lpoly = \lambda_6(\theta,\tilde{p}),
	\nonumber \\
	&
	\lpolone = \lambda_7(\theta,\tilde{p}) \} 
	\in (0,\infty)^{N_d}, 
	\quad N_d = 7, 
	\label{eqn:lambda_PBEmodel_latex}
	\end{align}	
		
\noindent leading to negligible weights of the integral terms in \eqref{eqn:PBE_latex}. 
		
	To achieve these goals, first, it is necessary to prove that the condition $\laggr \to 0$ (see \eqref{eqn:PBE_latex}-\eqref{eqn:aggr_rate}) implies diminishing the integral terms. Then the scaling procedure supporting such a condition on $\laggr$ and maintaining similar orders of magnitude for other scaled coefficients has to be established. 
	 	
\subsection{Discarding Integral Terms in the LPMF PBM}  	
\label{sec:prop_regime}

	We consider the following regime for coefficients $\lambda(\theta,\tilde{p})$ \eqref{eqn:lambda_PBEmodel_latex}:
  		
  	\begin{equation}
  	\begin{aligned}
	& \laggr \ll 1
	\quad \mbox{and the deviation of} \quad
	\{ \lcrit,\ldiff,\lphtr,
	\lnucl,\lpoly,\lpolone \}
	\quad \mbox{from} \quad \{1,1,1,1,1,1\}
	\\
	& 
	\mbox{is minimal, regardless of the values of}
	\quad \laggr.	
	\end{aligned}
	\label{eqn:lambdas_desired_with_v0ge1}
	\end{equation}	
		
		
	Next we demonstrate that given $a \le 0$, $0 < b < 1$ and $v_0 > 0$, the integral terms in \eqref{eqn:PBE_latex} tend to $0$ when $\laggr \to 0$ and $\lcrit,\ldiff,\lphtr,	\lnucl,\lpoly,\lpolone$ are constants with magnitudes $\approx 10^0$, as specified in \eqref{eqn:lambdas_desired_with_v0ge1}. In this section, we briefly explain the ideas behind the formal proof, leaving the detailed arguments for Appendices \ref{sec:time_dep_factors_rates}-\ref{sec:approx_arg}. 

	In order to investigate the limit for $\laggr \to 0$ of solutions to \eqref{eqn:PBE_latex}-\eqref{eqn:Sigma_m,w}, we multiply $\laggr$ in \eqref{eqn:PBE_latex}-\eqref{eqn:aggr_rate} by $\eps>0$ and consider the limit for $\eps \to 0$ of solutions to the arising equation with the constant $\laggr,\lcrit,\ldiff,\lphtr,\lnucl,\lpoly,\lpolone$:
	
	\begin{align}\label{eq:meps}
    \frac{ \partial \me (v,t) }	{ \partial t } 
	& =
	- 
	\frac{ 
	\partial ( \, g(v,t) \, \me (v,t) \, ) }
	{\partial v }
	\, +
	\, n(v,t)
	\, - 
	\, \changegreen{\mu} \, \me(v,t) 
	- 
	\, \me (v,t) \,
	\int_{0}^{\infty}
	\! \eps \, \changegreen{\alpha}(v,u,t) 
	\, \me (u,t) \, du \, +
	\non \\ 
	& \quad 
	+
	\frac{1}{2} \, 
	\int_0^v
	\! \eps \, \changegreen{\alpha}(v-u,u,t) 
	\, \me(v-u,t) \, \me(u,t) \, du,
	\quad \quad \quad
	\forall v,t \in \mathbb{R}^+. 
	\end{align}

\noindent It can be shown that  $\me$ is bounded in a suitable manner, independent of $\eps$. Thus, one expects that as $\eps\to 0$ the integral terms in the above equation will have a vanishing influence. Moreover, it is natural to presume that the corresponding solutions $\me$ tend to the limit $m_0$, that is also a solution of the equation formally obtained by setting $\eps=0$. Given that  $\me$ is bounded uniformly in $\eps$, the danger in this setting are the possible increasing oscillations as $\eps\to 0$. These can be controlled using the equation, namely the fact that the time oscillations, i.e. the time derivative of $\me$, is also controlled (using the terms on the right hand side). The existing bounds on $\me$ allow a very weak control on the terms on the right hand side and thus the convergence will take place in a suitably weak sense, namely 

\begin{equation}\label{lim:mepsm0}
\sup_{t\in [0,T]} \int_0^V 
\! (m_{\eps_k}(v,t)-m_0(v,t)) \, \varphi(v,t) \, dv \to 0,
\textrm{ as } k\to\infty,
\end{equation} for some  sequence $(\eps_k)_{k\in\N}$ with $\eps_k\to 0$ as $k\to \infty$ and for any function $\varphi:(0,V)\times [0,T]\to \mathbb{R}$ which is at least differentiable in both variables and for each $t\in [0,T]$ is zero on intervals of type $(0,\delta_t)\cup (V-\delta_t,V)$. 
A similar argument can be applied to the corresponding $w_\eps$ equations, but we shall omit it.
The details and proof of this convergence are presented in \autoref{sec:approx_arg}, whereas the essential conditions for such convergence, namely the boundedness of PBE solutions and  the finiteness of PBE solutions' moments are proved in \autoref{sec:Bounded_PBE_Sol} and \autoref{sec:finite_moments} respectively. 
	
\subsection{Optimal Scaling with Constraints (OSC)}
\label{sec:Opt_Scal_Constr}
	
	We first summarise the Optimal Scaling (OS) procedure developed in \cite{RUSCONI2019106944} and then show how to modify the OS\footref{foot:3OS} methodology in order to achieve the regime \eqref{eqn:lambdas_desired_with_v0ge1}.	

	The coefficients $\lambda(\theta,\tilde{p})$ \eqref{eqn:lambda_PBEmodel_latex} depend on the scaling factors $\theta := \{ \theta_j \}_{j=1}^{N_x} \in (0,\infty)^{N_x}$, where
	
	\begin{equation}
	\theta =
	\{ \changegreen{\nu_0, t_0, d_0} \}
	\in (0,\infty)^{N_x}, 
	\quad N_x = \changegreen{3},
	\label{eqn:theta_def}
	\end{equation}

\noindent and physical parameters 	
 	 
 	\begin{equation}
	\tilde{p} = 
	\{ \kaggr, \kdiff, \kpoly, \ksepar, \kphtr, 
	\vcrit, \NUMparticles, R, 
	V_{\mathrm{pol1}}, 
	\bar{V}_{\mathrm{mon2}}, \bar{V}_{\mathrm{pol2}} \}
	\in (0,\infty)^{\NUMparam},
	\quad
	\NUMparam = 11.
	\label{eqn:dimens_param_def}
	\end{equation}
	
\noindent The explicit formulas for $\lambda = \lambda(\theta,\tilde{p})$ and the experimental values $\tilde{p}_{\mathrm{exp}}$ of parameters $\tilde{p}$ are provided in \autoref{tab:PBE_lambdas_def} and \autoref{tab:p_exp_def}. We notice that the units of measure of $\theta$ and $\tilde{p}$ are discarded as explained in \autoref{sec:PBEmodel_latex}.

	As in most physical situations, the coefficients $\lambda$ (see for instance the case in \eqref{eqn:lambda_PBEmodel_latex}, i.e. \autoref{tab:PBE_lambdas_def}) can be presented in terms of governing parameters \cite{Barenblatt1996_book} as

	\begin{equation}
	\lambda_i(\theta,\tilde{p}) = 
	\kappa_i \, \theta^{\alpha^i_1}_1 \dots \theta^{\alpha^i_{N_x}}_{N_x}, 
	\quad
	\kappa_i = \kappa_i(\tilde{p}):
	(0,\infty)^{\NUMparam} \to (0,\infty),
	\quad
	\alpha^i_j \in \mathbb{R},
	\label{eqn:lambda_form}
	\end{equation}

\noindent for all $i=1, \dots, N_d$ and $j=1, \dots, N_x$. Given $\tilde{p}$, $\{\alpha^i_j\}_{i,j=1}^{N_d,N_x}$ and $\Theta := \{ \Theta_i \}_{i=1}^{N_d} \in \mathbb{R}^{N_d}$, the OS\footref{foot:3OS} procedure designed in \cite{RUSCONI2019106944} identifies the factors $\theta$ leading to the minimal distance between the orders of magnitude of coefficients $\lambda$ and the vector $\Theta$. In particular, such factors $\theta$ are obtained as the optimum $\theta_{\mathrm{opt}}$ of the cost function $C(\theta): (0,\infty)^{N_x} \to [0,\infty)$, i.e.
	
	\begin{equation}
	\theta_{\mathrm{opt}}
	:= 
	\mathop{\mathrm{argmin}}_{\theta \in (0,\infty)^{N_x}}
	C(\theta),
	\quad
	C(\theta)
	:=	
	\sum_{i=1}^{N_d}
	\left[ \log_{10}(\lambda_i(\theta,\tilde{p})) - \Theta_i \right]^2,	
	\label{eqn:theta_opt_def}
	\end{equation}	
	
\noindent with $\lambda_i$ being the $i$-th component of vector $\lambda$ \eqref{eqn:lambda_PBEmodel_latex} and $\Theta_i$ chosen as the desired order of magnitude of $\lambda_i$, $\forall i = 1, \dots, N_d$. 

	As demonstrated in \cite{RUSCONI2019106944}, the problem \eqref{eqn:theta_opt_def} can be solved as
	
	\begin{equation}
	\left( \sum_{i=1}^{N_d} \alpha^i_1 \alpha^i_j \right)
	\, \rho^\star_1
	\, + \,
	\dots
	\, + \,
	\left( \sum_{i=1}^{N_d} \alpha^i_{N_x} \alpha^i_j \right)
	\, \rho^\star_{N_x}
	=
	\log_{10}(\hat{\kappa}_j),
	\quad
	\forall j=1, \dots, N_x,
	\label{eqn:OS_lin_system}	
	\end{equation}
	
	\noindent where $\rho^\star_1,\dots,\rho^\star_{N_x}$ are defined as $\theta_{\mathrm{opt}} = \left\{ 10^{\rho^\star_j} \right\}_{j=1}^{N_x}$ and $\hat{\kappa}_j := \prod_{i=1}^{N_d} \kappa_i^{-\alpha^i_j} \, 10^{\sum_{i=1}^{N_d} \alpha^i_j \Theta_i}$, $\forall j=1, \dots, N_x$.
	
	In order to achieve the regime \eqref{eqn:lambdas_desired_with_v0ge1} of coefficients $\lambda$ proposed in \autoref{sec:prop_regime}, the just described OS\footref{foot:3OS} procedure is modified as follows. Given $\tilde{p}$, $\{\alpha^i_j\}_{i,j=1}^{N_d,N_x}$ and $\changegreen{\Theta_2}, \dots, \Theta_{N_d}$, we consider the metric $C$ \eqref{eqn:theta_opt_def} to be a function of the augmented argument $\tilde{\theta}$:
		
	\begin{equation}
	\tilde{\theta} := 
	\{ \theta, \Theta_1 \} 
	\in (0,\infty)^{N_x} \times \mathbb{R},
	\quad
	\theta := 
	\{ \theta_1, \dots, \theta_{N_x} \}
	\in (0,\infty)^{N_x}.
	\label{eqn:Augmented_Arg_tilde_theta}
	\end{equation}

\noindent Then, the values of $\tilde{\theta}$ are selected by minimising the distance $C(\tilde{\theta})$ between the magnitudes of coefficients $\lambda$ and the vector $\Theta$, subjected to the following constraint:

	\begin{equation}
	\tilde{\theta}_{\mathrm{opt}} 
	:=  
	\mathop{\mathrm{argmin}}_{
	\tilde{\theta} = \{\theta,\Theta_1\} }
	C(\tilde{\theta}),
	\quad
	\Theta_1 \le q_1,
	\label{eqn:OSC_constraints}
	\end{equation}
	
\noindent with $q_1 \in \mathbb{R}$ a predefined parameter. 

	In what follows, we explain how to find a solution to \eqref{eqn:OSC_constraints} and, thus, to determine the factors
	
	\begin{equation}
	\tilde{\theta}^{N_x}_{\mathrm{opt}} 
	:= \{ \theta^\star_1, \dots, \theta^\star_{N_x} \}
	\in (0,\infty)^{N_x}
	\label{eqn:OSC_solution}
	\end{equation}

\noindent corresponding to $\tilde{\theta}_{\mathrm{opt}} = \{ \theta^\star_1, \dots, \theta^\star_{N_x}, \Theta^\star_1 \}$ \eqref{eqn:OSC_constraints}. 

	The functional shape \eqref{eqn:lambda_form} of the coefficients $\lambda$ and the change of variables $\rho_j := \log_{10}(\theta_j)$, $\forall j=1,\dots,N_x$, allow us to write the function $C$ \eqref{eqn:theta_opt_def} of the augmented argument $\tilde{\theta}$ \eqref{eqn:Augmented_Arg_tilde_theta} as
	
	\begin{equation}
	C(\tilde{\rho}) =
	\sum_{i=1}^{N_d}
	\left[ \log_{10}(\kappa_i) 
	+ \sum_{j=1}^{N_x} \alpha^i_j \rho_j 
	- \Theta_i \right]^2,	
	\quad
	\forall \tilde{\rho} :=
	\{  \rho_1, \dots, \rho_{N_x}, \Theta_1 \} 
	\in \mathbb{R}^{N_x+\changegreen{1}},
	\label{eqn:Augmented_Arg_tilde_rho}
	\end{equation}

\noindent with $\kappa_i>0$ and $\alpha^i_j, \changegreen{\Theta_2}, \dots, \Theta_{N_d} \in \mathbb{R}$ being fixed parameters $\forall i,j = 1, \dots, N_{d,x}$. Then, \eqref{eqn:OSC_constraints} reads as
	 
	\begin{equation}
	\begin{cases}
	\tilde{\rho}_{\mathrm{opt}} & := 
	\mathop{\mathrm{argmin}}_{\tilde{\rho} \in \Omega}
	C(\tilde{\rho}),
	\\
	\Omega & :=
	\{ \tilde{\rho} \in \mathbb{R}^{N_x+\changegreen{1}} \, |
	\, g_1(\tilde{\rho}) \ge 0 \},
	\\
	g_1(\tilde{\rho}) & := - \Theta_1 + q_1,
	\end{cases}
	\label{eqn:rho_opt_def_constr}
	\end{equation}	
	
\noindent where $C(\tilde{\rho})$ and $\tilde{\rho}$ are given in \eqref{eqn:Augmented_Arg_tilde_rho}, $g_1 : \mathbb{R}^{N_x+\changegreen{1}} \to \mathbb{R}$ and $q_1 \in \mathbb{R}$ is a predefined parameter. \changegreen{Since the equation $c \, \nabla g_1 = 0$ admits the only solution $c=0$ (where $c$ is a scalar), i.e. the linear
independence constraint qualification (LICQ) holds at any $\tilde{\rho} \in \mathbb{R}^{N_x+1}$,} there exists $l_1 \in \mathbb{R}$ such that the following Karush-Kuhn-Tucker conditions (see \cite{Nocedal_book_2006} Chapter 12) are satisfied by any local minimum of $C$ in $\Omega$ \eqref{eqn:rho_opt_def_constr}: 

	\begin{equation}
	\begin{cases}
	& \nabla_{\tilde{\rho}} 
	\mathcal{L}(\tilde{\rho}, l_1) = 0,
	\\
	& l_1 \, g_1(\tilde{\rho}) = 0,
	\\	
	& g_1(\tilde{\rho}), \, l_1 \ge 0.
	\end{cases}
 	\label{eqn:KKT_conditions}
	\end{equation}

\noindent Here the Lagrangian function $\mathcal{L}: \mathbb{R}^{N_x+\changegreen{2}} \to \mathbb{R}$ is defined as

	\begin{equation}
	\mathcal{L}(\tilde{\rho}, l_1)
	:=
	C(\tilde{\rho}) - l_1 \, g_1(\tilde{\rho}),
	\quad
	\forall \tilde{\rho} 
	\in \mathbb{R}^{N_x+\changegreen{1}},
	\quad
	\forall l_1 \in \mathbb{R},
	\label{eqn:Lagr_fun_def}
	\end{equation}

\noindent with $l_1$ known as Lagrange multiplier \cite{doi:10.1080/0025570X.2009.11953617}. 
	
	Denoting $x_1 := g_1(\tilde{\rho})$, one gets from \eqref{eqn:KKT_conditions} the following system of $N_x+\changegreen{3}$ equations with $N_x+\changegreen{3}$ unknowns $\{ \rho_1, \dots, \rho_{N_x}, \Theta_1, x_1, l_1 \} \in \mathbb{R}^{N_x+\changegreen{3}}$:

	\begin{equation}
	\left\{	
 	\begin{aligned}
 	\Theta_1 + x_1 & = q_1, 
 	\\
 	\sum_{k=1}^{N_x} 
	\left( \sum_{i=1}^{N_d} \alpha^i_j \, \alpha^i_k \right) \rho_k
	- \alpha^{\changegreen{1}}_j 
	\, \Theta_\changegreen{1} 
	& =
	\sum_{i=\changegreen{2}}^{N_d} \alpha^i_j \, \Theta_i
	- \sum_{i=1}^{N_d} \alpha^i_j \, \log_{10}(\kappa_i),
	\quad j=1,\dots,N_x, 
	\\
	\sum_{k=1}^{N_x} \alpha^{\changegreen{1}}_k 
	\, \rho_k - \Theta_{\changegreen{1}} 
	- l_{\changegreen{1}}/2 
	& = - \log_{10}(\kappa_{\changegreen{1}}), 
	\\
	l_1 \, x_1 & = 0, 
 	\end{aligned}
	\right.
	\label{eqn:OSC_system_to_solve}
	\end{equation}

\noindent whose only admissible solutions are those corresponding to $ x_1, l_1 \ge 0$. 

	Equations \eqref{eqn:OSC_system_to_solve} along with the predefined $q_1$, $\Theta_{\changegreen{2}},\dots,\Theta_{N_d} \in \mathbb{R}$, $\alpha^i_j \in \mathbb{R}$ and $\kappa_i > 0$, $\forall i,j = 1,\dots,N_{d,x}$, give rise to two linear systems if $l_1 \, x_1 = 0$ is replaced with one of the two possible solutions

\begin{equation}
l_1 = 0,
\label{eqn:OSC_lin_system_S1}
\end{equation}	 

\begin{equation}
x_1 = 0.
\label{eqn:OSC_lin_system_S2}
\end{equation}	 

\noindent From now on we shall refer to the system \eqref{eqn:OSC_system_to_solve}-\eqref{eqn:OSC_lin_system_S1} as $(S_1)$ and to the system comprising \eqref{eqn:OSC_system_to_solve}, \eqref{eqn:OSC_lin_system_S2} as $(S_2)$. 

	Then $\tilde{\rho}_{\mathrm{opt}} = \{ \rho^\star_1, \dots, \rho^\star_{N_x}, \Theta^\star_1 \}$ can be calculated as solution to either $(S_1)$ or $(S_2)$ in such a way that $(i)$ a system admits a solution, $(ii)$ $ x_1, l_1 \ge 0$ and $(iii)$ the cost $C$ \eqref{eqn:Augmented_Arg_tilde_rho} is minimal. Finally, we compute $\tilde{\theta}_{\mathrm{opt}}$ \eqref{eqn:OSC_constraints} and the factors $\tilde{\theta}^{N_x}_{\mathrm{opt}}$ \eqref{eqn:OSC_solution} as

	\begin{equation}
	\tilde{\theta}_{\mathrm{opt}}
	=
	\{ 10^{\rho^\star_1}, \dots, 10^{\rho^\star_{N_x}}, 
	\Theta^\star_1 \},
	\quad
	\tilde{\theta}^{N_x}_{\mathrm{opt}}
	=
	\{ 10^{\rho^\star_1}, \dots, 10^{\rho^\star_{N_x}} \}. 
	\label{eqn:OSC_numerical_solution}
	\end{equation}	 

\noindent The computed $\tilde{\rho}_{\mathrm{opt}}$ is a global solution of the constrained minimisation problem \eqref{eqn:rho_opt_def_constr}, because $C(\tilde{\rho})$ and $-g_1(\tilde{\rho})$ are convex functions of $\tilde{\rho} \in \mathbb{R}^{N_x+\changegreen{1}}$, as shown in \cite{Martin1985,HANSON1981545}. In particular, the Hessian $\nabla_{\tilde{\rho}} \nabla_{\tilde{\rho}} \, C$ is a positive semi-definite matrix $\forall \tilde{\rho} \in \mathbb{R}^{N_x+\changegreen{1}}$ and $-g_1(\tilde{\rho})$ is an affine, thus, convex function of $\tilde{\rho} \in \mathbb{R}^{N_x+\changegreen{1}}$. Then, $\tilde{\theta}_{\mathrm{opt}}$ \eqref{eqn:OSC_numerical_solution} provides a global optimum for the problem 	\eqref{eqn:OSC_constraints}.

	To sum up, we adapted the OS\footref{foot:3OS} methodology to the problems with a constrained parameter, such as the scaling regime \eqref{eqn:lambdas_desired_with_v0ge1}, and reformulated the OS\footref{foot:3OS} minimisation problem \eqref{eqn:theta_opt_def} as \eqref{eqn:OSC_constraints}. 	
	The optimal factors \eqref{eqn:OSC_solution} are then  computed by solving a linear system selected between $(S_1)$ and $(S_2)$. 
	We remark that the same formulation can be used if in \eqref{eqn:OSC_constraints} a constraint is applied to any of the coefficients $\lambda$ \eqref{eqn:lambda_PBEmodel_latex}. Moreover, it should be possible to derive linear systems of the form $(S_1)$ or $(S_2)$ in a similar manner, when required constraints differ from \eqref{eqn:OSC_constraints}. 
	
	In what follows we shall refer to the new scaling procedure as Optimal Scaling with Constraints or OSC\footref{foot:4OSC}.

\subsection{Computation of Dimensionless Coefficients of the LPMF PBM using OSC}
\label{sec:OSC_num_res}	

	With the aim of achieving the regime \eqref{eqn:lambdas_desired_with_v0ge1} in the LPMF PBM\footref{foot:1LPMFPBM} \eqref{eqn:PBE_latex}-\eqref{eqn:Sigma_m,w}, we apply the OSC\footref{foot:4OSC} procedure proposed in \autoref{sec:Opt_Scal_Constr} to the coefficients $\lambda$ \eqref{eqn:lambda_PBEmodel_latex} defined in \autoref{tab:PBE_lambdas_def}. We remark that the results provided in this section correspond to the physically grounded values of $a=-1/3$ and $b=2/3$, as specified in \autoref{tab:PBE_lambdas_def}.

	Hence, we consider $N_d=7$ dimensionless coefficients $\lambda$ and $N_x=3$ scaling factors $\theta=\{\nu_0,t_0,d_0\}$ (see \autoref{tab:PBE_lambdas_def}). Let us denote $\rho_1:=\log_{10}(\nu_0)$, $\rho_2:=\log_{10}(t_0)$, $\rho_3:=\log_{10}(d_0)$ and assign $\Theta_2=\Theta_3=\dots=\Theta_7=0$ in order to satisfy the second condition in \eqref{eqn:lambdas_desired_with_v0ge1}. Then, $(S_1)$ and $(S_2)$, each admitting a unique solution $\forall q_1 \in \mathbb{R}$, read as: 
		
\begin{equation}
(S_1) :
\begin{cases}
\nu_0 & = \frac{ \kappa_2^{9/16} \left( \kappa_3 \, \kappa_6 \, \kappa_7 \right)^{3/16} } { \kappa_5^{3/8} },
\\
t_0 & = \frac{\kappa_7^{1/4}}
{ \left( \kappa_3^3 \, \kappa_4^3 \, \kappa_5 \, \kappa_6 \right)^{1/8} },
\\
d_0 & = \frac{ \kappa_5^{7/8} \, \kappa_7^{3/16} }
{ \kappa_4^{1/4} \left( \kappa_2^{15} \, \kappa_3^9 \, \kappa_6 \right)^{1/16} },
\\
\Theta_1 & = \log_{10}\left( \kappa _1
\sqrt[16]{ \frac{ \kappa_5^8 \, \kappa_7^9}
{ \kappa_2^9 \, \kappa_3^{13} \, \kappa_4^{10} \, \kappa _6 } } \right),
\\
x_1 & = q_1 + 
\log_{10}\left(
\frac{1}{\kappa _1}
\sqrt[16]{ \frac{ \kappa_2^9 \, \kappa_3^{13} \, \kappa_4^{10} \, \kappa _6 } { \kappa_5^8 \, \kappa_7^9} } \right),
\\
l_1 & = 0,
\end{cases}
\quad
(S_2) :
\begin{cases}
\nu_0 & = 10^{-\frac{9 q_1}{47}} 
\frac{ \kappa_1^{9/47} \, \kappa_2^{171/376} \, \kappa_3^{3/94} \, \kappa_6^{33/188} \, \kappa_7^{111/376} }
{ \kappa_4^{45/376} \, \kappa_5^{105/376} },
\\
t_0 & = 10^{\frac{10 q_1}{47}} 
\frac{ \kappa_2^{45/376} \, \kappa_7^{49/376} }
{ \kappa_1^{10/47} \, \kappa_3^{19/94} \, \kappa_4^{91/376}
\, \kappa_5^{87/376} \, \kappa_6^{21/188} }, 
\\
d_0 & = 10^{\frac{27 q_1}{47}}
\frac{ \kappa_4^{41/376} \, \kappa_5^{221/376} }
{ \kappa_1^{27/47} \, \kappa_2^{231/376} \, \kappa_3^{9/94} \, \kappa_6^{5/188} \, \kappa_7^{51/376} }, 
\\
\Theta_1 & = q_1, 
\\
x_1 & = 0, 
\\
l_1 & = -\frac{2}{47} 
\left[
16 q_1 +
\log_{10}\left(
\frac{ \kappa_2^9 \, \kappa_3^{13} \, \kappa_4^{10} \, \kappa_6}{ \kappa_1^{16} \, \kappa_5^8 \, \kappa_7^9 }
\right)
\right],
\end{cases}
\label{eqn:OSC_sol_PBE_latex}
\end{equation}

\noindent with $\kappa_1, \dots, \kappa_7$ provided in \autoref{tab:PBE_lambdas_def} and $x_1, l_1 \ge 0$ (see \autoref{sec:Opt_Scal_Constr}). We remark that the solutions of $(S_1)$ and $(S_2)$ in \eqref{eqn:OSC_sol_PBE_latex} are equivalent when $\pqtaggr = 10^{q_1}$, where we denote

\begin{equation}
\pqtaggr := \kappa_1 \sqrt[16]{ \frac{ \kappa_5^8 \, \kappa_7^9}
{ \kappa_2^9 \, \kappa_3^{13} \, \kappa_4^{10} \, \kappa_6 } }
=
\frac{ ( 36 \pi )^{-13/48} \, \kaggr \, \ksepar^{1/2} \, V_{\mathrm{pol1}}^{9/16} \, \bar{V}_{\mathrm{mon2}}^{1/16} }
{ \NUMparticles \, \vcrit^{17/16} \, \kdiff^{13/16} \, \kphtr^{5/8} 
\, \left( \, \kpoly \, R \, \bar{V}_{\mathrm{pol2}} \, \right)^{1/16} },
\label{eqn:Pi_a_def}
\end{equation}

\noindent with $\kappa_1, \dots, \kappa_7$ from \autoref{tab:PBE_lambdas_def} and the other parameters given in \autoref{tab:p_exp_def}. Since $x_1$ and $l_1$ must be always non-negative, the solution of $(S_1)$ is admissible only if $\pqtaggr \le 10^{q_1}$, whereas when $\pqtaggr > 10^{q_1}$, the solution of $(S_2)$ is permitted.

	Then, for any given $q_1 \in \mathbb{R}$, the optimal scaling factors $\tilde{\theta}^{N_x}_{\mathrm{opt}} =\{\nu_0, t_0, d_0\}$ correspond to $(S_1)$ if 
$\pqtaggr \le 10^{q_1}$, or $(S_2)$ otherwise:	
				
\begin{equation}
\begin{pmatrix}
\nu_0 \\
t_0 \\
d_0
\end{pmatrix}
=
\begin{cases}
\begin{pmatrix}
\frac{ \kappa_2^{9/16} \left( \kappa_3 \, \kappa_6 \, \kappa_7 \right)^{3/16} } { \kappa_5^{3/8} }
\\
\frac{\kappa_7^{1/4}}
{ \left( \kappa_3^3 \, \kappa_4^3 \, \kappa_5 \, \kappa_6 \right)^{1/8} }
\\
\frac{ \kappa_5^{7/8} \, \kappa_7^{3/16} }
{ \kappa_4^{1/4} \left( \kappa_2^{15} \, \kappa_3^9 \, \kappa_6 \right)^{1/16} }
\end{pmatrix}
&
\mbox{if} \quad \pqtaggr \le 10^{q_1},
\\ \\
\begin{pmatrix}
10^{-\frac{9 q_1}{47}} 
\frac{ \kappa_1^{9/47} \, \kappa_2^{171/376} \, \kappa_3^{3/94} \, \kappa_6^{33/188} \, \kappa_7^{111/376} }
{ \kappa_4^{45/376} \, \kappa_5^{105/376} }
\\
10^{\frac{10 q_1}{47}} 
\frac{ \kappa_2^{45/376} \, \kappa_7^{49/376} }
{ \kappa_1^{10/47} \, \kappa_3^{19/94} \, \kappa_4^{91/376}
\, \kappa_5^{87/376} \, \kappa_6^{21/188} }
\\
10^{\frac{27 q_1}{47}}
\frac{ \kappa_4^{41/376} \, \kappa_5^{221/376} }
{ \kappa_1^{27/47} \, \kappa_2^{231/376} \, \kappa_3^{9/94} \, \kappa_6^{5/188} \, \kappa_7^{51/376} } 
\end{pmatrix}
& 
\mbox{otherwise},
\end{cases}
\label{eqn:OSC_optimal_scal_factors}
\end{equation}

\noindent with $\kappa_1, \dots, \kappa_7$ defined in \autoref{tab:PBE_lambdas_def} and $\pqtaggr$ in \eqref{eqn:Pi_a_def}. As a result, the following are the coefficients $\lambda(\tilde{\theta}^{N_x}_{\mathrm{opt}},\tilde{p})$ \eqref{eqn:lambda_PBEmodel_latex} provided by OSC\footref{foot:4OSC} (for any given $q_1 \in \mathbb{R}$):
	
\begin{equation}
\begin{pmatrix}
\laggr 
\\
\lcrit
\\ 
\ldiff 
\\
\lphtr 
\\
\lnucl 
\\
\lpoly 
\\
\lpolone
\end{pmatrix}
=
\begin{pmatrix}
\lambda_1(\tilde{\theta}^{N_x}_{\mathrm{opt}},\tilde{p}) 
\\
\lambda_2(\tilde{\theta}^{N_x}_{\mathrm{opt}},\tilde{p}) 
\\
\lambda_3(\tilde{\theta}^{N_x}_{\mathrm{opt}},\tilde{p}) 
\\ 
\lambda_4(\tilde{\theta}^{N_x}_{\mathrm{opt}},\tilde{p}) 
\\ 
\lambda_5(\tilde{\theta}^{N_x}_{\mathrm{opt}},\tilde{p}) 
\\
\lambda_6(\tilde{\theta}^{N_x}_{\mathrm{opt}},\tilde{p}) 
\\
\lambda_7(\tilde{\theta}^{N_x}_{\mathrm{opt}},\tilde{p}) 
\end{pmatrix}
=
\begin{cases}
\begin{pmatrix}
\pqtaggr
\\
\frac{ \kappa_2^{7/16} \, \kappa_5^{3/8} }{
\left( \kappa_3 \, \kappa_6 \, \kappa _7 \right)^{3/16} }
\\
\frac{ \kappa_3^{9/16} \, \kappa_7^{3/16} }
{ \kappa_2^{3/16} \, \kappa _4^{3/8} \, \kappa_6^{3/16} }
\\
\frac{ \kappa_4 \, \kappa_7^{1/4} }
{ \left( \kappa_3^3 \, \kappa_4^3 \, \kappa_5 \, \kappa_6 \right)^{1/8} }
\\
\frac{ \left( \kappa_2 \, \kappa_5 \right)^{3/8} }
{ \kappa_6^{1/4} \left( \kappa_4 \, \kappa_7 \right)^{1/8} }
\\
\frac{ \kappa_6^{9/16} }
{ \kappa_4^{1/8} \, \kappa_5^{1/4} \left( \kappa_2^3 \, \kappa_3^3 \, \kappa_7^5 \right)^{1/16} }
\\
\frac{ \kappa_4^{1/4} \left( \kappa_3^3 \, \kappa_7^7 \right)^{1/16} }{ \kappa_5^{1/8} \, \left( \kappa_2^3 \, \kappa_6^5 \right)^{1/16} }
\end{pmatrix}
& \mbox{if} \quad \pqtaggr \le 10^{q_1},
\\ \\
\begin{pmatrix}
10^{\frac{31 q_1}{47}} \pqtaggr^{16/47}
\\
10^{\frac{9 q_1}{47}}
\frac{ \kappa_2^{205/376} \kappa_4^{45/376} \kappa_5^{105/376} }
{ \kappa_1^{9/47} \kappa_3^{3/94} \kappa_6^{33/188}
\kappa_7^{111/376} }
\\
10^{\frac{13 q_1}{47}} 
\frac{ \kappa_3^{37/47} \kappa_7^{3/94} } 
{ \kappa_1^{13/47} \kappa_2^{3/94} \kappa_4^{19/94} \kappa_5^{13/94} \kappa_6^{8/47} }
\\
10^{\frac{10 q_1}{47}}
\frac{ \kappa_2^{45/376} \kappa_4^{285/376} \kappa_7^{49/376} }
{ \kappa_1^{10/47} \kappa_3^{19/94} \kappa_5^{87/376} \kappa_6^{21/188} }
\\
10^{-\frac{8 q_1}{47}}
\frac{ \kappa_1^{8/47} \kappa_2^{105/376} \kappa_5^{173/376} }
{ \kappa_3^{13/94} \kappa_4^{87/376} \kappa_6^{49/188} \kappa_7^{11/376} }
\\
10^{\frac{q_1}{47}}
\frac{ \kappa _6^{53/94} }
{ \kappa_1^{1/47} \kappa _2^{33/188} \kappa_3^{8/47} \kappa_4^{21/188} \kappa_5^{49/188} \kappa_7^{61/188} }
\\
10^{-\frac{9 q_1}{47}}
\frac{ \kappa_1^{9/47} \kappa_3^{3/94} \kappa_4^{49/376} \kappa_7^{205/376} }
{ \kappa_2^{111/376} \kappa_5^{11/376} \kappa_6^{61/188} }
\end{pmatrix}
& \mbox{otherwise},
\end{cases}
\label{eqn:lambdas_computed_OSC}
\end{equation}
		
\noindent with $\pqtaggr$ defined in \eqref{eqn:Pi_a_def}.		
	
	When $\pqtaggr \ge 10^{q_1}$, the coefficients $\lambda$ \eqref{eqn:lambdas_computed_OSC} are equivalent to $\lambda(\theta_{\mathrm{opt}},\tilde{p})$ provided by the conventional OS\footref{foot:3OS} \cite{RUSCONI2019106944} with $\Theta_1=q_1$ and $\Theta_2=\Theta_3=\dots=\Theta_7=0$ in 	\eqref{eqn:theta_opt_def}. Indeed, if $\pqtaggr \ge 10^{q_1}$, the solution of $(S_2)$ must be considered in \eqref{eqn:OSC_sol_PBE_latex} leading to $\Theta_1=q_1$ and $x_1=0$. As a consequence, the first and last equations in \eqref{eqn:OSC_system_to_solve} are satisfied with any $l_1 \in \mathbb{R}$. Then, such a multiplier $l_1$ can assure the second-last equation in \eqref{eqn:OSC_system_to_solve} for any possible $\rho_1, \dots, \rho_{N_x}$ and $\Theta_1$. Finally, the variables $\rho_1, \dots, \rho_{N_x}$ must only satisfy the equations labeled with $j=1,\dots,N_x$ in \eqref{eqn:OSC_system_to_solve}. Since the latter equations coincide with the ones fulfilled by $\rho^{\star}_1, \dots, \rho^{\star}_{N_x}$ in \eqref{eqn:OS_lin_system}, it follows the equivalence between the OS' coefficients $\lambda(\theta_{\mathrm{opt}},\tilde{p})$ and the OSC's coefficients $\lambda$ \eqref{eqn:lambdas_computed_OSC} when $\pqtaggr \ge 10^{q_1}$, $\Theta_1=q_1$ and $\Theta_2=\dots=\Theta_7=0$.	 

	The important consequence of such an equivalence is that the solution obtained from $(S_2)$ cannot satisfy the second condition of the regime \eqref{eqn:lambdas_desired_with_v0ge1} as all the coefficients $\lambda$, including $\laggr$, are involved in the minimisation step of the OS\footref{foot:3OS} procedure. Thus, the solution of $(S_2)$ should be excluded from the further consideration.
	
	Let us inspect the case of $\pqtaggr \le 10^{q_1}$, when the coefficients $\lambda$ arise from system $(S_1)$. In this scenario, \eqref{eqn:lambdas_computed_OSC} coincides with the result of applying the conventional OS\footref{foot:3OS} approach \cite{RUSCONI2019106944} to the LPMF PBM\footref{foot:1LPMFPBM} \eqref{eqn:PBE_latex}-\eqref{eqn:Sigma_m,w} without integral terms in \eqref{eqn:PBE_latex}, i.e. $\alpha \equiv 0$. Indeed, it is possible to consider the problem \eqref{eqn:theta_opt_def}, where the cost function $C$ does not contain the term with index $i=1$, i.e. the aggregation term. Then, by solving such a problem, with the choice of $\Theta_i=0$, $i>1$, one can recover the optimal factors $\theta_{\mathrm{opt}} = \{\nu_0,t_0,d_0\}$ in \eqref{eqn:OSC_optimal_scal_factors} and the corresponding $\lambda(\theta_{\mathrm{opt}},\tilde{p})$ in \eqref{eqn:lambdas_computed_OSC} for $\pqtaggr \le 10^{q_1}$. Since by design the OS\footref{foot:3OS} procedure ensures that the distance of the coefficients $\{ \lcrit, \ldiff, \lphtr, \lnucl, \lpoly, \lpolone \}$ in \eqref{eqn:lambdas_computed_OSC} from the unitary vector $\{1,\dots,1\}$ is minimal, it is safe to conclude that when $\pqtaggr \le 10^{q_1}$ the second condition of the targeted regime \eqref{eqn:lambdas_desired_with_v0ge1} is satisfied.
 
	One immediate observation from the analysis of $(S_1)$ and $(S_2)$ is that the OSC\footref{foot:4OSC} can be viewed as a guided OS\footref{foot:3OS} with a clear criterion on a choice of $\Theta_i$ to satisfy a chosen regime.

	Our next step is to check if the first condition of \eqref{eqn:lambdas_desired_with_v0ge1} can be fulfilled when $\pqtaggr \le 10^{q_1}$. The answer follows straight away from the expression of $\laggr$ in \eqref{eqn:lambdas_computed_OSC}, which implies that $\laggr < 1$ when $\pqtaggr \le 10^{q_1}$ and $q_1 < 0$. In other words, in order to achieve the first condition of \eqref{eqn:lambdas_desired_with_v0ge1}, one must at least have $\pqtaggr < 1$. Thus, given $\kappa_1, \dots, \kappa_7$, the condition $\pqtaggr < 1$ can serve as an indicator of the feasibility of neglecting aggregation (integral) terms in the LPMF PBM\footref{foot:1LPMFPBM}, whereas  $\pqtaggr = 1$ helps to distinguish between regions of ``fast'' and ``low'' aggregation. Clearly, the smaller $\pqtaggr$ means a better approximation of the LPMF PBM\footref{foot:1LPMFPBM} by the model without aggregation terms.	

\begin{figure}[!h]
	\centering
	\begin{subfigure}[h]{.495\linewidth}
	\centering
	\includegraphics[scale=0.65]{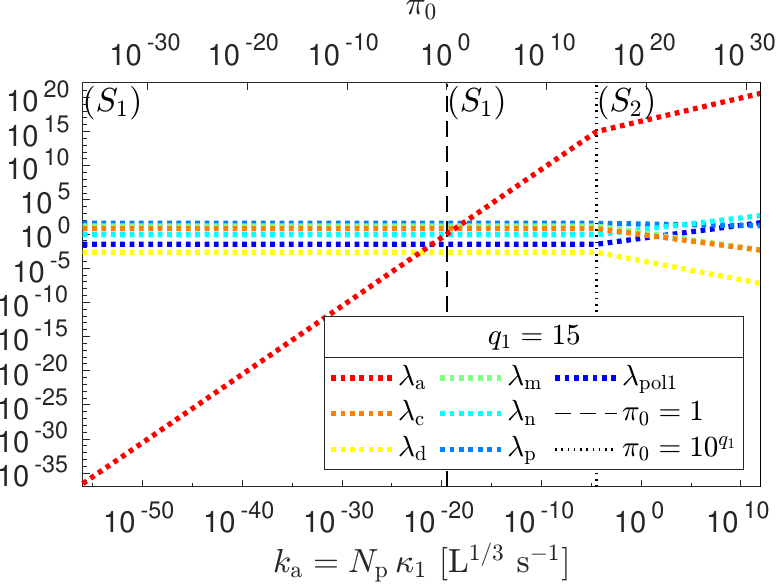}
	\caption{Coefficients $\lambda$ \eqref{eqn:lambdas_computed_OSC} when $q_1=15$.}
	\label{fig:lambdas_OSC_(a)}
	\end{subfigure}	
	\begin{subfigure}[h]{.495\linewidth}
	\centering
	\includegraphics[scale=0.65]{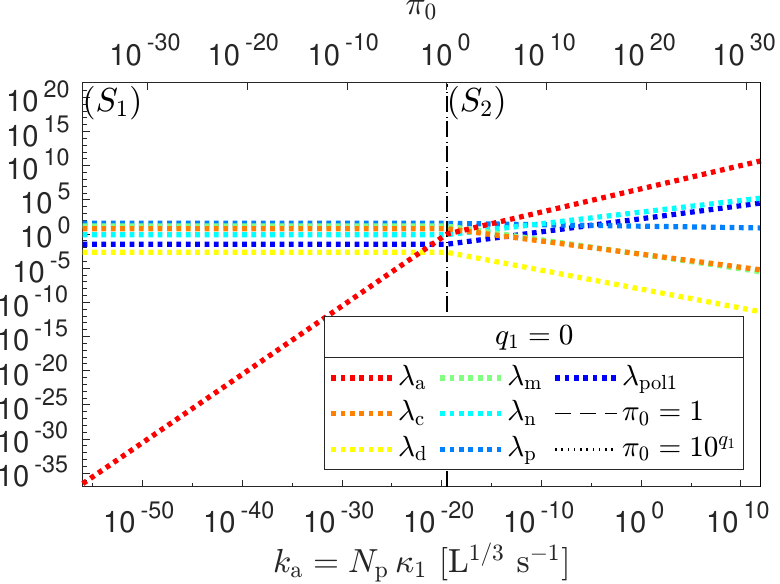}
	\caption{Coefficients $\lambda$ \eqref{eqn:lambdas_computed_OSC} when $q_1=0$.}
	\label{fig:lambdas_OSC_(b)}
	\end{subfigure}	
	\begin{subfigure}[h]{.495\linewidth}
	\centering
	\includegraphics[scale=0.65]{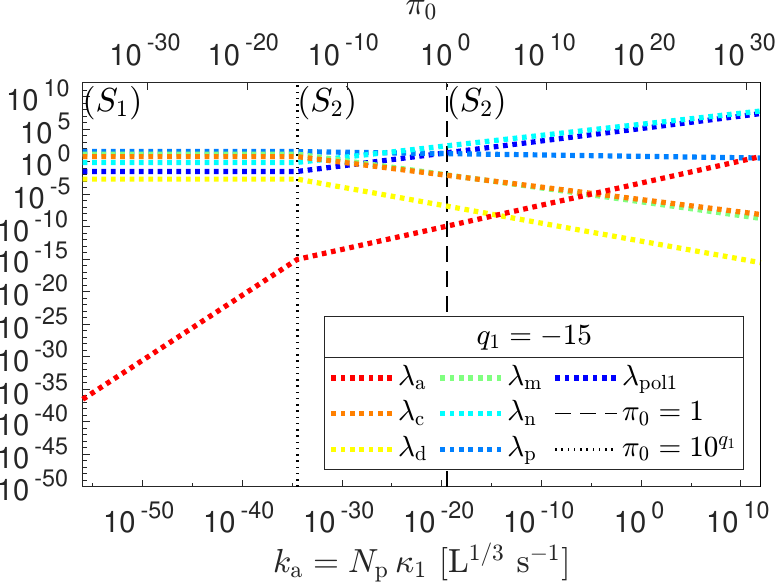}
	\caption{Coefficients $\lambda$ \eqref{eqn:lambdas_computed_OSC} when $q_1=-15$.}
	\label{fig:lambdas_OSC_(c)}
	\end{subfigure}	
	\begin{subfigure}[h]{.495\linewidth}
	\centering
	\includegraphics[scale=0.65]{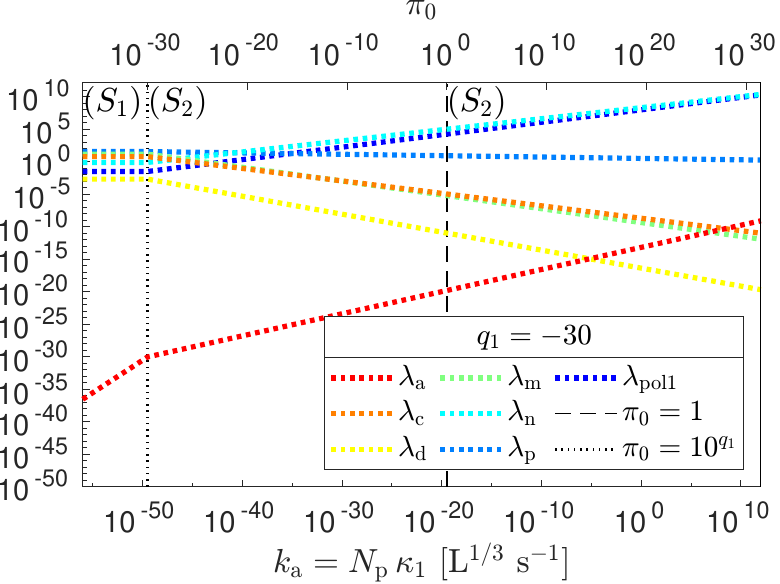}
	\caption{Coefficients $\lambda$ \eqref{eqn:lambdas_computed_OSC} when $q_1=-30$.}
	\label{fig:lambdas_OSC_(d)}
	\end{subfigure}	
	\caption{Coefficients $\lambda(\tilde{\theta}^{N_x}_{\mathrm{opt}},\tilde{p})$ \eqref{eqn:lambdas_computed_OSC}, with $\kappa_1,\dots,\kappa_7$ defined in \autoref{tab:PBE_lambdas_def}, $\tilde{p}=\tilde{p}_{\mathrm{exp}}$ given in \autoref{tab:p_exp_def}, i.e. $\NUMparticles=2.8 \times 10^{17}$, and $q_1 \in \mathbb{R}$. The black vertical dotted lines delimit the ranges of admissibility for solutions corresponding to $(S_1)$ and $(S_2)$ respectively, while the black vertical dashed lines correspond to $\pqtaggr=1$. The coefficients $\lambda$ \eqref{eqn:lambdas_computed_OSC} achieve the regime \eqref{eqn:lambdas_desired_with_v0ge1} when they arise from $(S_1)$ and $\pqtaggr$ \eqref{eqn:Pi_a_def} is $<1$. The values of $\kaggr$ are shown in $\LITREunits^{1/3}$ $\SECONDunits^{-1}$, with $\LITREunits$ for Litres and $\SECONDunits$ for seconds.}
	\label{fig:lambdas_OSC}
	\end{figure}

	Finally, to illustrate our analysis we plot the coefficients $\lambda$ \eqref{eqn:lambdas_computed_OSC} for some choices of $q_1 \in \mathbb{R}$ and the range of $\kappa_1$ values, hence $\pqtaggr$, in \autoref{fig:lambdas_OSC}. Figures \ref{fig:lambdas_OSC_(a)}-\ref{fig:lambdas_OSC_(d)} reveal that the regime  \eqref{eqn:lambdas_desired_with_v0ge1} is achieved when $\pqtaggr < 1$ and the solutions are taken from $(S_1)$. In this scenario, $\laggr < 1$ (with $\laggr$ decreasing towards $0$ as $\kappa_1 \to 0$) and the coefficients $ \lcrit, \ldiff, \lphtr, \lnucl, \lpoly, \lpolone $ do not change when $\kappa_1 \to 0$ ($\kappa_2,\dots,\kappa_7$ are  fixed). On the contrary, if the coefficients $\lambda$ are computed according to $(S_2)$, $\lcrit, \ldiff, \lphtr, \lnucl, \lpoly, \lpolone$ vary when $\laggr$ is reduced by decreasing values of $\kappa_1$. We also notice that $\pqtaggr < 1$ provides the largest range of $\kappa_1$'s values fulfilling \eqref{eqn:lambdas_desired_with_v0ge1} ($\kappa_2,\dots,\kappa_7$ are fixed). This is because $\pqtaggr \ge 1$ implies that either $\laggr = \pqtaggr \ge 1$ (if $\pqtaggr \le 10^{q_1}$ $(S_1)$ and $q_1 \ge 0$) or the coefficients $\lcrit, \ldiff, \lphtr, \lnucl, \lpoly, \lpolone$ do not satisfy the second condition of the regime \eqref{eqn:lambdas_desired_with_v0ge1} (if $\pqtaggr \ge 10^{q_1}$ $(S_2)$). 

	We conclude that the equations \eqref{eqn:OSC_optimal_scal_factors} and \eqref{eqn:lambdas_computed_OSC} for $\pqtaggr \le 10^{q_1}$ along with the condition $\pqtaggr < 1$ pave the way for the LPMF PBM\footref{foot:1LPMFPBM} without aggregation (integral) terms. The inequality $\pqtaggr < 1$ suggests a condition for physical parameters to meet \eqref{eqn:lambdas_desired_with_v0ge1}. In other words, this is a necessary condition for dropping the aggregation terms in LPMF PBM\footref{foot:1LPMFPBM} without significantly changing the solution. In \autoref{sec:Num_Test}, we shall test its ability to correctly predict the values of parameters $\kappa_1,\dots,\kappa_7$ that allow discarding the integral terms in \eqref{eqn:PBE_latex}. For that, we shall compare the solutions $m$ and $w$ to \eqref{eqn:PBE_latex}-\eqref{eqn:Sigma_m,w} with the solutions of the corresponding
equations formally obtained by setting $\alpha$ in \eqref{eqn:aggr_rate} to $0$.

\section{Approximate LPMF PBM}
\label{sec:Approx_Model} 

	In this section we exploit the regime \eqref{eqn:lambdas_desired_with_v0ge1} to derive LPMF PBM\footref{foot:1LPMFPBM} without integral terms, but with the capability to accurately approximate \eqref{eqn:PBE_latex}-\eqref{eqn:Sigma_m,w}. We remark that the analysis presented in this section is also valid for the whole family of such models with parameters
	
\begin{equation}
a \le 0, \quad 0 < b < 1, \quad v_0 > 0.
\label{eqn:param_abv0}	
\end{equation}		
	 
	We show that in addition to the obvious computational advantages expected from the elimination of integral terms, for some choices of parameters $b$ within $0<b<1$, the approximate LPMF PBM\footref{foot:1LPMFPBM} allows for uncoupling the computation of the time-dependent factors $\TDFdiffusion(t)$, $\TDFpolymeriz(t)$ and $\tilde{\eta}_0(t)$ \eqref{eqn:growth_rate}-\eqref{eqn:nucl_rate} from the solutions $m$ and $w$ to \eqref{eqn:PBE_latex}. As a result, several potential advantages are expected, such as improved computational performance and prospects for advancement of integration algorithms for solving the model.
	
\subsection{Validation}
\label{sec:Ground_Approx_IT}
 
	When \eqref{eqn:lambdas_desired_with_v0ge1} holds, it is possible to consider, in the sense discussed in \autoref{sec:prop_regime}, the model \eqref{eqn:PBE_latex}-\eqref{eqn:Sigma_m,w} without integral terms, i.e. with $\laggr=0$, as an approximation of the original model with very small $\alpha(v,u,t)$, $\forall v,u,t \in \mathbb{R}^+$. 
	
	However, such an approximation cannot be valid if the discarded terms, i.e. the Smoluchowski's coagulation equation \cite{Smoluchowski1916}, undergo singularity formation at a finite time $T^*>0$. Several kinds of singular behaviour are shown in the literature. For instance, as reviewed in \cite{WATTIS20061}, solutions to coagulation equations may lose ``mass'' after a finite time $T^*$ by the formation of particles with ``infinite size'' (gelation phenomenon). 
	
	Another undesired situation is when solutions blow up to infinity at a finite time $T^*$, as a possible (but not necessary) consequence of their self-similarity \cite{Kevrekidis2017}. If a finite-time singularity arises, the solution at times $t \ge T^*$ of the approximate model cannot provide an accurate approximation of the solution to \eqref{eqn:PBE_latex}, despite the choice of very small $\alpha(v,u,t)$, $\forall v,u,t \in \mathbb{R}^+$. 
	
	In the light of discussed singular behaviours, it is necessary to ground the approximation made by discarding the integral terms in \eqref{eqn:PBE_latex}. In particular, we want to exclude the possibility of such scenarios as finite-time gelation and blow-ups for the model \eqref{eqn:PBE_latex}-\eqref{eqn:Sigma_m,w} when \eqref{eqn:param_abv0} holds. 	
	
	Gelation occurs through formation of particles with ``infinite size''. Such a possibility can be excluded if the total sizes $C_m(t)$ and $C_w(t)$ produced by coagulation are finite $\forall t \in \mathbb{R}^+$ \cite{Dubovskii1994}, i.e.
	
	\begin{equation}
	\begin{cases}
	C_m(t) & :=
	\int_0^{\infty} \! \int_0^{\infty} (v+u) 
	\, \changegreen{\alpha}(v,u,t) 
	\, m(v,t) \, m(u,t) \, du \, dv
	< \infty,
	\quad \forall t \in \mathbb{R}^+,
	\\	
	C_w(t) & :=
	\int_0^{\infty} \! \int_0^{\infty} (v+u) 
	\, \changegreen{\alpha}(v,u,t) 
	\, w(v,t) \, w(u,t) \, du \, dv
	< \infty,
	\quad \forall t \in \mathbb{R}^+.
	\end{cases}
	\label{eqn:total_size_coag_finite}
	\end{equation}
	
\noindent When $\changegreen{\alpha}$ in \eqref{eqn:total_size_coag_finite} is given by \eqref{eqn:aggr_rate} and $m,w$ are solutions to \eqref{eqn:PBE_latex}, one obtains

	\begin{equation}
	\begin{cases}
	C_m(t) & = 
	2 \changegreen{\tilde{\alpha}_0}(t)
	\left[ M^{1+a}(t) M^0(t) + M^1(t) M^a(t) \right],
	\quad \forall t \in \mathbb{R}^+,
	\\
	C_w(t) & = 
	2 \changegreen{\tilde{\alpha}_0}(t)
	\left[ W^{1+a}(t) W^0(t) + W^1(t) W^a(t) \right],
	\quad \forall t \in \mathbb{R}^+,
	\end{cases}	
	\end{equation}

\noindent where $M^x(t) = \int_0^{\infty} \! v^x \, m(v,t) \, dv$ and $W^x(t) = \int_0^{\infty} \! v^x \, w(v,t) \, dv$, $\forall x \in \mathbb{R}$, $\forall t \in \mathbb{R}^+$. With help of \eqref{eqn:param_abv0} and the boundedness of $\changegreen{\tilde{\alpha}_0}(t) \in (0,\changegreen{\KAPPAalpha}]$, $\changeorange{\KAPPAalpha<\infty}$, $\forall t \in \mathbb{R}^+$ (\autoref{sec:time_dep_factors_rates}), we apply \autoref{prop:rel_order_mom_latex} to obtain:

	\begin{equation}
	\begin{cases}
	C_m(t) & \le  
	\changegreen{ 2 \KAPPAalpha 
	\, \left[
	(K^{1+a}_m+M^1(t)) (K^0_m+M^1(t))
	+
	(K^a_m+M^1(t)) M^1(t)
	\right],
	\quad
	\forall t \in [0,T],}	
	\\
	C_w(t) & \le  
	\changegreen{ 2 \KAPPAalpha 
	\, \left[
	(K^{1+a}_w+W^1(t)) (K^0_w+W^1(t))
	+
	(K^a_w+W^1(t)) W^1(t)
	\right],
	\quad
	\forall t \in [0,T],}
	\end{cases}
	\label{eqn:finite_sizes}
	\end{equation}

\noindent where $K^x_m,K^x_w<\infty$, $\forall x \in \mathbb{R}$, and $T$ can be arbitrary prescribed, as long as it is finite. Since the first-order moments $M^1(t)$ and $W^1(t)$ are finite for all time $t \in \mathbb{R}^+$ under \eqref{eqn:param_abv0} (see \autoref{prop:finite_M,W1_PBE_latex} in \autoref{sec:finite_moments}), equations \eqref{eqn:finite_sizes} assure that finite-time gelation does not affect the solutions to the PBE system \eqref{eqn:PBE_latex}-\eqref{eqn:Sigma_m,w}.

	The same requirements on parameters $a$, $b$ and $v_0$, i.e. \eqref{eqn:param_abv0}, allow us to show the boundedness of $m(v,t)$ and $w(v,t)$ when $v \in \mathbb{R}^+$, $t \in [0,T]$ and $T < \infty$ is an arbitrary fixed number (see \autoref{prop:bounded_sol_PBElatex} in \autoref{sec:Bounded_PBE_Sol}), and thus to exclude the existence of the solutions $m,w$ to \eqref{eqn:PBE_latex}-\eqref{eqn:Sigma_m,w} which blow up to infinity at a finite time. 
	 
	In conclusion, finite-time singularities do not emerge in the PBE system \eqref{eqn:PBE_latex}-\eqref{eqn:Sigma_m,w} and hence the approximation discarding the integral terms in \eqref{eqn:PBE_latex} is valid. 	 
	 	 
	Next it will be shown that if $(1-b)^{-1} \in \mathbb{N}$ (like, for example, in \autoref{tab:PBE_lambdas_def}) then an approximate model can be simplified further through uncoupling the computation of time-dependent factors in \eqref{eqn:growth_rate}-\eqref{eqn:nucl_rate} from the solutions $m$ and $w$ to \eqref{eqn:PBE_latex}. We shall call a model derived in such a way a LPMF PBM\footref{foot:1LPMFPBM} of reduced complexity, or r-LPMF PBM\footref{foot:5rLPMFPBM}.	 

\subsection{r-LPMF PBM}
\label{sec:Approx_Model_form} 

	By setting $\changegreen{\alpha}(v,u,t) \equiv 0$ in \eqref{eqn:PBE_latex} one obtains: 

	\begin{equation}
	\begin{cases}	
	\partial_t m(v,t)  
	& =
	- \, \partial_v \, ( \, g(v,t) \, m(v,t) \, ) 
	\, + \, n(v,t)
	\, - \, \changegreen{\mu} \, m(v,t),
	\quad 
	\forall v,t \in \mathbb{R}^+, 
	\\  	
	\partial_t w(v,t)  
	& =
	- \, \partial_v \, ( \, g(v,t) \, w(v,t) \, ) 
	\, + \, \changegreen{\mu} \, m(v,t),
	\quad 
	\forall v,t \in \mathbb{R}^+, 
	\\     
    m(v,0) & = w(v,0) = m(0,t) = w(0,t) = 0,
	\quad 
	\forall v,t \in \mathbb{R}^+.		 
	\end{cases}
	\label{eqn:PBE_latex_approx}
	\end{equation}
	
\noindent Here the rates $g$, $n$ and $\changegreen{\mu}$ in \eqref{eqn:PBE_latex_approx} are defined as before (\eqref{eqn:growth_rate}-\eqref{eqn:phase_tr_rate}), i.e.	
	
	\begin{equation}
	g(v,t) =
	\, \TDFdiffusion(t) \, v^b	 
	+ 
	\, \TDFpolymeriz(t) \, v,
	\quad	
	\TDFpolymeriz(t) = 
	\, \lpoly
	\, \frac{ \Psi(t) }{ V_p(t) },
	\quad
	\TDFdiffusion(t) =
	\, \ldiff 
	\, \Phi(t) 
	\, (\Psi(t)+1)^{2/3},
	\quad
	0<b<1,
	\label{eqn:growth_rate_approx}
	\end{equation}
	
	\begin{equation}
    n(v,t) =
	\, \tilde{\eta}_0(t) \, \delta(v-v_0),
	\quad
	\tilde{\eta}_0(t) = \, \lnucl \, \Phi(t),
	\quad v_0 = \lcrit,
	\quad \mbox{with } \delta(x) \mbox{ the Dirac delta},
	\label{eqn:nucl_rate_approx}
	\end{equation}
   
	\begin{equation}
	\changegreen{\mu = \lphtr}.
	\label{eqn:phase_tr_rate_approx}
	\end{equation} 
    
\noindent The \changered{time-dependent} functions $\Psi(t)$ and $\Phi(t)$ in \eqref{eqn:growth_rate_approx}-\eqref{eqn:nucl_rate_approx} are computed as in \eqref{eqn:Psi}-\eqref{eqn:Phi}:

	\begin{equation}
	\begin{cases}
	\frac{d\Psi(t)}{dt}
	& = 
	- \changegreen{\lpoly} 
	\, \frac{ \Psi(t) }{ \Psi(t) + 1 }
	\, \frac{ \Psi(t) + \Psi_r }
	{ V_{\mathrm{pol2}}(t) + 
	\changegreen{\lpolone} },
	\quad \forall t \in \mathbb{R}^+, 
	\\
	\frac{dV_{\mathrm{pol2}}(t)}{dt} 
	& = 
	\, \changegreen{\lpoly} 
	\, \frac{ \Psi(t) }{ \Psi(t) + 1 },
	\quad \forall t \in \mathbb{R}^+, 
	\\
	\Psi(0) & = \, \bar{\Psi},
	\quad
	V_{\mathrm{pol2}}(0) = \, 0,
	\end{cases} 
	\label{eqn:Psi_Vpol2_approx}
	\end{equation}	
		
	\begin{equation}
    \Phi(t) =    
	\max \left\{
    \frac{ V^{\mathrm{mat}}_{\mathrm{pol2}}(t) }
    { ( \Psi(t) + 1 ) 
    ( V^{\mathrm{mat}}_{\mathrm{pol2}}(t) 
    + 
    \changegreen{\lpolone} ) } 
    - 
    \Phi_s,
    0 \right\},    
    \quad \forall t \in \mathbb{R}^+.
    \label{eqn:Phi_approx}
    \end{equation} 
    
\noindent The variable $V_p(t)$ in \eqref{eqn:growth_rate_approx} can be expressed as \eqref{eqn:V_p_alternative} in \autoref{sec:time_dep_factors_rates}:

	\begin{equation}
	V_p(t) = 
	\left( \Psi(t)+1 \right)
	\changegreen{\left[ V_{\mathrm{pol2}}(t) + 
	\lpolone \right]},
	\quad \forall t \in \mathbb{R}^+,
	\end{equation}

\noindent and $V^{\mathrm{mat}}_{\mathrm{pol2}}(t)$ in \eqref{eqn:Phi_approx} as (see \eqref{eqn:sum_V_x_pol2} in \autoref{sec:time_dep_factors_rates}):

	\begin{equation}
	V^{\mathrm{mat}}_{\mathrm{pol2}}(t)
	=
	\changegreen{V_{\mathrm{pol2}}(t) - M^1(t) - W^1(t)},
	\quad		
	\forall t \in \mathbb{R}^+,
	\label{eqn:V_mat_pol2_approx}
	\end{equation}    
    	
\noindent where $M^1(t) := \int_0^{\infty} \! v \, m(v,t) \, dv$ and $W^1(t) := \int_0^{\infty} \! v \, w(v,t) \, dv$. We remark that \eqref{eqn:V_mat_pol2_approx} makes use of the equivalence between the variables $V^{\mathrm{c_m}}_{\mathrm{pol2}},V^{\mathrm{c_w}}_{\mathrm{pol2}}$ \eqref{eqn:V_cm_pol2}-\eqref{eqn:V_cw_pol2} and the first-order moments $M^1,W^1$ of the solutions $m,w$ to \eqref{eqn:PBE_latex_approx}. Such an equivalence is unveiled by integrating \eqref{eqn:PBE_latex_approx} times $v$ over $\mathbb{R}^+$ that leads to evolution equations equivalent to \eqref{eqn:V_cm_pol2}-\eqref{eqn:V_cw_pol2}, i.e. the equations \eqref{eqn:ODE_M,Wx_PBE_latex_noIT_simplified} at $x=1$, shown below. 

	If $M^1,W^1$ in \eqref{eqn:V_mat_pol2_approx} are obtained by integration of $m,w$, the calculation of $\TDFdiffusion(t)$, $\TDFpolymeriz(t)$ and $\tilde{\eta}_0(t)$ in \eqref{eqn:growth_rate_approx}-\eqref{eqn:nucl_rate_approx} requires the knowledge of $m(v,t)$ and $w(v,t)$. However, it is possible to uncouple such a computation, as we show below. In particular, we aim to demonstrate that for specific values of a parameter $b$, $0< b <1$ and $(1-b)^{-1} \in \mathbb{N}$, the evolution equations of moments $M^x, W^x$, $x = k \, (1-b)$ and $k = 0, 1, \dots, (1-b)^{-1}$ of solutions to \eqref{eqn:PBE_latex_approx} are closed-form ODE equations:	
	
	\begin{equation}
	\begin{cases}
	\frac{d}{dt} M^{k (1-b)}(t)
	& =
	( k  \, (1-b) \, \TDFpolymeriz(t) 
	- \changegreen{\mu} ) 
	\, M^{k (1-b)}(t)
	\, + \, 		
	k \, (1-b) \, \TDFdiffusion(t) \, M^{(k-1)(1-b)}(t)
	\, + \\
	& \quad
	\, + \,
	\tilde{\eta}_0(t) \, v_0^{k(1-b)},
	\quad k = 0, 1, \dots, (1-b)^{-1},
	\quad \forall t \in \mathbb{R}^+,
	\\
	\frac{d}{dt} W^{k (1-b)}(t)
	& =
	k \, (1-b) \, \TDFpolymeriz(t) \, W^{k (1-b)}(t)
	\, + \, 		
	k \, (1-b) \, \TDFdiffusion(t) \, W^{(k-1)(1-b)}(t)
	\, + \\
	& \quad
	\, + \,
	\changegreen{\mu} \, M^{k(1-b)}(t),
	\quad k = 0, 1, \dots, (1-b)^{-1},
	\quad \forall t \in \mathbb{R}^+,	
	\\
	M^{k(1-b)}(0) & = W^{k(1-b)}(0) = 0,
	\quad k = 0, 1, \dots, (1-b)^{-1}.
	\end{cases}
	\label{eqn:ODE_M,Wx_PBE_latex_no_IT_x=k(1-b)}
	\end{equation} 	
		
	Let us consider the evolution equations for the moments $M^x(t) := \int_0^{\infty} \! v^x \, m(v,t) \, dv$ and $W^x(t) := \int_0^{\infty} \! v^x \, w(v,t) \, dv$ with any fixed order $x > - b$, $0 < b < 1$, of solutions $m,w$ to \eqref{eqn:PBE_latex_approx}:
	
	\begin{equation}
	\begin{cases}
	\frac{dM^x(t)}{dt}
	& =
	- \, \lim_{v \to \infty} \, g^x_m(v,t) 
	\, + \,	
	( x \TDFpolymeriz(t) - \changegreen{\mu} ) \, M^x(t) 
	\, + \,
	x \TDFdiffusion(t) \, M^{x+b-1}(t) 
	\, + \, 
	\tilde{\eta}_0(t) \, v_0^x,
	\\
	\frac{dW^x(t)}{dt}
	& =
	- \, \lim_{v \to \infty} \, g^x_w(v,t) 
	\, + \,	
	x \TDFpolymeriz(t) \, W^x(t) 
	\, + \,
	x \TDFdiffusion(t) \, W^{x+b-1}(t) 
	\, + \, 
	\changegreen{\mu} \, M^x(t),
	\\
	M^x(0) & = W^x(0) = 0,
	\end{cases}
	\label{eqn:ODE_M,Wx_PBE_latex_noIT_x_real}
	\end{equation}

\noindent where

	\begin{equation}
	g^x_y(v,t) := 
	( \TDFdiffusion(t) v^{x+b} 
	+ \TDFpolymeriz(t) v^{x+1} )
	\, y(v,t),
	\quad y=m,w.
	\label{eqn:gx_PBElatex_noIT}
	\end{equation}
	
	First, we intend to show that $g^x_{m,w}(v,t) \to 0$ when $v \to \infty$, for any fixed $x \in \mathbb{R}$ and $t \in \mathbb{R}^+$, and thus equations \eqref{eqn:ODE_M,Wx_PBE_latex_noIT_x_real} can be further simplified. 
	
	Following the derivation presented in \autoref{sec:Bounded_PBE_Sol}, one can demonstrate that the solution $m$ to \eqref{eqn:PBE_latex_approx} is \eqref{eqn:sol_m_PBE_latex} with $m(v,0)=\mathcal{I}_m(v,t)=0$, $\forall v \in \mathbb{R}$, $\forall t \in \mathbb{R}^+$, and a similar expression can be obtained for $w$. Then, it follows that the solutions $m(v,t)$ and $w(v,t)$ to \eqref{eqn:PBE_latex_approx} are equal to $0$ if $\varphi_*(t,v) > v_0$. Moreover, the condition $\varphi_*(t,v) > v_0$ is equivalent to $v = \varphi(t,\varphi_*(t,v)) > \varphi(t,v_0)$, due to $\varphi(t,v_*)$ \eqref{eq:transportODE} being strictly increasing function of its second argument $v_*$ (see \autoref{sec:Bounded_PBE_Sol}). Thus, $m(v,t)=w(v,t)=0$ for all $v > \varphi(t,v_0)$ at any fixed time $t \in \mathbb{R}^+$. 
	
	As discussed in \autoref{sec:Bounded_PBE_Sol}, the ODE \eqref{eq:transportODE} provides a well-defined (global) solution $\varphi(t,v_0)<\infty$ and, hence, one obtains $\lim_{v \to \infty} v^y \, z(v,t) = 0$ for $z=m,w$ with any fixed $y=x+b,x+1 \in \mathbb{R}$ and $t \in \mathbb{R}^+$. Recalling that $\TDFdiffusion, \TDFpolymeriz$ are bounded functions of time $t$ (see \autoref{sec:time_dep_factors_rates}), we conclude that $g^x_{m,w}(v,t) \to 0$ as $v \to \infty$, for any fixed $x \in \mathbb{R}$ and $t \in \mathbb{R}^+$. This property allows for rewriting \eqref{eqn:ODE_M,Wx_PBE_latex_noIT_x_real} as
	
	\begin{equation}
	\begin{cases}
	\frac{dM^x(t)}{dt}
	& =
	( x \TDFpolymeriz(t) - 
	\changegreen{\mu} ) \, M^x(t) 
	\, + \,
	x \TDFdiffusion(t) \, M^{x+b-1}(t) 
	\, + \, 
	\tilde{\eta}_0(t) \, v_0^x,
	\\
	\frac{dW^x(t)}{dt}
	& =
	x \TDFpolymeriz(t) \, W^x(t) 
	\, + \,
	x \TDFdiffusion(t) \, W^{x+b-1}(t) 
	\, + \, 
	\changegreen{\mu} \, M^x(t),
	\\
	M^x(0) & = W^x(0) = 0.
	\end{cases}
	\label{eqn:ODE_M,Wx_PBE_latex_noIT_simplified}
	\end{equation}
	
	The ODE system \eqref{eqn:ODE_M,Wx_PBE_latex_noIT_simplified} for $x = 1$ is defined by the set of moments with orders $1$ and $1+j(b-1)$ for all integers $j\ge1$ and, in general, is an \changered{infinite-dimensional} system ($b-1\ne0$). However, the dimensionality of such a system can be reduced to a finite value by $(i)$ exploiting the behaviour of \eqref{eqn:ODE_M,Wx_PBE_latex_noIT_simplified} at $x=0$ and $(ii)$ restricting further the admissible values of $b$. 

	Let us consider first \eqref{eqn:ODE_M,Wx_PBE_latex_noIT_simplified} for $x=0$. We notice that $\tilde{\eta}_0(t)$ in \eqref{eqn:ODE_M,Wx_PBE_latex_noIT_simplified} depends on $M^1(t)$ and $W^1(t)$ through \eqref{eqn:nucl_rate_approx}-\eqref{eqn:V_mat_pol2_approx}. Then the time derivatives of the zero-order moments $M^0(t)$ and $W^0(t)$ are functions of the moments of orders $0$ and $1$ only. In particular, since $x=0$ in \eqref{eqn:ODE_M,Wx_PBE_latex_noIT_simplified}, the moments of order $b-1 < 0$ vanish (here we use the finiteness of function $\TDFdiffusion(t)$ and PBE solutions' moments shown in \autoref{sec:time_dep_factors_rates} and \autoref{sec:finite_moments} respectively). 
	
	When $x=1$ the moments $M^0(t)$ and $W^0(t)$ can appear in \eqref{eqn:ODE_M,Wx_PBE_latex_noIT_simplified} only if there exists an integer $j \ge 1$ such that $1+j(b-1)=0$, i.e. if $(1-b)^{-1} \in \mathbb{N}$. In this case the number of state variables of the ODE system \eqref{eqn:ODE_M,Wx_PBE_latex_noIT_simplified} is not infinite anymore. In particular, the ODE system state is the finite set of moments $M^y(t)$ and $W^y(t)$, with $y=1+j(b-1), 1$ and integer $j=1,\dots,(1-b)^{-1}$. This is because, when $j = (1-b)^{-1} \in \mathbb{N}$, i.e. $y=0$, the time derivatives of $M^0(t)$ and $W^0(t)$ depend only on variables already belonging to the state of the ODE system, i.e. the moments of orders $0$ and $1$. 
	
	Finally, we notice that such a \changered{finite-dimensional} ODE system corresponds to \eqref{eqn:ODE_M,Wx_PBE_latex_no_IT_x=k(1-b)}. We also remark that the ODE system \eqref{eqn:ODE_M,Wx_PBE_latex_no_IT_x=k(1-b)} is written in a closed-form. Indeed, the only unknowns in the RHS of \eqref{eqn:ODE_M,Wx_PBE_latex_no_IT_x=k(1-b)} are the moments $M^{k (1-b)}(t)$ and $W^{k (1-b)}(t)$, with $k =  0, 1, \dots, (1-b)^{-1}$, whose time derivatives constitute the LHS of \eqref{eqn:ODE_M,Wx_PBE_latex_no_IT_x=k(1-b)}. In particular, the moments of order $(k-1)(1-b)$ are cancelled out when $k=0$.
	
	Once coupled with \eqref{eqn:growth_rate_approx}-\eqref{eqn:V_mat_pol2_approx}, the ODE system \eqref{eqn:ODE_M,Wx_PBE_latex_no_IT_x=k(1-b)} can be integrated numerically in time. Such a numerical solution provides the values of moments $M^1(t)$ and $W^1(t)$ (when $k=(1-b)^{-1} \in \mathbb{N}$) required for the calculation of $\TDFdiffusion(t)$, $\TDFpolymeriz(t)$ and $\tilde{\eta}_0(t)$ \eqref{eqn:growth_rate_approx}-\eqref{eqn:nucl_rate_approx}, without the need for accessing values of densities $m(v,t)$ and $w(v,t)$. 

	In conclusion, we have derived the r-LPMF PBM\footref{foot:5rLPMFPBM} \eqref{eqn:PBE_latex_approx}-\eqref{eqn:ODE_M,Wx_PBE_latex_no_IT_x=k(1-b)} whose coefficients are defined in \autoref{tab:PBE_lambdas_def} and \autoref{tab:p_exp_def}. Such a model approximates the LPMF PBM\footref{foot:1LPMFPBM} \eqref{eqn:PBE_latex}-\eqref{eqn:Sigma_m,w} under the coefficients' regime \eqref{eqn:lambdas_desired_with_v0ge1}. When $a \le 0$, $0<b<1$, $(1-b)^{-1} \in \mathbb{N}$, $v_0>0$, the r-LPMF PBM\footref{foot:5rLPMFPBM} offers a significant reduction in model complexity compared with its predecessor LPMF PBM\footref{foot:1LPMFPBM}. In particular, it avoids the computation of integral terms, while allows for uncoupling the computation of functions $\TDFdiffusion(t)$, $\TDFpolymeriz(t)$ and $\tilde{\eta}_0(t)$ from the solutions to the PBE of interest. 
 	 				
\section{Numerical Testing}
\label{sec:Num_Test} 

	Our next objectives are $(i)$ to provide numerical evidence for the feasibility and accuracy of the approximation discussed above and $(ii)$ to investigate the advantages of the r-LPMF PBM\footref{foot:5rLPMFPBM} over the LPMF PBM\footref{foot:1LPMFPBM}. Though the r-LPMF PBM\footref{foot:5rLPMFPBM} represents the family of models with $a \le 0$, $0 < b < 1$ and $(1-b)^{-1} \in \mathbb{N}$, up to now, only the model \eqref{eqn:PBE_latex}-\eqref{eqn:Sigma_m,w} with $a = -1/3$ and $b=2/3$ was physically supported \cite{DDPM_2016,RUSCONI2019106944}. Therefore, we limit our tests to this particular model and the corresponding approximate model \eqref{eqn:PBE_latex_approx}-\eqref{eqn:ODE_M,Wx_PBE_latex_no_IT_x=k(1-b)} with $b=2/3$. Moreover, we employ the physically grounded values of $\bar{\Psi}$, $\Psi_r$ and $\Phi_s$ provided in \autoref{tab:PBE_lambdas_def}.
	
	We use the GMOC method \cite{RUSCONI2019106944} for computing numerical solutions to the PBE systems \eqref{eqn:PBE_latex}-\eqref{eqn:phase_tr_rate} and \eqref{eqn:PBE_latex_approx}-\eqref{eqn:phase_tr_rate_approx}. The Runge-Kutta method (RK4) \cite{Tan2012} is applied to integrate in time the ODE systems \eqref{eqn:V_mat_pol2}-\eqref{eqn:Sigma_m,w} and \eqref{eqn:Psi_Vpol2_approx}-\eqref{eqn:ODE_M,Wx_PBE_latex_no_IT_x=k(1-b)}. As suggested in \cite{RUSCONI2019106944}, the pointwise evaluation of the function $n(v,t)$ in \eqref{eqn:nucl_rate} and \eqref{eqn:nucl_rate_approx} relies on the approximation

\begin{equation}
\delta(v-v_0) \approx \mathcal{N}(v;v_0,\sigma_0), 
\quad \forall v \in \mathbb{R},
\quad v_0 \gg \sigma_0 > 0,
\label{eqn:approx_GMOC_Dirac_delta}
\end{equation}	 

\noindent where $\mathcal{N}(x;x_0,s_0)$ denotes the probability density function at $x \in \mathbb{R}$ of a Gaussian random variable with mean $x_0 \in \mathbb{R}$ and standard deviation $s_0>0$.

\subsection{Feasibility and Accuracy of r-LPMF PBM}	 
\label{sec:accuracy_approx}

	As discussed in \autoref{sec:Approx_Model}, the r-LPMF PBM\footref{foot:5rLPMFPBM} \eqref{eqn:PBE_latex_approx}-\eqref{eqn:ODE_M,Wx_PBE_latex_no_IT_x=k(1-b)} is derived by exploiting the regime \eqref{eqn:lambdas_desired_with_v0ge1} of coefficients $\lambda$ \eqref{eqn:lambda_PBEmodel_latex} defined in \autoref{tab:PBE_lambdas_def}. Given $a=-1/3$ and $b=2/3$ from \autoref{tab:PBE_lambdas_def}, \autoref{sec:OSC_num_res} provides a condition on physical parameters to meet \eqref{eqn:lambdas_desired_with_v0ge1}. Specifically, the regime \eqref{eqn:lambdas_desired_with_v0ge1} can be achieved when the following inequality holds:

\begin{equation}
\pqtaggr<1,
\quad \mbox{with} \quad 
\pqtaggr 
\quad \mbox{defined as} \quad 
\eqref{eqn:Pi_a_def}
\quad \mbox{when} \quad
a=-1/3
\quad \mbox{and} \quad
b=2/3.
\label{eqn:Pi_a<1}
\end{equation}

\noindent Therefore, we expect the solutions of the LPMF PBM\footref{foot:1LPMFPBM} \eqref{eqn:PBE_latex}-\eqref{eqn:Sigma_m,w} to be reproduced by the r-LPMF PBM\footref{foot:5rLPMFPBM} \eqref{eqn:PBE_latex_approx}-\eqref{eqn:ODE_M,Wx_PBE_latex_no_IT_x=k(1-b)} when the parameters from \autoref{tab:p_exp_def} satisfy \eqref{eqn:Pi_a<1}.

\begin{figure}[!hp]
	\centering
	\includegraphics[scale=1.325]{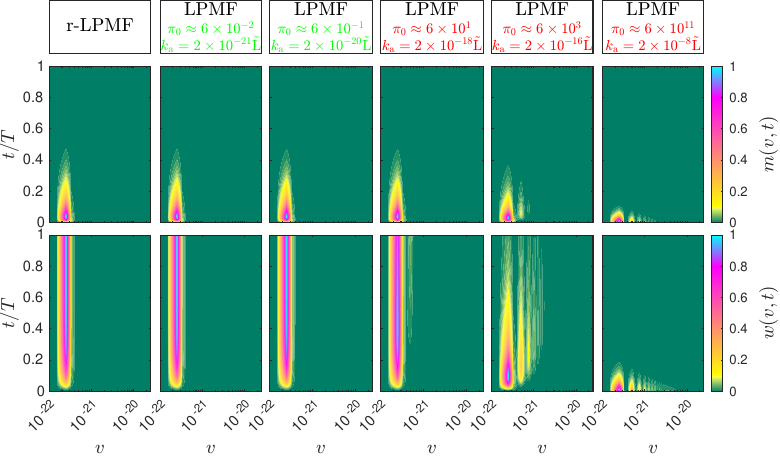}
	\caption{Solutions $m(v,t)$ and $w(v,t)$ of the r-LPMF PBM \eqref{eqn:PBE_latex_approx}-\eqref{eqn:ODE_M,Wx_PBE_latex_no_IT_x=k(1-b)} (left column) and the LPMF PBM\eqref{eqn:PBE_latex}-\eqref{eqn:Sigma_m,w}, with coefficients $\lambda(\theta,\tilde{p})$ (\autoref{tab:PBE_lambdas_def}), $\theta=\{\nu_0,t_0,d_0\}$, $\nu_0=1/d_0=1 \, \LITREunits$, $t_0=1 \, \SECONDunits$ and $\tilde{p}=\tilde{p}_{\mathrm{exp}}$ (\autoref{tab:p_exp_def}). The values of $\kaggr$ are chosen in such a way that both scenarios, $(i)$ $\pqtaggr$ \eqref{eqn:Pi_a_def} $<1$ ($\kaggr$ and $\pqtaggr$ are shown in green) and $(ii)$ $\pqtaggr \ge 1$ ($\kaggr$ and $\pqtaggr$ in red), are present. The tested values of $\kaggr$ are reported in $\AGGRRATEunits := \LITREunits^{1/3} \, \SECONDunits^{-1}$. The shown $y=m,w$ are scaled as $y(v,t) \gets [ y(v,t) - \min_{v,t} y(v,t) ] / [ \max_{v,t} y(v,t) - \min_{v,t} y(v,t) ]$, whereas the time $t$ is normalized to $T=10^6$. Computational settings and parameter values are detailed in \autoref{tab:settings_feas_Pia<1}.}
	\label{fig:comp_sol_rLPMF-LPMF}
	\end{figure}

	First, we verify such a statement for the parameters $\tilde{p}=\tilde{p}_{\mathrm{exp}}$ given in \autoref{tab:p_exp_def} and a set of the aggregation rate $\kaggr$'s values. Since $\pqtaggr \propto \kaggr$ (see \eqref{eqn:Pi_a_def} and \autoref{tab:PBE_lambdas_def}), decreasing values of $\kaggr$ ultimately will lead to $\pqtaggr<1$, i.e. a slow aggregation, while increasing values of $\kaggr$ will give $\pqtaggr \ge 1$, i.e. fast aggregation. In agreement, \autoref{fig:comp_sol_rLPMF-LPMF} shows that the solutions $m(v,t)$ and $w(v,t)$ of the r-LPMF PBM\footref{foot:5rLPMFPBM} (first column from the left) are indistinguishable from the solutions of the LPMF PBM\footref{foot:1LPMFPBM} with aggregation rates $\kaggr$ leading to $\pqtaggr<1$ ($\kaggr$ and $\pqtaggr$ in green). On the other hand, a difference is appreciable for larger $\kaggr$ giving $\pqtaggr \ge 1$ ($\kaggr$ and $\pqtaggr$ in red). 

	
		
	In order to quantify the deviation between compared solutions, we define the metric
	
\begin{equation}
e_y(t) :=
\frac{ \max_v 
| y_{\mbox{\footnotesize{LPMF}}}(v,t)	
- y_{\mbox{\footnotesize{r-LPMF}}}(v,t) | }
{ \max_v | y_{\mbox{\footnotesize{r-LPMF}}}(v,t) | },
\quad y=m,w,
\quad \forall t \in \mathbb{R}^+,
\label{eqn:metric_comp_rLPMF-LPMF}
\end{equation}	 

\noindent where $y_{\mbox{\footnotesize{LPMF}}}, y_{\mbox{\footnotesize{r-LPMF}}}$ correspond to solutions either $m$ or $w$ of the models LPMF \eqref{eqn:PBE_latex}-\eqref{eqn:Sigma_m,w} and r-LPMF \eqref{eqn:PBE_latex_approx}-\eqref{eqn:ODE_M,Wx_PBE_latex_no_IT_x=k(1-b)} respectively.

\begin{figure}[!hp]
	\centering
	\begin{subfigure}[h]{.495\linewidth}
	\centering
	\includegraphics[scale=0.65]{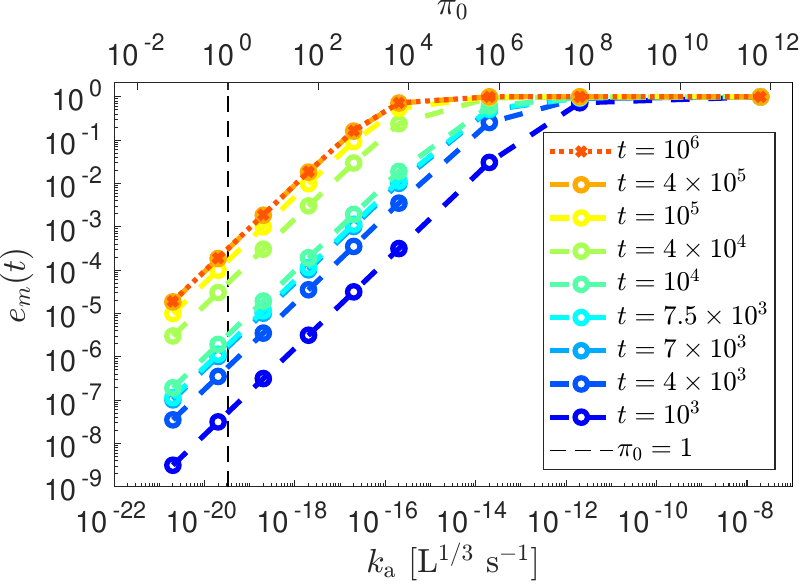}
	\end{subfigure}	
	\begin{subfigure}[h]{.495\linewidth}
	\centering
	\includegraphics[scale=0.65]{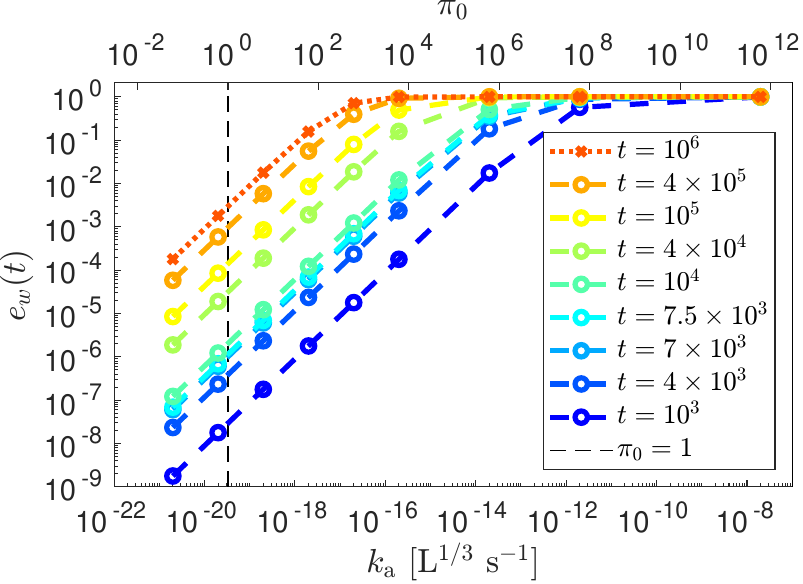}
	\end{subfigure}	
	\caption{Dependence of the metric $e_y(t)$ \eqref{eqn:metric_comp_rLPMF-LPMF}, for $y=m$ (left) and $y=w$ (right), on the aggregation rate $\kaggr$ (\autoref{tab:p_exp_def}) and, thus, $\pqtaggr$ \eqref{eqn:Pi_a_def} $\propto \kaggr$ (see \autoref{tab:PBE_lambdas_def}). Values of $\kaggr$ leading to $\pqtaggr<1$ ensure $e_m(t),e_w(t) \lesssim 10^{-3}$ for all tested times $t$. The plots show $\kaggr$ in $\LITREunits^{1/3}$ $\SECONDunits^{-1}$, while the remaining physical parameters are provided in \autoref{tab:p_exp_def}. Computational settings and parameter values are reported in \autoref{tab:settings_feas_Pia<1}.}
	\label{fig:metric_comp_rLPMF-LPMF}
	\end{figure}

	\autoref{fig:metric_comp_rLPMF-LPMF} demonstrates that the relative approximation errors $e_m,e_w$ \eqref{eqn:metric_comp_rLPMF-LPMF} are reduced when aggregation rates $\kaggr$ and, thus, $\pqtaggr \propto \kaggr$ decrease. In particular, values of $\kaggr$ corresponding to $\pqtaggr<1$ ensure $e_m(t),e_w(t) \lesssim 10^{-3}$ for all tested times $t$. We remark that the chosen range of simulated times $t$ allows for monitoring the dynamical processes accounted by the r-LPMF PBM\footref{foot:5rLPMFPBM} up to the point when solutions $m$ and $w$ reach stationary states (see first column from the left in \autoref{fig:comp_sol_rLPMF-LPMF}). Computational settings and parameter values are reported in \autoref{tab:settings_feas_Pia<1}. 
				
\clearpage

	\begin{table}[!ht]
	\centering
	\begin{tabular}{c c c}
		
	\hline
	& \textbf{LPMF PBM\footref{foot:1LPMFPBM}} & \textbf{r-LPMF PBM\footref{foot:5rLPMFPBM}} \\
	\hline
		
	\rowcolor[HTML]{EFEFEF}
	\textbf{PBE} & 
	\eqref{eqn:PBE_latex}-\eqref{eqn:phase_tr_rate} &
	\eqref{eqn:PBE_latex_approx}-\eqref{eqn:phase_tr_rate_approx} \\
		
	\textbf{ODE} & 
	\eqref{eqn:V_mat_pol2}-\eqref{eqn:Sigma_m,w} &   	\eqref{eqn:Psi_Vpol2_approx}-\eqref{eqn:ODE_M,Wx_PBE_latex_no_IT_x=k(1-b)}\\
		
	\rowcolor[HTML]{EFEFEF}	
	\textbf{Coefficients} $\lambda$
	&
	\multicolumn{2}{c}{$\lambda=\lambda(\theta,\tilde{p})$ (\autoref{tab:PBE_lambdas_def}), $\theta=\{\nu_0,t_0,d_0\}$}
	\\
	
	\rowcolor[HTML]{EFEFEF}	
	&
	\multicolumn{2}{c}{$\nu_0=1/d_0=1 \, \LITREunits$, $t_0=1 \, \SECONDunits$}
	\\
	
	\textbf{Parameters}  $\tilde{p}$
	& 
	\multicolumn{2}{c}{Figures \ref{fig:comp_sol_rLPMF-LPMF}-\ref{fig:metric_comp_rLPMF-LPMF}: $\tilde{p}=\tilde{p}_{\mathrm{exp}}$ (\autoref{tab:p_exp_def})}
	\\	
	&
	\multicolumn{2}{c}{Figures \ref{fig:sol_rLPMF-LPMF_Kd}-\ref{fig:err_rLPMF-LPMF_Kd}: $\tilde{p}$ set as \eqref{eqn:feas_param_exp1}}
	\\
	& 
	\multicolumn{2}{c}{Figures \ref{fig:sol_rLPMF-LPMF_Ks}-\ref{fig:err_rLPMF-LPMF_Ks}: $\tilde{p}$ set as \eqref{eqn:feas_param_exp2}}
	\\
	& 
	\multicolumn{2}{c}{Figures \ref{fig:sol_rLPMF-LPMF_Vc}-\ref{fig:err_rLPMF-LPMF_Vc}: $\tilde{p}$ set as \eqref{eqn:feas_param_exp3}}
	\\	
	
	\rowcolor[HTML]{EFEFEF}
	\textbf{PBE solver}
	&
	\multicolumn{2}{c}{GMOC \cite{RUSCONI2019106944} with \eqref{eqn:approx_GMOC_Dirac_delta} and $\sigma_0 = v_0 / 10$}
	\\
		
	\textbf{ODE solver}
	&
	\multicolumn{2}{c}{Runge-Kutta (RK4) \cite{Tan2012}}
	\\
	
	\rowcolor[HTML]{EFEFEF}	
	\textbf{Volume Grid $\textbf{v}$}
	&
	\multicolumn{2}{c}{$\textbf{v} = \{ k h \}_{k=0}^N$, $h = V / N$, $V = 10^2 \, v_0$, $N=10^3$}	
	\\

	\textbf{Time Grid $\textbf{t}$}
	&
	\multicolumn{2}{c}{$\textbf{t} = \{ k \tau \}_{k=0}^M$, $\tau = T / M$}
	\\
	&
	\multicolumn{2}{c}{Figures \ref{fig:comp_sol_rLPMF-LPMF}-\ref{fig:metric_comp_rLPMF-LPMF}: $T=10^6$, $M=5 \times 10^5$ }
	\\	
	& 
	\multicolumn{2}{c}{Figures \ref{fig:sol_rLPMF-LPMF_Kd}-\ref{fig:err_rLPMF-LPMF_Kd}: $T=5\times10^5$, $M=5\times10^5$ }
	\\
	& 
	\multicolumn{2}{c}{Figures \ref{fig:sol_rLPMF-LPMF_Ks}-\ref{fig:err_rLPMF-LPMF_Ks}: $T=10^6$, $M=10^6$ }
	\\
	& 
	\multicolumn{2}{c}{Figures \ref{fig:sol_rLPMF-LPMF_Vc}-\ref{fig:err_rLPMF-LPMF_Vc}: $T=10^6$, $M=5 \times 10^5$ }
	\\	
	
	\rowcolor[HTML]{EFEFEF}				
	\textbf{SW/HW} 
	&
	\multicolumn{2}{c}{C++ \textbf{BCAM code}, 64-bit Linux OS, 2.40GHz proc.}
	\\
		
	\hline
		
	\end{tabular}
	\caption{Computational settings and parameter values for integration of the LPMF PBM \eqref{eqn:PBE_latex}-\eqref{eqn:Sigma_m,w} and r-LPMF PBM \eqref{eqn:PBE_latex_approx}-\eqref{eqn:ODE_M,Wx_PBE_latex_no_IT_x=k(1-b)} within the volume domain $[0,V]$ and the time interval $[0,T]$. The resulting solutions and relative errors \eqref{eqn:metric_comp_rLPMF-LPMF} are shown in Figures \ref{fig:comp_sol_rLPMF-LPMF}-\ref{fig:err_rLPMF-LPMF_Vc}. The scaling factors $\theta$ are reported in Litres [$\LITREunits$] and seconds [$\SECONDunits$]. \textbf{SW/HW} stands for Software \& Hardware, while \textbf{BCAM code} refers to the in-house package.}	
	\label{tab:settings_feas_Pia<1}
	\end{table}	
	
	In what follows, we provide further evidence that the inequality $\pqtaggr < 1$ \eqref{eqn:Pi_a<1} can serve as a quantitative criterion for the feasibility of r-LPMF PBM\footref{foot:5rLPMFPBM}. In particular, we verify that the solutions of LPMF PBM\footref{foot:1LPMFPBM} can be reproduced by the r-LPMF PBM\footref{foot:5rLPMFPBM}, when the condition \eqref{eqn:Pi_a<1} is met by parameters $\tilde{p}$ that take values in a broader range than the one proposed in \autoref{tab:p_exp_def}. More specifically, sets of $\kdiff$, $\ksepar$ and $\vcrit$'s values are explored.
			
\begin{figure}[!hp]
	\centering
	\begin{subfigure}[h]{1\linewidth}
	\centering
	\includegraphics[width=16.75cm,keepaspectratio]{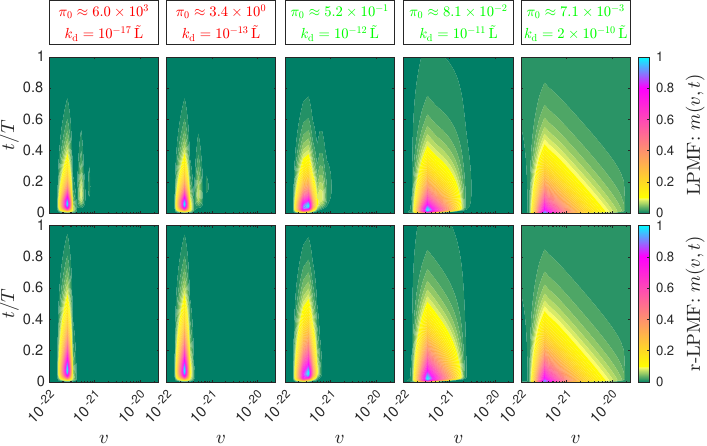}
	\end{subfigure}	
	\begin{subfigure}[h]{1\linewidth}
	\centering
	\includegraphics[width=16.75cm,keepaspectratio]{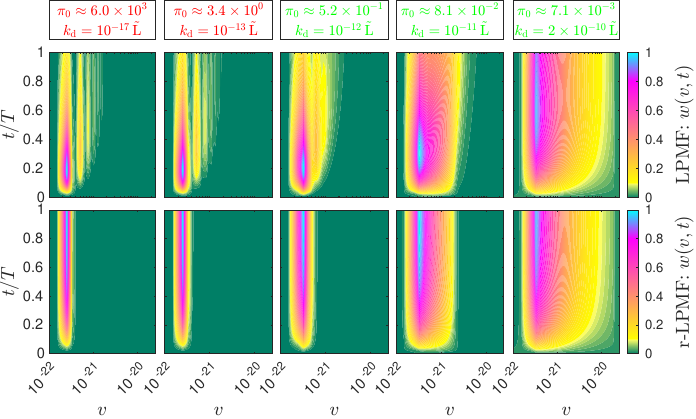}
	\end{subfigure}	
	\caption{Solutions $m(v,t)$ and $w(v,t)$ of the r-LPMF PBM \eqref{eqn:PBE_latex_approx}-\eqref{eqn:ODE_M,Wx_PBE_latex_no_IT_x=k(1-b)} and the LPMF PBM \eqref{eqn:PBE_latex}-\eqref{eqn:Sigma_m,w}, with coefficients $\lambda(\theta,\tilde{p})$ (\autoref{tab:PBE_lambdas_def}), $\theta=\{\nu_0,t_0,d_0\}$, $\nu_0=1/d_0=1 \, \LITREunits$, $t_0=1 \, \SECONDunits$ and $\tilde{p}$ given as \eqref{eqn:feas_param_exp1}. The values of $\kdiff$ are chosen in such a way that both scenarios, $(i)$ $\pqtaggr$ \eqref{eqn:Pi_a_def} $< 1$ ($\kdiff$ and $\pqtaggr$ are shown in green) and $(ii)$ $\pqtaggr \ge 1$ ($\kdiff$ and $\pqtaggr$ in red), are present. The plot provides $\kdiff$ in $\DIFFRATEunits := \LITREunits^{1/3} \, \SECONDunits^{-1}$. The shown $y=m,w$ are scaled as $y(v,t) \gets [ y(v,t) - \min_{v,t} y(v,t) ] / [ \max_{v,t} y(v,t) - \min_{v,t} y(v,t) ]$, whereas the time $t$ is normalized to $T=5 \times 10^5$. Computational settings and parameter values are detailed in \autoref{tab:settings_feas_Pia<1}.}
	\label{fig:sol_rLPMF-LPMF_Kd}
	\end{figure}

	The first experiment considers the transport rate $\kdiff$ ranging in the interval
	
\begin{equation}
\begin{aligned}
& \kdiff \in [10^{-17},2\times10^{-10}] \, \DIFFRATEunits,
\quad \mbox{with the fixed} \quad
\kaggr = 2 \times 10^{-16} \, \AGGRRATEunits
\\
& \mbox{and the remaining} 
\quad \tilde{p} = \tilde{p}_{\mathrm{exp}} \quad
\mbox{as defined in \autoref{tab:p_exp_def}},
\end{aligned}
\label{eqn:feas_param_exp1}
\end{equation}	 

\noindent where $\AGGRRATEunits := \LITREunits^{1/3} \, \SECONDunits^{-1}$. \autoref{fig:sol_rLPMF-LPMF_Kd} compares the LPMF PBM\footref{foot:1LPMFPBM} and r-LPMF PBM\footref{foot:5rLPMFPBM} solutions, indicating in red and green the values of $\kdiff$ leading to $\pqtaggr \ge 1$ and $\pqtaggr < 1$ respectively.

	Since $\pqtaggr \propto \kdiff^{-13/16}$ (see \eqref{eqn:Pi_a_def} and \autoref{tab:PBE_lambdas_def}), it follows that $\pqtaggr \ge 1$, i.e. a regime of fast aggregation, can be achieved by decreasing values of $\kdiff$. Indeed, we observe that the LPMF PBM\footref{foot:1LPMFPBM} and r-LPMF PBM\footref{foot:5rLPMFPBM} solutions compared in \autoref{fig:sol_rLPMF-LPMF_Kd} are quite different when $\pqtaggr \ge 1$ ($\kdiff$ and $\pqtaggr$ in red). 
	
	On the other hand, a larger $\kdiff$ promotes the transport process in such a way that aggregation has a lower importance in the overall dynamics of LPMF PBM\footref{foot:1LPMFPBM} solutions. \autoref{fig:sol_rLPMF-LPMF_Kd} shows that the condition $\pqtaggr<1$ allows detecting regimes of moderate aggregation, when $\tilde{p}$ are set as in \eqref{eqn:feas_param_exp1}.

\begin{figure}[!ht]
	\centering
	\begin{subfigure}[h]{.495\linewidth}
	\centering
	\includegraphics[scale=0.65]{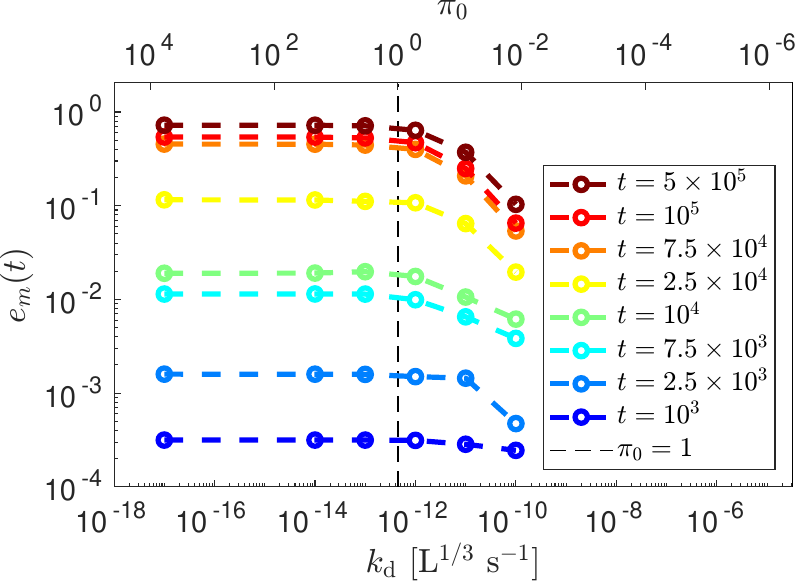}
	\end{subfigure}	
	\begin{subfigure}[h]{.495\linewidth}
	\centering
	\includegraphics[scale=0.65]{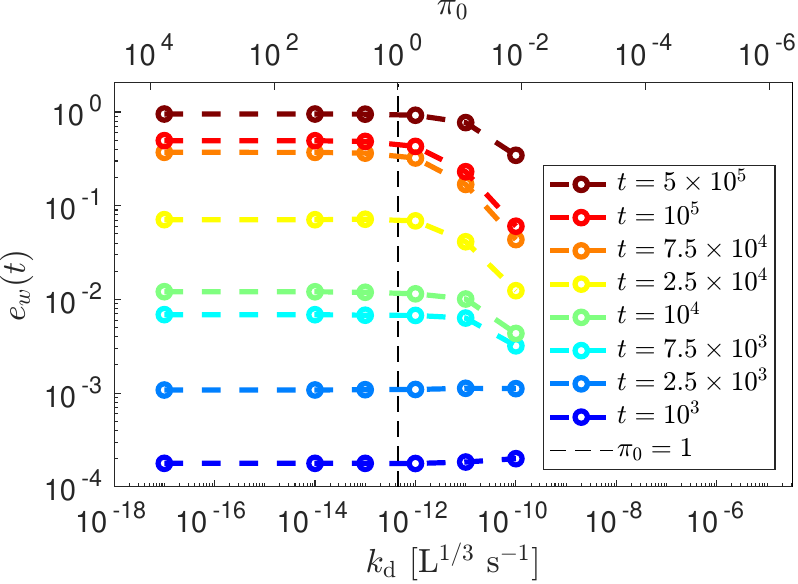}
	\end{subfigure}	
	\caption{Metric $e_y(t)$ \eqref{eqn:metric_comp_rLPMF-LPMF}, with $y=m$ (left) and $y=w$ (right), versus transport rate $\kdiff$. The quantity $\pqtaggr$ \eqref{eqn:Pi_a_def} decreases with increasing $\kdiff$, i.e. $\pqtaggr \propto \kdiff^{-13/16}$ (see \eqref{eqn:Pi_a_def} and \autoref{tab:PBE_lambdas_def}). The errors $e_m,e_w$ can be reduced by enlarging $\kdiff$ to meet $\pqtaggr<1$. The plots provide $\kdiff$ in $\LITREunits^{1/3}$ $\SECONDunits^{-1}$, while the remaining physical parameters are presented in \eqref{eqn:feas_param_exp1}. Computational settings and parameter values are reported in \autoref{tab:settings_feas_Pia<1}.}
	\label{fig:err_rLPMF-LPMF_Kd}
	\end{figure}

	These observations can be quantified by the relative errors \eqref{eqn:metric_comp_rLPMF-LPMF} between LPMF PBM and r-LPMF PBM solutions compared in \autoref{fig:sol_rLPMF-LPMF_Kd}. In particular, \autoref{fig:err_rLPMF-LPMF_Kd} shows that the difference \eqref{eqn:metric_comp_rLPMF-LPMF} can be decreased by enlarging $\kdiff$ to meet $\pqtaggr<1$ \eqref{eqn:Pi_a<1}. The approximation errors span the range of magnitudes from $10^{-4}$ to $10^{0}$ when polymerization time increases from $\approx 17$ min to $\approx 139$ hours.
	
\clearpage		
		
	Next, we consider the nucleation rate $\ksepar$ assuming values in the interval
	
\begin{equation}
\begin{aligned}
& \ksepar \in [2.5 \times 10^{-17}, 2.5 \times 10^{-5}] 
\, \LITREunits \, \SECONDunits^{-1},
\quad \mbox{with the fixed} \quad
\kaggr = 2 \times 10^{-16} 
\, \LITREunits^{1/3} \, \SECONDunits^{-1},
\\
& \mbox{and the remaining} 
\quad \tilde{p} = \tilde{p}_{\mathrm{exp}} \quad
\mbox{as defined in \autoref{tab:p_exp_def}}.
\end{aligned}
\label{eqn:feas_param_exp2}
\end{equation}	 
	
\begin{figure}[!hp]
	\centering
	\begin{subfigure}[h]{1\linewidth}
	\centering
	\includegraphics[width=13.5cm,keepaspectratio]{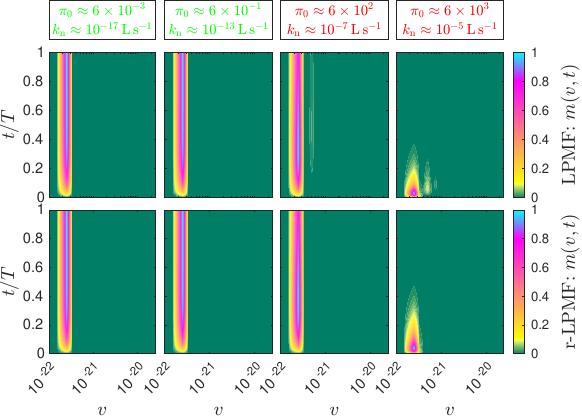}
	\end{subfigure}	
	\begin{subfigure}[h]{1\linewidth}
	\centering
	\includegraphics[width=13.5cm,keepaspectratio]{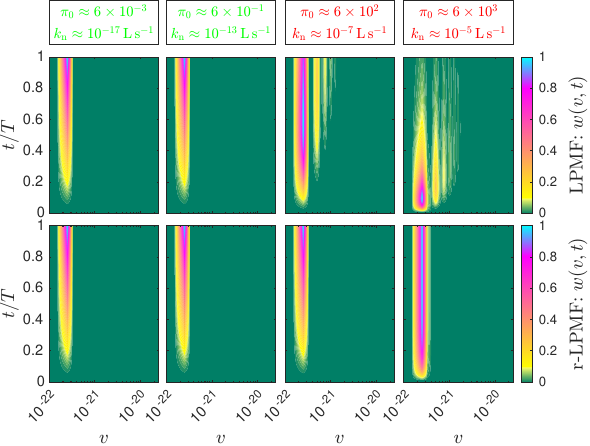}
	\end{subfigure}	
	\caption{Solutions $m(v,t)$ and $w(v,t)$ of the r-LPMF PBM \eqref{eqn:PBE_latex_approx}-\eqref{eqn:ODE_M,Wx_PBE_latex_no_IT_x=k(1-b)} and the LPMF PBM \eqref{eqn:PBE_latex}-\eqref{eqn:Sigma_m,w}, with coefficients $\lambda(\theta,\tilde{p})$ (\autoref{tab:PBE_lambdas_def}), $\theta=\{\nu_0,t_0,d_0\}$, $\nu_0=1/d_0=1 \, \LITREunits$, $t_0=1 \, \SECONDunits$ and $\tilde{p}$ given as \eqref{eqn:feas_param_exp2}. The values of $\ksepar$ are chosen in such a way that both scenarios, $(i)$ $\pqtaggr$ \eqref{eqn:Pi_a_def} $< 1$ ($\ksepar$ and $\pqtaggr$ are shown in green) and $(ii)$ $\pqtaggr \ge 1$ ($\ksepar$ and $\pqtaggr$ in red), are present. The shown $y=m,w$ are scaled as $y(v,t) \gets [ y(v,t) - \min_{v,t} y(v,t) ] / [ \max_{v,t} y(v,t) - \min_{v,t} y(v,t) ]$, whereas the time $t$ is normalized to $T=10^6$. Computational settings and parameter values are detailed in \autoref{tab:settings_feas_Pia<1}.}
	\label{fig:sol_rLPMF-LPMF_Ks}
	\end{figure}

\noindent \autoref{fig:sol_rLPMF-LPMF_Ks} compares solutions of the LPMF PBM\footref{foot:1LPMFPBM} and r-LPMF PBM\footref{foot:5rLPMFPBM} corresponding to a set of physical parameters chosen as in \eqref{eqn:feas_param_exp2}. The values of $\ksepar$ leading to $\pqtaggr \ge 1$ (in red) result in the significant deviation of the LPMF PBM\footref{foot:1LPMFPBM} solutions from the corresponding r-LPMF PBM\footref{foot:5rLPMFPBM} solutions. On the contrary, when $\ksepar$ satisfies $\pqtaggr < 1$ (in green), it is not possible to spot any visual difference between LPMF PBM\footref{foot:1LPMFPBM} and r-LPMF PBM\footref{foot:5rLPMFPBM} solutions. This confirms that the inequality $\pqtaggr<1$ is a reliable threshold for detecting regimes of slow aggregation.

\begin{figure}[!hp]
	\centering
	\begin{subfigure}[h]{.495\linewidth}
	\centering
	\includegraphics[scale=0.675]{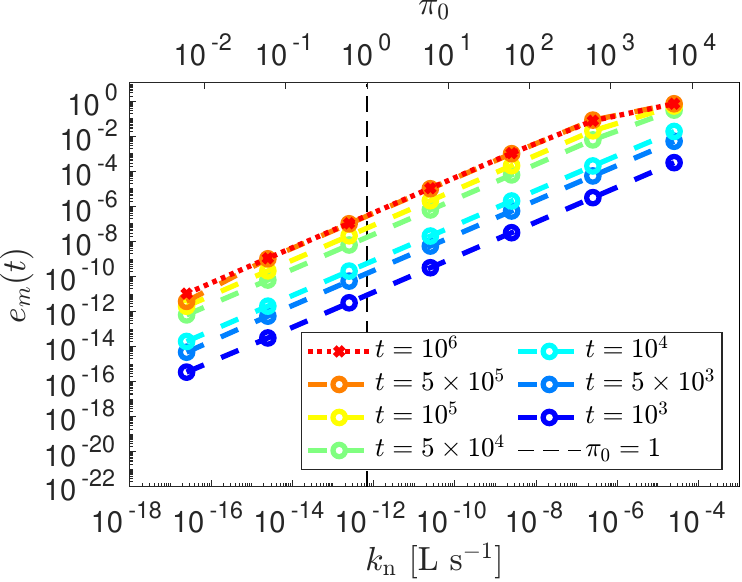}
	\end{subfigure}	
	\begin{subfigure}[h]{.495\linewidth}
	\centering
	\includegraphics[scale=0.675]{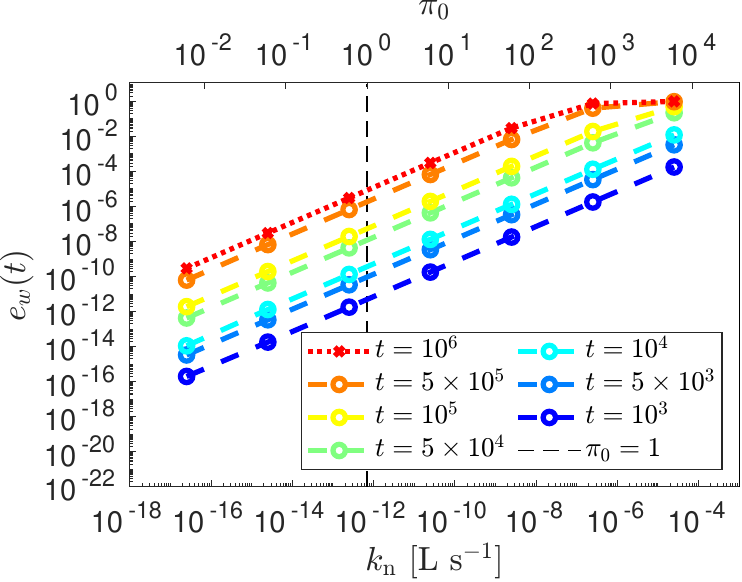}
	\end{subfigure}	
	\caption{Metric $e_y(t)$ \eqref{eqn:metric_comp_rLPMF-LPMF}, with $y=m$ (left) and $y=w$ (right), versus nucleation rate $\ksepar$. The dimensionless quantity $\pqtaggr$ \eqref{eqn:Pi_a_def} is directly proportional to $\ksepar$, i.e. $\pqtaggr \propto \ksepar^{1/2}$ (see \eqref{eqn:Pi_a_def} and \autoref{tab:PBE_lambdas_def}). The plots provide $\ksepar$ in $\LITREunits$ $\SECONDunits^{-1}$, while the remaining physical parameters are set in \eqref{eqn:feas_param_exp2}. Computational settings and parameter values are reported in \autoref{tab:settings_feas_Pia<1}.}
	\label{fig:err_rLPMF-LPMF_Ks}
	\end{figure}

	\autoref{fig:err_rLPMF-LPMF_Ks} illustrates how the relative errors \eqref{eqn:metric_comp_rLPMF-LPMF} vary with respect to the nucleation rate $\ksepar$ \eqref{eqn:feas_param_exp2}. In particular, the difference between LPMF PBM\footref{foot:1LPMFPBM} and r-LPMF PBM\footref{foot:5rLPMFPBM} solutions decreases with reducing $\ksepar$ and, consequently, $\pqtaggr$, since $\pqtaggr \propto \ksepar^{1/2}$ (see \eqref{eqn:Pi_a_def} and \autoref{tab:PBE_lambdas_def}).
	
	Finally, we perform an experiment by assuming the nucleation size $\vcrit$ ranging in
				
\begin{equation}
\begin{aligned}
& \vcrit \in [2.5 \times 10^{-22}, 2.5 \times 10^{-16}] 
\, \LITREunits,
\quad \mbox{with the fixed} \quad
\kaggr = 2 \times 10^{-16} 
\, \LITREunits^{1/3} \, \SECONDunits^{-1},
\\
& \mbox{and the remaining} 
\quad \tilde{p} = \tilde{p}_{\mathrm{exp}} \quad
\mbox{as defined in \autoref{tab:p_exp_def}}.
\end{aligned}
\label{eqn:feas_param_exp3}
\end{equation}	 

\begin{figure}[!hp]
	\centering
	\begin{subfigure}[h]{1\linewidth}
	\centering
	\includegraphics[width=13.5cm,keepaspectratio]{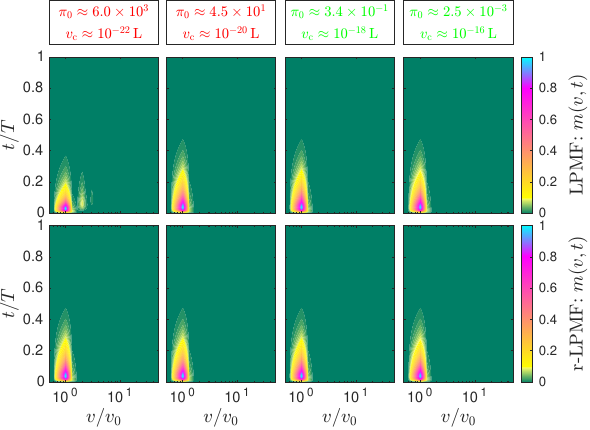}
	\end{subfigure}	
	\begin{subfigure}[h]{1\linewidth}
	\centering
	\includegraphics[width=13.5cm,keepaspectratio]{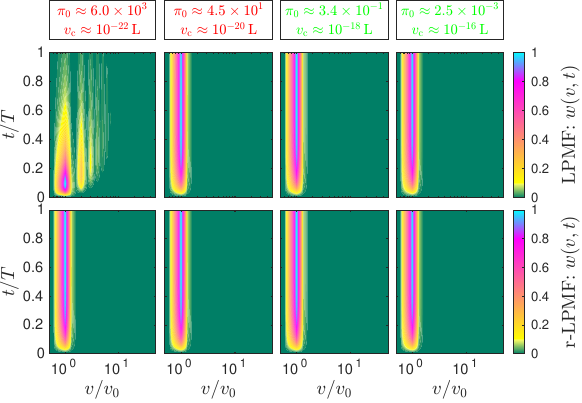}
	\end{subfigure}	
	\caption{Solutions $m(v,t)$ and $w(v,t)$ of the r-LPMF PBM \eqref{eqn:PBE_latex_approx}-\eqref{eqn:ODE_M,Wx_PBE_latex_no_IT_x=k(1-b)} and the LPMF PBM\eqref{eqn:PBE_latex}-\eqref{eqn:Sigma_m,w}, with coefficients $\lambda(\theta,\tilde{p})$ (\autoref{tab:PBE_lambdas_def}), $\theta=\{\nu_0,t_0,d_0\}$, $\nu_0=1/d_0=1 \, \LITREunits$, $t_0=1 \, \SECONDunits$ and $\tilde{p}$ given as \eqref{eqn:feas_param_exp3}. The values of $\vcrit$ are chosen in such a way that both scenarios, $(i)$ $\pqtaggr$ \eqref{eqn:Pi_a_def} $< 1$ ($\vcrit$ and $\pqtaggr$ are shown in green) and $(ii)$ $\pqtaggr \ge 1$ ($\vcrit$ and $\pqtaggr$ in red), are present. The shown $y=m,w$ are scaled as $y(v,t) \gets [ y(v,t) - \min_{v,t} y(v,t) ] / [ \max_{v,t} y(v,t) - \min_{v,t} y(v,t) ]$, whereas the volume $v$ and time $t$ are normalized to $v_0=\vcrit/\nu_0$ and $T=10^6$ respectively. Computational settings and parameter values are detailed in \autoref{tab:settings_feas_Pia<1}.}
	\label{fig:sol_rLPMF-LPMF_Vc}
	\end{figure}

\begin{figure}[!hp]
	\centering
	\begin{subfigure}[h]{.495\linewidth}
	\centering
	\includegraphics[scale=0.675]{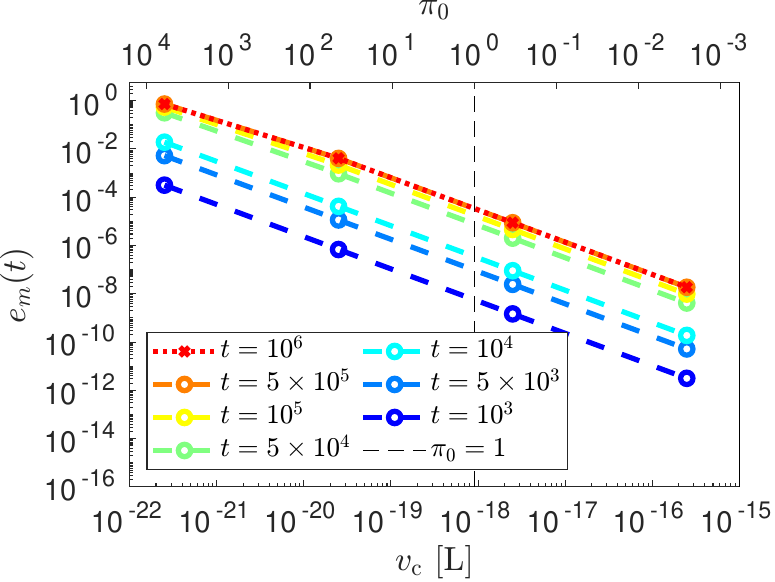}
	\end{subfigure}	
	\begin{subfigure}[h]{.495\linewidth}
	\centering
	\includegraphics[scale=0.675]{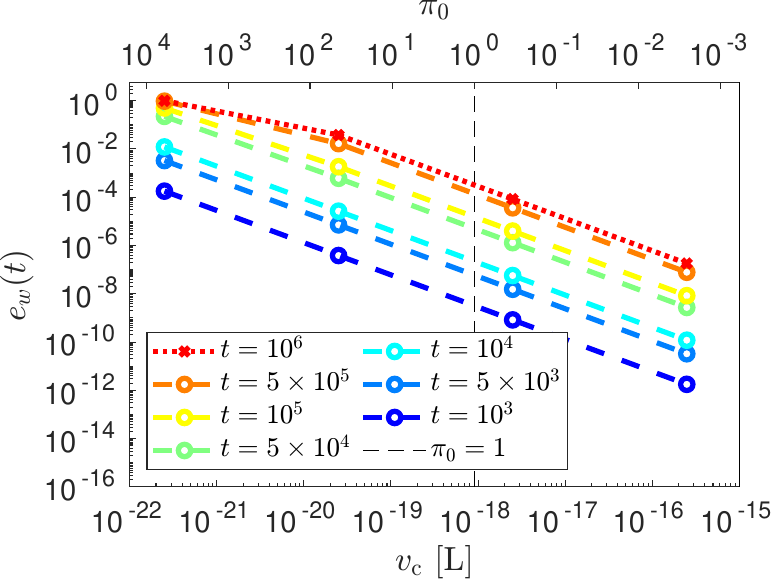}
	\end{subfigure}	
	\caption{Metric $e_y(t)$ \eqref{eqn:metric_comp_rLPMF-LPMF}, with $y=m$ (left) and $y=w$ (right), versus nucleation size $\vcrit$ shown in Litres [$\LITREunits$]. The remaining physical parameters are provided in \eqref{eqn:feas_param_exp3}. The dimensionless quantity $\pqtaggr$ \eqref{eqn:Pi_a_def} is inversely proportional to $\vcrit$, i.e. $\pqtaggr \propto \vcrit^{-17/16}$ (see \eqref{eqn:Pi_a_def} and \autoref{tab:PBE_lambdas_def}). Computational settings and parameter values are reported in \autoref{tab:settings_feas_Pia<1}.}
	\label{fig:err_rLPMF-LPMF_Vc}
	\end{figure}

\noindent \autoref{fig:sol_rLPMF-LPMF_Vc} compares solutions of the LPMF PBM\footref{foot:1LPMFPBM} and r-LPMF PBM\footref{foot:5rLPMFPBM}, showing that they are indistinguishable when $\vcrit$ \eqref{eqn:feas_param_exp3} gives $\pqtaggr<1$ ($\vcrit$ and $\pqtaggr$ in green). In addition, \autoref{fig:err_rLPMF-LPMF_Vc} provides the relative difference \eqref{eqn:metric_comp_rLPMF-LPMF} as a function of $\vcrit$ \eqref{eqn:feas_param_exp3} or $\pqtaggr$ \eqref{eqn:Pi_a_def}. Since $\pqtaggr \propto \vcrit^{-17/16}$ (see \eqref{eqn:Pi_a_def} and \autoref{tab:PBE_lambdas_def}), the approximation error \eqref{eqn:metric_comp_rLPMF-LPMF} is reduced by increasing $\vcrit$ and, thus, decreasing $\pqtaggr$.

	In conclusion, we have validated that the inequality \eqref{eqn:Pi_a<1} can successfully serve as a criterion for applicability of the approximate model r-LPMF PBM\footref{foot:5rLPMFPBM}. Indeed, our experiments indicate that \eqref{eqn:Pi_a<1} can predict ranges of parameters promoting slow aggregation and, thus, the conditions when it is feasible to neglect the integral terms in the LPMF PBM\footref{foot:1LPMFPBM}. We remark that \eqref{eqn:Pi_a<1} can be checked prior to modelling with minimal computational effort.
	 	
\clearpage

\subsection{Computational Efficiency of the r-LPMF PBM}	
\label{sec:efficiency_approx}

	Lastly, we investigate the computational advantages of the r-LPMF PBM\footref{foot:5rLPMFPBM} over the LPMF PBM\footref{foot:1LPMFPBM}. In particular, we compare the computational times required by the scaled r-LPMF PBM\footref{foot:5rLPMFPBM} and LPMF PBM\footref{foot:1LPMFPBM} models to obtain ``a true solution''. Here ``a true solution'' is generated by the original unscaled LPMF PBM\footref{foot:1LPMFPBM} run for a long time with the safe choices of time and volume step sizes. The choices of simulation parameters for such runs were carefully verified. The performed experiments make use of parameters $\tilde{p} = \hat{p}_{\mathrm{exp}}$,
	
\begin{equation}
\hat{p}_{\mathrm{exp}}
:= \tilde{p}_{\mathrm{exp}}
\quad \mbox{(\autoref{tab:p_exp_def})},
\quad \mbox{with} \quad
\kaggr = 2 \times 10^{-20} 
\, \LITREunits^{1/3}
\, \SECONDunits^{-1}.
\label{eqn:hat_p_exp_def}
\end{equation}	
	
\noindent Such a choice of parameters \eqref{eqn:hat_p_exp_def} allows discarding the integral terms in \eqref{eqn:PBE_latex}-\eqref{eqn:Sigma_m,w} without a substantial loss of accuracy (see Figures \ref{fig:comp_sol_rLPMF-LPMF}-\ref{fig:metric_comp_rLPMF-LPMF} in \autoref{sec:accuracy_approx}). 
		
	The first experiment consists in monitoring relative errors and required computational times, while the volume grid size $N$ is increased and the time grid size $M$ is fixed (see \autoref{tab:settings_vol_acc} for values of $N$ and $M$). We take as a reference the solutions of the unscaled LPMF PBM\footref{foot:1LPMFPBM} (see description below \eqref{eqn:comp_models_X}) computed with a volume grid size $N=N_{\mathrm{ref}}$ (see \autoref{tab:settings_vol_acc}). Then, the relative errors between compared solutions are estimated as
	
\begin{equation}
\varepsilon^y_X(N) :=
\frac{ \max_v
| y_X^N(v,T) -
y_Z^{N_{\mathrm{ref}}}(v,T) | }
{ \max_v | y_Z^{N_{\mathrm{ref}}}(v,T) | },
\quad Z = \mbox{\footnotesize{unscaled LPMF PBM\footref{foot:1LPMFPBM}}},
\label{eqn:error_wrt_volume}
\end{equation}

\noindent where $y_X^N(v,T)$ is the solution $y=m,w$ at volume $v$ and time $t=T$ (see \autoref{tab:settings_vol_acc}) of a model $X$ integrated by using a volume grid with size $N$. The models considered in \eqref{eqn:error_wrt_volume} are labeled as

\begin{equation}
X =
\begin{cases}
\mbox{scaled LPMF PBM\footref{foot:1LPMFPBM}:} &
\mbox{\eqref{eqn:PBE_latex}-\eqref{eqn:Sigma_m,w}, $\lambda=\lambda(\tilde{\theta}^{N_x}_{\mathrm{opt}},\tilde{p})$ \eqref{eqn:lambdas_computed_OSC},
$q_1=0$, $\tilde{p}=\hat{p}_{\mathrm{exp}}$ \eqref{eqn:hat_p_exp_def}},
\\
\mbox{scaled r-LPMF PBM\footref{foot:5rLPMFPBM}:} &
\mbox{\eqref{eqn:PBE_latex_approx}-\eqref{eqn:ODE_M,Wx_PBE_latex_no_IT_x=k(1-b)}, $\lambda=\lambda(\tilde{\theta}^{N_x}_{\mathrm{opt}},\tilde{p})$ \eqref{eqn:lambdas_computed_OSC},
$q_1=0$, $\tilde{p}=\hat{p}_{\mathrm{exp}}$ \eqref{eqn:hat_p_exp_def}}.
\end{cases}
\label{eqn:comp_models_X}
\end{equation}

\noindent The unscaled LPMF PBM\footref{foot:1LPMFPBM} model or $Z$ is defined in \eqref{eqn:PBE_latex}-\eqref{eqn:Sigma_m,w}, with $\lambda=\lambda(\theta,\tilde{p})$ (\autoref{tab:PBE_lambdas_def}), $\theta=\theta_{\mathrm{unit}}$ (\autoref{tab:settings_vol_acc}) and $\tilde{p}=\hat{p}_{\mathrm{exp}}$ \eqref{eqn:hat_p_exp_def}. Solutions $y_Z^{N_{\mathrm{ref}}}(v,T)$ in \eqref{eqn:error_wrt_volume} are suitably scaled. 

\begin{figure}[!hp]
	\centering
	\begin{subfigure}[h]{.495\linewidth}
	\centering
	\includegraphics[height=7.5cm,keepaspectratio]{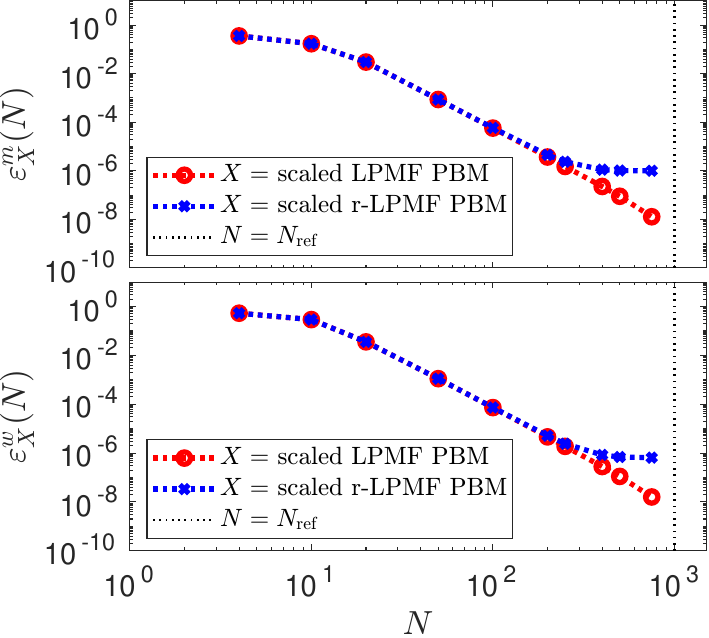}
	\caption{Errors $\varepsilon^{m,w}_X(N)$ \eqref{eqn:error_wrt_volume} vs. volume grid size $N$.}
	\label{fig:errors_vs_vol_rLPMF}
	\end{subfigure}	
	\begin{subfigure}[h]{.495\linewidth}
	\centering
	\includegraphics[height=7.5cm,keepaspectratio]{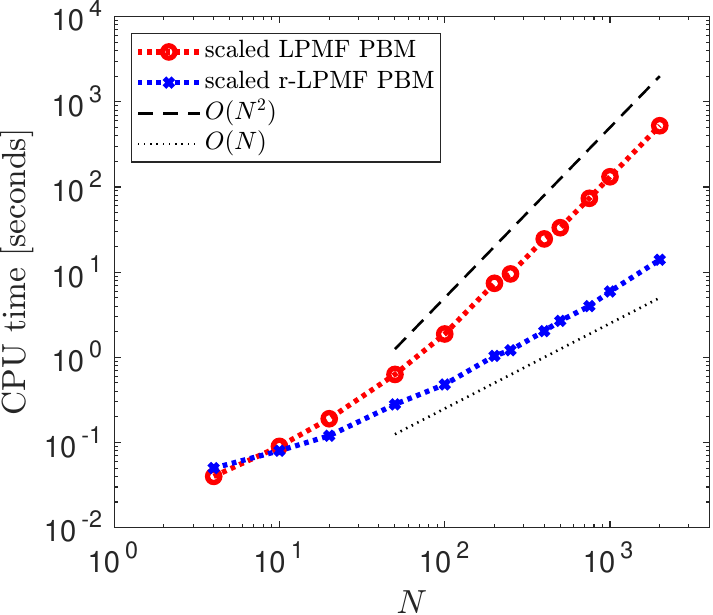}
	\caption{CPU time vs. volume grid size $N$.}
	\label{fig:CPUtime_vs_vol_rLPMF}
	\end{subfigure}	
	\caption{\autoref{fig:errors_vs_vol_rLPMF}: Relative errors $\varepsilon^m_X,\varepsilon^w_X$ \eqref{eqn:error_wrt_volume} of the scaled r-LPMF PBM (blue crosses) and LPMF PBM (red circles) with respect to the unscaled LPMF PBM. Computational settings and parameter values are described in detail in \autoref{tab:settings_vol_acc}. \autoref{fig:CPUtime_vs_vol_rLPMF}: CPU time required for integration of the scaled models in \eqref{eqn:comp_models_X} under conditions summarised in \autoref{tab:settings_vol_acc}.}
	\label{fig:efficiency_vol_rLPMF}
	\end{figure}

	\autoref{fig:efficiency_vol_rLPMF} shows relative errors $\varepsilon^{m,w}_X(N)$ \eqref{eqn:error_wrt_volume} and computational times required by the scaled LPMF PBM\footref{foot:1LPMFPBM} and r-LPMF PBM\footref{foot:5rLPMFPBM}, while \autoref{tab:settings_vol_acc} reports computational settings and parameter values. If $N \lesssim 2 \times 10^2$, the scaled r-LPMF PBM\footref{foot:5rLPMFPBM} and LPMF PBM\footref{foot:1LPMFPBM} show almost identical errors. This confirms the negligible role of aggregation in the simulated process (see \autoref{fig:errors_vs_vol_rLPMF}). In addition, achieving such a level of accuracy requires up to an order of magnitude smaller computational time for r-LPMF PBM\footref{foot:5rLPMFPBM} than it is needed by LPMF PBM\footref{foot:1LPMFPBM} (see \autoref{fig:CPUtime_vs_vol_rLPMF}). Thus, if $N \lesssim 2 \times 10^2$, replacing LPMF PBM\footref{foot:1LPMFPBM} with r-LPMF PBM\footref{foot:5rLPMFPBM} improves the computational efficiency by up to an order of magnitude. Moreover, \autoref{fig:CPUtime_vs_vol_rLPMF} indicates that the computational time required by LPMF PBM\footref{foot:1LPMFPBM} increases quadratically when $N \to \infty$, i.e. it is $O(N^2)$, while the r-LPMF PBM\footref{foot:5rLPMFPBM}'s cost scales as $O(N)$ for large $N$.
	
	For $N > 2 \times 10^2$, the accuracy achieved by the scaled r-LPMF PBM\footref{foot:5rLPMFPBM} approaches the asymptotic value $\approx 10^{-6}$ (blue crosses in \autoref{fig:errors_vs_vol_rLPMF}). This is in agreement with the results plotted in \autoref{fig:metric_comp_rLPMF-LPMF}. Indeed, at the evaluated time $t_0 \, T = 7.2 \times 10^3 \, \SECONDunits$ (see \autoref{tab:settings_vol_acc}), the relative difference between the compared r-LPMF PBM\footref{foot:5rLPMFPBM} and LPMF PBM\footref{foot:1LPMFPBM} is $\approx 10^{-6}$, as quantified in \autoref{fig:metric_comp_rLPMF-LPMF} by the second turquoise circle from the left at time $t_0 \, t \in [7,7.5] \times 10^3 \, \SECONDunits$, where $t_0 = 1 \, \SECONDunits$ (see \autoref{tab:settings_feas_Pia<1}).
	
	\begin{table}[!hp]
	\centering
	\begin{tabular}{c c c c}
		
	\hline
	& \textbf{unscaled} & \textbf{scaled} 
	& \textbf{scaled} \\
	& \textbf{LPMF PBM\footref{foot:1LPMFPBM}} & \textbf{LPMF PBM\footref{foot:1LPMFPBM}} 
	& \textbf{r-LPMF PBM\footref{foot:5rLPMFPBM}} \\
	\hline
		
	\rowcolor[HTML]{EFEFEF}
	\textbf{PBE} &
	\eqref{eqn:PBE_latex}-\eqref{eqn:phase_tr_rate} &
	\eqref{eqn:PBE_latex}-\eqref{eqn:phase_tr_rate} &
	\eqref{eqn:PBE_latex_approx}-\eqref{eqn:phase_tr_rate_approx} \\
		
	\textbf{ODE} & 
	\eqref{eqn:V_mat_pol2}-\eqref{eqn:Sigma_m,w} &
	\eqref{eqn:V_mat_pol2}-\eqref{eqn:Sigma_m,w} &   	\eqref{eqn:Psi_Vpol2_approx}-\eqref{eqn:ODE_M,Wx_PBE_latex_no_IT_x=k(1-b)}\\
		
	\rowcolor[HTML]{EFEFEF}	
	\textbf{Coefficients} $\lambda$ 
	&
	\multicolumn{3}{c}{
	$\lambda=\lambda(\theta,\tilde{p})$ (\autoref{tab:PBE_lambdas_def}),
	$\tilde{p}=\hat{p}_{\mathrm{exp}}$ \eqref{eqn:hat_p_exp_def}}
	\\
		
	\textbf{Factors $\theta$}
	&
	$\theta=\theta_{\mathrm{unit}}=\{\nu_0,t_0,d_0\}$
	&
	\multicolumn{2}{c}{$\theta=\tilde{\theta}^{N_x}_{\mathrm{opt}}$ \eqref{eqn:OSC_optimal_scal_factors}, $q_1=0$}
	\\		
	&
	$ \nu_0 = 1 \, \LITREunits $ 
	&
	\multicolumn{2}{c}{$\nu_0 \approx 4.0 \times 10^{-23} \, \LITREunits$}
	\\		
	& 
	$ t_0 = 1 \, \SECONDunits $
	&
	\multicolumn{2}{c}{$t_0 \approx 1.5 \times 10^6 \, \SECONDunits$}
	\\
	& 
	$ d_0 = 1 \, \LITREunits^{-1} $
	&
	\multicolumn{2}{c}{$d_0 \approx 4.7 \times 10^{45} \, \LITREunits^{-1}$}
	\\
	
	\rowcolor[HTML]{EFEFEF}
	\textbf{PBE solver}
	&
	\multicolumn{3}{c}{GMOC \cite{RUSCONI2019106944} with \eqref{eqn:approx_GMOC_Dirac_delta} and $\sigma_0 = v_0 / 10$}
	\\
		
	\textbf{ODE solver}
	&
	\multicolumn{3}{c}{Runge-Kutta (RK4) \cite{Tan2012}}
	\\
	
	\rowcolor[HTML]{EFEFEF}
	\textbf{Volume Grid $\textbf{v}$}
	&
	\multicolumn{3}{c}{$\textbf{v} = \{ k h \}_{k=0}^N$, $h = V/N$, $V=2v_0$}
	\\

	\textbf{Grid Sizes $N$ and $N_{\mathrm{ref}}$}
	&
	\multicolumn{3}{c}{$N \in 2\mathbb{N} \cap [4, 2 \times 10^3]$, $N_{\mathrm{ref}} = 10^3$}
	\\ 
		
	\rowcolor[HTML]{EFEFEF}
	\textbf{Time Grid $\textbf{t}$}
	&
	\multicolumn{3}{c}{$\textbf{t} = \{ k \tau \}_{k=0}^M$, $\tau =T/M$, $M = 10^3$}
	\\ 

	\textbf{Evaluated time $t=T$}
	&
	\multicolumn{3}{c}{$T = (7.2 \times 10^3 \, \SECONDunits)/t_0$}
	\\	
	
	\rowcolor[HTML]{EFEFEF}
	\textbf{SW/HW} 
	&
	\multicolumn{3}{c}{C++ \textbf{BCAM code}, 64-bit Linux OS, 2.40GHz proc.}
	\\
		
	\hline
		
	\end{tabular}
	\caption{Computational settings and parameter values for integration of the LPMF PBM \eqref{eqn:PBE_latex}-\eqref{eqn:Sigma_m,w} and r-LPMF PBM \eqref{eqn:PBE_latex_approx}-\eqref{eqn:ODE_M,Wx_PBE_latex_no_IT_x=k(1-b)} within the volume domain $[0,V]$ and the time interval $[0,T]$. The symbols $\LITREunits$ and $\SECONDunits$ stand for Litre and second respectively, while $2\mathbb{N}$ labels the set of even natural numbers. Relative errors \eqref{eqn:error_wrt_volume} and computational times are compared in \autoref{fig:efficiency_vol_rLPMF}. \textbf{SW/HW} stands for Software \& Hardware, while \textbf{BCAM code} refers to the in-house package.}	
	\label{tab:settings_vol_acc}
	\end{table}	

\clearpage

	Next, we compare numerical solutions of the scaled r-LPMF PBM\footref{foot:5rLPMFPBM} and LPMF PBM\footref{foot:1LPMFPBM} obtained with the volume grid size $N$ fixed as specified in \autoref{tab:settings_time_acc} and the time grid size $M$ increasing up to the reference value $M=M_{\mathrm{ref}}$ reported in \autoref{tab:settings_time_acc}. We take as a reference the solutions of the unscaled LPMF PBM\footref{foot:1LPMFPBM} (see description below \eqref{eqn:comp_models_X}) computed with such a time grid size $M=M_{\mathrm{ref}}$. The relative errors between compared solutions are defined as

\begin{equation}
\xi^y_X(M) :=
\frac{ \max_v
| y_X^M(v,T) -
y_Z^{M_{\mathrm{ref}}}(v,T) | }
{ \max_v | y_Z^{M_{\mathrm{ref}}}(v,T) | },
\quad
Z = \mbox{\footnotesize{unscaled LPMF PBM\footref{foot:1LPMFPBM}}},
\label{eqn:error_wrt_time}
\end{equation}

\noindent where $y_X^M(v,T)$ denotes the solution $y=m,w$ at volume $v$ and time $t=T$ (see \autoref{tab:settings_time_acc}) of a model $X$ \eqref{eqn:comp_models_X} integrated by using a time grid of size $M$. As in \eqref{eqn:error_wrt_volume}, the solutions $y_Z^{M_{\mathrm{ref}}}(v,T)$ are suitably scaled.

\begin{figure}[!hp]
	\centering
	\begin{subfigure}[h]{.495\linewidth}
	\centering
	\includegraphics[height=7.5cm,keepaspectratio]{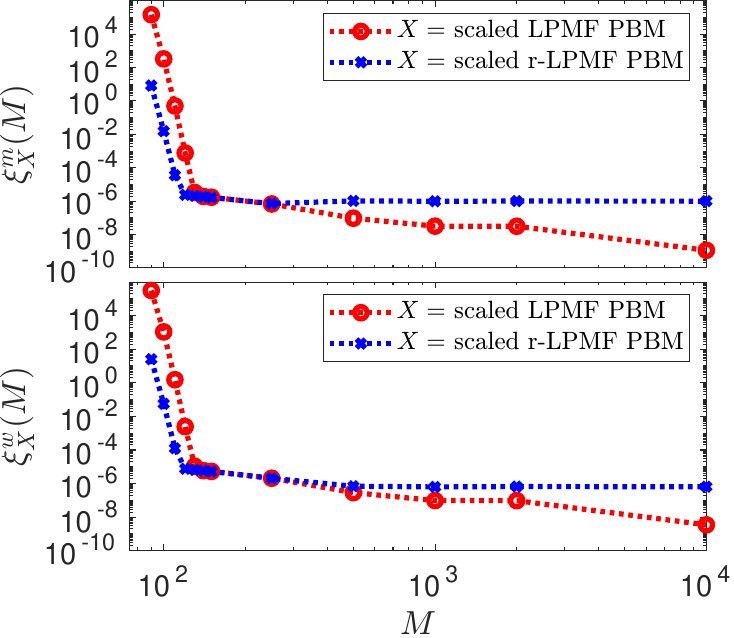}
	\caption{Errors $\xi^{m,w}_X(M)$ \eqref{eqn:error_wrt_time} vs. time grid size $M$.}
	\label{fig:errors_vs_time_rLPMF}
	\end{subfigure}	
	\begin{subfigure}[h]{.495\linewidth}
	\centering
	\includegraphics[height=7.5cm,keepaspectratio]{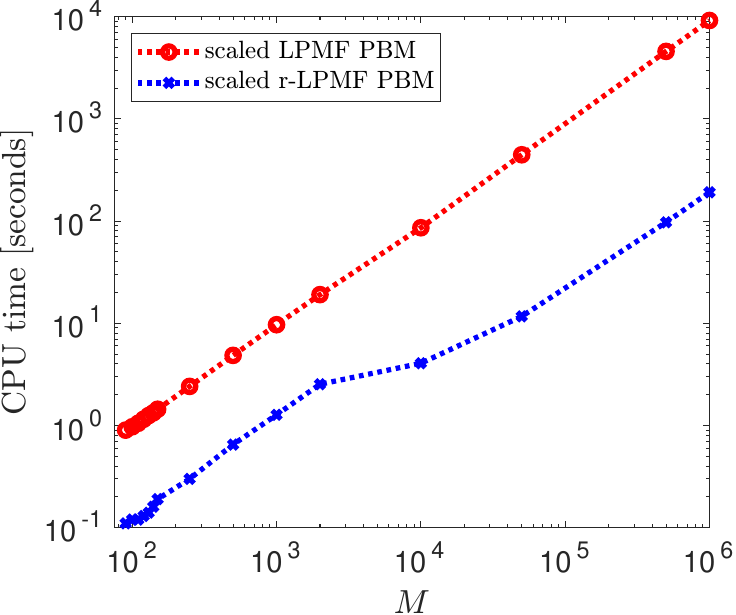}
	\caption{CPU time vs. time grid size $M$.}
	\label{fig:CPUtime_vs_time_rLPMF}
	\end{subfigure}	
	\caption{\autoref{fig:errors_vs_time_rLPMF}: Relative errors $\xi^m_X,\xi^w_X$ \eqref{eqn:error_wrt_time} of the scaled r-LPMF PBM (blue crosses) and LPMF PBM (red circles) with respect to the unscaled LPMF PBM. Computational settings and parameter values are described in detail in \autoref{tab:settings_time_acc}. \autoref{fig:CPUtime_vs_time_rLPMF}: CPU time required for integration of the scaled models in \eqref{eqn:comp_models_X} under conditions summarised in \autoref{tab:settings_time_acc}.}
	\label{fig:efficiency_time_rLPMF}
	\end{figure}

	\autoref{fig:efficiency_time_rLPMF} shows relative errors $\xi^{m,w}_X(M)$ \eqref{eqn:error_wrt_time} and computational times required by the scaled LPMF PBM\footref{foot:1LPMFPBM} and r-LPMF PBM\footref{foot:5rLPMFPBM}. \autoref{tab:settings_time_acc} details computational settings and parameter values.
	
	When $M \lesssim 2.5 \times 10^2$, the relative errors $\xi^{m,w}_{\mbox{\footnotesize{r-LPMF}}}$ are smaller than or equal to $\xi^{m,w}_{\mbox{\footnotesize{LPMF}}}$ (see \autoref{fig:errors_vs_time_rLPMF}). This means that the scaled r-LPMF PBM\footref{foot:5rLPMFPBM} is not only a good approximation of the unscaled LPMF PBM\footref{foot:1LPMFPBM} (for the suggested set of the parameters $\tilde{p}$) but its simplified nature also allows for achieving a better accuracy than it is possible with the scaled LPMF PBM\footref{foot:1LPMFPBM} using the same time/volume grid sizes. Moreover, the computational times required by the scaled r-LPMF PBM\footref{foot:5rLPMFPBM} are up to an order of magnitude smaller than the computational times associated with the scaled LPMF PBM\footref{foot:1LPMFPBM} (see \autoref{fig:CPUtime_vs_time_rLPMF}). For larger values of $M$, i.e. $M \gtrsim 10^4$, the difference between the elapsed CPU times is even more pronounced, reaching up to two orders of magnitude. This is because the measured CPU times also account for time spent on writing data to files, whose contribution in the total CPU time decreases (in percentage) with increasing grid sizes.
	
	Finally, we notice that the errors produced by r-LPMF PBM\footref{foot:5rLPMFPBM} for $M \gtrsim 10^3$ (blue crosses in \autoref{fig:errors_vs_time_rLPMF}) tend to the same asymptotic value $\approx 10^{-6}$ that was observed in the previous experiments (see \autoref{fig:errors_vs_vol_rLPMF} and \autoref{fig:metric_comp_rLPMF-LPMF}).  	

	\begin{table}[!hp]
	\centering
	\begin{tabular}{c c c c}
		
	\hline
	& \textbf{unscaled} & \textbf{scaled} 
	& \textbf{scaled} \\
	& \textbf{LPMF PBM\footref{foot:1LPMFPBM}} & \textbf{LPMF PBM\footref{foot:1LPMFPBM}} 
	& \textbf{r-LPMF PBM\footref{foot:5rLPMFPBM}\footref{foot:1LPMFPBM}} \\
	\hline
		
	\rowcolor[HTML]{EFEFEF}
	\textbf{PBE} &
	\eqref{eqn:PBE_latex}-\eqref{eqn:phase_tr_rate} &
	\eqref{eqn:PBE_latex}-\eqref{eqn:phase_tr_rate} &
	\eqref{eqn:PBE_latex_approx}-\eqref{eqn:phase_tr_rate_approx} \\
		
	\textbf{ODE} & 
	\eqref{eqn:V_mat_pol2}-\eqref{eqn:Sigma_m,w} &
	\eqref{eqn:V_mat_pol2}-\eqref{eqn:Sigma_m,w} &   	\eqref{eqn:Psi_Vpol2_approx}-\eqref{eqn:ODE_M,Wx_PBE_latex_no_IT_x=k(1-b)}\\
		
	\rowcolor[HTML]{EFEFEF}	
	\textbf{Coefficients} $\lambda$
	&
	\multicolumn{3}{c}{
	$\lambda=\lambda(\theta,\tilde{p})$ (\autoref{tab:PBE_lambdas_def}),
	$\tilde{p}=\hat{p}_{\mathrm{exp}}$ \eqref{eqn:hat_p_exp_def}}
	\\
		
	\textbf{Factors $\theta$}
	&
	$\theta=\theta_{\mathrm{unit}}=\{\nu_0,t_0,d_0\}$
	&
	\multicolumn{2}{c}{$\theta=\tilde{\theta}^{N_x}_{\mathrm{opt}}$ \eqref{eqn:OSC_optimal_scal_factors}, $q_1=0$}
	\\		
	&
	$ \nu_0 = 1 \, \LITREunits $
	&
	\multicolumn{2}{c}{$\nu_0 \approx 4.0 \times 10^{-23} \, \LITREunits$}
	\\		
	& 
	$ t_0 = 1 \, \SECONDunits $
	&
	\multicolumn{2}{c}{$t_0 \approx 1.5 \times 10^6 \, \SECONDunits$}
	\\
	& 
	$ d_0 = 1 \, \LITREunits^{-1} $
	&
	\multicolumn{2}{c}{$d_0 \approx 4.7 \times 10^{45} \, \LITREunits^{-1}$}
	\\
	
	\rowcolor[HTML]{EFEFEF}
	\textbf{PBE solver}
	&
	\multicolumn{3}{c}{GMOC \cite{RUSCONI2019106944} with \eqref{eqn:approx_GMOC_Dirac_delta} and $\sigma_0 = v_0 / 10$}
	\\
		
	\textbf{ODE solver}
	&
	\multicolumn{3}{c}{Runge-Kutta (RK4) \cite{Tan2012}}
	\\
	
	\rowcolor[HTML]{EFEFEF}
	\textbf{Volume Grid $\textbf{v}$}
	&
	\multicolumn{3}{c}{$\textbf{v} = \{ k h \}_{k=0}^N$, $h = V/N$, $V=2v_0$, $N=250$}
	\\
	
	\textbf{Time Grid $\textbf{t}$}
	&
	\multicolumn{3}{c}{$\textbf{t} = \{ k \tau \}_{k=0}^M$, $\tau =T/M$}
	\\ 

	\rowcolor[HTML]{EFEFEF}
	\textbf{Grid Sizes $M$ and $M_{\mathrm{ref}}$}
	&
	\multicolumn{3}{c}{$M \in [90,10^6]$, $M_{\mathrm{ref}} = 10^6$}
	\\ 

	\textbf{Evaluated time $t=T$}
	&
	\multicolumn{3}{c}{$T = (7.2\times10^3 \, \SECONDunits)/t_0$}
	\\	
	
	\rowcolor[HTML]{EFEFEF}
	\textbf{SW/HW} 
	&
	\multicolumn{3}{c}{C++ \textbf{BCAM code}, 64-bit Linux OS, 2.40GHz proc.}
	\\
		
	\hline
		
	\end{tabular}
	\caption{Computational settings and parameter values for integration of the LPMF PBM \eqref{eqn:PBE_latex}-\eqref{eqn:Sigma_m,w} and r-LPMF PBM \eqref{eqn:PBE_latex_approx}-\eqref{eqn:ODE_M,Wx_PBE_latex_no_IT_x=k(1-b)} within the volume domain $[0,V]$ and the time interval $[0,T]$. The symbols $\LITREunits$ and $\SECONDunits$ stand for Litre and second respectively. Relative errors \eqref{eqn:error_wrt_time} and computational times are compared in \autoref{fig:efficiency_time_rLPMF}. \textbf{SW/HW} stands for Software \& Hardware, while \textbf{BCAM code} refers to the in-house package.}	
	\label{tab:settings_time_acc}
	\end{table}		

\clearpage

\section{Conclusions} 
\label{sec:concl_disc}
	
	In this paper, we explore the feasibility of the reduction of complexity of the Population Balance Model for Latex Particles Morphology Formation, LPMF PBM\footref{foot:1LPMFPBM}, through disregard of the aggregation terms of the model. The recent nondimensionalization procedure Optimal Scaling (OS) \cite{RUSCONI2019106944} is employed under the mathematically supported constraints to reveal  a quantitative criterion for locating regions of slow and fast aggregation. Such a criterion helps to recognize those physical parameters that suppress aggregation and enable the use of dimensionless population models of reduced complexity, denoted as r-LPMF PBM\footref{foot:5rLPMFPBM} in this work. The models are derived by means of the OS\footref{foot:3OS} with constraints (OSC) and are mathematically justified.

	In comparison with the LPMF PBM\footref{foot:1LPMFPBM}, the new models are lacking integral terms, and all their solely time-dependent components are computed without resorting to preceding or successive solutions of the governing population balance equations. These features result in improved computational performance and also open the opportunity to facilitate numerical integration algorithms for solving effectively the proposed models. 
	
	Comparative performance of the r-LPMF PBM\footref{foot:5rLPMFPBM} and LPMF PBM\footref{foot:1LPMFPBM} with the physically validated parameters values \cite{DDPM_2016} is investigated. The numerical tests reveal that in order to maintain the same level of accuracy, the r-LPMF PBM\footref{foot:5rLPMFPBM} requires up to 2 orders of magnitude less computational effort than it is needed by the LPMF PBM\footref{foot:1LPMFPBM}. 

	The results presented in this paper can be beneficial to particle aggregation studies, e.g., \cite{YU2001392,C5CP07238G}, investigating slow and fast aggregation regimes, as well as estimating the corresponding rate constants. In particular, our scaling procedure allows determining an upper bound for rates leading to slow aggregation. Under such a regime, the reduced complexity of the r-LPMF PBM\footref{foot:5rLPMFPBM} permits a more accessible insight into the system's properties. In this light, our future direction is to explore the possibilities for enhancing the numerical treatment of the introduced r-LPMF PBM\footref{foot:5rLPMFPBM}, taking full advantages of its simplified structure, in order to enable on-the-fly recommendations for technological conditions in the synthesis of new multiphase morphologies.

\section*{Acknowledgements}

	We acknowledge the financial support by the Ministerio de Econom\'ia y Competitividad \\ (MINECO) of the Spanish Government through BCAM Severo Ochoa accreditation SEV-2017-0718, MTM2017-82184-R, PID2019-104927GB-C22 and PID2020-114189RB-I00 grants funded by AEI/FEDER, UE. This work was supported by the BERC 2022-2025 Program and by the ELKARTEK Programme, grants KK-2020/00049, KK-2020/00008, KK-2021/00022 and KK-2021/00064, funded by the Basque Government. This work has been possible thanks to the support of the computing infrastructure of the i2BASQUE academic network, the in-house BCAM-MSLMS group's cluster Monako and the technical and human support provided by IZO-SGI SGIker of UPV/EHU and European funding (ERDF and ESF).				

\appendix

\counterwithin{equation}{section}
\renewcommand{\theequation}{\thesection.\arabic{equation}}

\counterwithin{table}{section}

\counterwithin{theorem}{section}
\renewcommand{\thetheorem}{\thesection.\arabic{theorem}}

\renewcommand{\thesection}{A}
\section{LPMF PBM: Time-Dependent Factors of Rate Functions}
\label{sec:time_dep_factors_rates}

	Here, we characterise the dependency on time of rate functions \changegreen{$\alpha(v,u,t)$} \eqref{eqn:aggr_rate}, $g(v,t)$ \eqref{eqn:growth_rate} and $n(v,t)$ \eqref{eqn:nucl_rate}. In particular, we focus on their time-dependent factors
	
	\begin{equation}
	\tilde{\textbf{p}}(t) := 
	\{ \changegreen{\tilde{\alpha}_0}(t), \TDFdiffusion(t), \TDFpolymeriz(t), \tilde{\eta}_0(t) \},
	\quad \forall t \in \mathbb{R}^+.
	\label{eqn:time-dep_fact_appx}
	\end{equation} 
	
	As shown in \autoref{sec:PBEmodel_latex}, the factors $\tilde{\textbf{p}}(t)$ depend on $\Psi(t)$ \eqref{eqn:Psi}, $\Phi(t)$ \eqref{eqn:Phi} and $V_p(t)$ \eqref{eqn:Vp}. First, it is possible to show that
	
	\begin{equation}
	\Psi(t) \in (0,\bar{\Psi}],
	\quad \forall t \in \mathbb{R}^+,
	\quad \mbox{and} \quad
	\lim_{t \to \infty} \Psi(t) = 0.
	\label{eqn:Psi_charact}	
	\end{equation}
	
\noindent Let us denote

	\begin{equation}
	z(t) := \Psi(t) \, e^{\int_0^t \! F(s) \, ds},
	\quad \mbox{with} \quad
	F(t) := 
	\frac{ \changegreen{\lpoly} \, ( \Psi(t) + \Psi_r ) }
	{ ( \Psi(t) + 1 ) ( V_{\mathrm{pol2}}(t) + \changegreen{\lpolone} ) }, 
	\quad \forall t \in \mathbb{R}^+,
	\end{equation}
	
\noindent where $\Psi(t)$ is a solution of the ODE \eqref{eqn:Psi}. We observe that $z'(t)=0$ and $z(t)=z(0)=\Psi(0)=\bar{\Psi}$, $\forall t \in \mathbb{R}^+$. By using the above definition of $z(t)$, one obtains an implicit relation for $\Psi(t)$:

	\begin{equation}
	\Psi(t) = \bar{\Psi} \, e^{- \int_0^t \! F(s) \, ds}, 
	\quad \forall t \in \mathbb{R}^+.	
	\label{eqn:Psi_implicit_sol}
	\end{equation}		 

\noindent Then, \eqref{eqn:Psi_implicit_sol} implies that $\Psi(t) > 0$, $\forall t \in \mathbb{R}^+$. From \eqref{eqn:V_pol2}, $V_{\mathrm{pol2}}(0) = 0$ and $V'_{\mathrm{pol2}}(t) > 0$, $\forall t \in \mathbb{R}^+$. Thus, $V_{\mathrm{pol2}}(t) \ge 0$ and $F(t) > 0$, $\forall t \in \mathbb{R}^+$. Then, $\Psi(t) \le \bar{\Psi}$, $V'_{\mathrm{pol2}}(t) < \changegreen{\lpoly} \, \bar{\Psi}$, $V_{\mathrm{pol2}}(t) < \changegreen{\lpoly} \, \bar{\Psi} \, t$ and $F(t) > ( \changegreen{\lpoly} \, \Psi_r ) / [ (\bar{\Psi}+1) ( \changegreen{\lpoly} \, \bar{\Psi} \, t + \changegreen{\lpolone} ) ]$, $\forall t \in \mathbb{R}^+$. As a consequence, $\int_0^t \! F(s) \, ds$ diverges to $\infty$, $\Psi(t)$ \eqref{eqn:Psi_implicit_sol} tends to $0$, as $t \to \infty$, and thus \eqref{eqn:Psi_charact} follows. 

	Then, we show that 

	\begin{equation}
	V_p(t) > \changegreen{\lpolone} > 0,
	\quad \forall t \in \mathbb{R}^+.	
	\label{eqn:Vp_charact}
	\end{equation}	 

\noindent \changegreen{By summing \eqref{eqn:V_mat_pol2}, \eqref{eqn:V_cm_pol2} and \eqref{eqn:V_cw_pol2}, one obtains \eqref{eqn:V_pol2}.} \changeorange{In other words, one has}

	\begin{equation}
	\changeorange{ \begin{cases}
	\frac{dV^{\mathrm{mat}}_{\mathrm{pol2}}(t)}{dt}
	+
	\frac{dV^{\mathrm{c_m}}_{\mathrm{pol2}}(t)}{dt}
	+
	\frac{dV^{\mathrm{c_w}}_{\mathrm{pol2}}(t)}{dt}
	& = 	
	\frac{dV_{\mathrm{pol2}}(t)}{dt},
	\quad
	\forall t \in \mathbb{R}^+, \\
	V^{\mathrm{mat}}_{\mathrm{pol2}}(0)
	+	
	V^{\mathrm{c_m}}_{\mathrm{pol2}}(0)
	+	
	V^{\mathrm{c_w}}_{\mathrm{pol2}}(0)
	& = 
	V_{\mathrm{pol2}}(0).
	\end{cases} }
	\label{eqn:proof_sumVpol2}
	\end{equation}	 

\noindent \changeorange{By integrating \eqref{eqn:proof_sumVpol2} over $[0,t]$, it follows}

	\begin{equation}
	\changegreen{
	V^{\mathrm{mat}}_{\mathrm{pol2}}(t)
	+	
	V^{\mathrm{c_m}}_{\mathrm{pol2}}(t)
	+	
	V^{\mathrm{c_w}}_{\mathrm{pol2}}(t)
	=
	V_{\mathrm{pol2}}(t),
	\quad
	\forall t \in \mathbb{R}^+.}
	\label{eqn:sum_V_x_pol2}
	\end{equation}
	
\noindent By plugging \eqref{eqn:sum_V_x_pol2} in \eqref{eqn:Vp}, one obtains

	\begin{equation}
	\changegreen{
	V_p(t) = \, \left( \Psi(t)+1 \right)
	\, \left[ 
	V_{\mathrm{pol2}}(t)	
	+ 
	\lpolone
	\right]
	> \lpolone > 0,
	\quad \forall t \in \mathbb{R}^+.}
	\label{eqn:V_p_alternative}
	\end{equation}
	
\noindent that concludes a proof of \eqref{eqn:Vp_charact}.
		
	Finally, we demonstrate that	
	
	\begin{equation}
	\Phi(t) \in [0,1), \quad \forall t \in \mathbb{R}^+.
	\label{eqn:Phi_charact}
	\end{equation}
	
\noindent Definition \eqref{eqn:Phi} implies that $\Phi(t)$ is non-negative for any $t \in \mathbb{R}^+$ and $\Phi(0)=0$. Then, \eqref{eqn:V_mat_pol2} shows that $dV^{\mathrm{mat}}_{\mathrm{pol2}}(t)/dt > 0$ when $V^{\mathrm{mat}}_{\mathrm{pol2}}(t)=0$. Thus, $V^{\mathrm{mat}}_{\mathrm{pol2}}(t) \ge 0$, $\forall t \in \mathbb{R}^+$. As a consequence

	\begin{equation}
	\Phi(t)  
	\le
	\frac{ V^{\mathrm{mat}}_{\mathrm{pol2}}(t) }
    { ( \Psi(t) + 1 ) 
    ( V^{\mathrm{mat}}_{\mathrm{pol2}}(t) 
    + 
    \changegreen{\lpolone} ) }
    < 1,
    \quad
    \forall t \in \mathbb{R}^+.	
	\end{equation}

	In conclusion, \eqref{eqn:Psi_charact}, \eqref{eqn:Vp_charact} and \eqref{eqn:Phi_charact} provide insight into properties of the time-dependent factors $\tilde{\textbf{p}}(t)$, as summarised in \autoref{tab:factors_charact}. In the following Appendices we prove the important properties of the LPMF PBM\footref{foot:1LPMFPBM} which make its approximation with the r-LPMF PBM\footref{foot:5rLPMFPBM} possible.  		

	\begin{table}[!h]
	\centering
	\begin{tabular}[t]{ll}

	\hline
	\textbf{Sign} & \textbf{Admits Zero-Value} \\
	\hline

	\rowcolor[HTML]{EFEFEF}	
	$\changegreen{\tilde{\alpha}_0}(t) > 0$, $\forall t \in \mathbb{R}^+$ 
	& No: $\changegreen{\tilde{\alpha}_0}(t) > \changegreen{\laggr} > 0$, $\forall t \in \mathbb{R}^+$ \\
		
	$\TDFdiffusion(t) \ge 0$, $\forall t \in \mathbb{R}^+$ & Yes: $\TDFdiffusion(t) = 0 \Leftrightarrow \Phi(t) = 0$, e.g. $t=0$ \\ 

	\rowcolor[HTML]{EFEFEF}	
	$\TDFpolymeriz(t) > 0$, $\forall t \in \mathbb{R}^+$ & No, but $\lim_{t \to \infty} \TDFpolymeriz(t) = 0$ \\
	
	$\tilde{\eta}_0(t) \ge 0$, $\forall t \in \mathbb{R}^+$ & Yes: $\tilde{\eta}_0(t) = 0 \Leftrightarrow \Phi(t) = 0$, e.g. $t=0$ \\ 
	
	\hline			
	\multicolumn{2}{l}{\textbf{Upper Bound}} \\	
	\hline
	
	\rowcolor[HTML]{EFEFEF}	
	\multicolumn{2}{l}{$\tilde{\alpha}_0(t) \le \KAPPAalpha := \laggr \, (\bar{\Psi}+1)^{\nicefrac{14}{3}}<\infty$, $\forall t \in \mathbb{R}^+$} \\
		
	\multicolumn{2}{l}{$\TDFdiffusion(t) < \KAPPAdi := \ldiff \, (\bar{\Psi}+1)^{\nicefrac{2}{3}}<\infty$, $\forall t \in \mathbb{R}^+$} \\
	
	\rowcolor[HTML]{EFEFEF}	
	\multicolumn{2}{l}{$\TDFpolymeriz(t) < \KAPPApi := \lpoly \, \bar{\Psi} / \changegreen{\lpolone}<\infty$, $\forall t \in \mathbb{R}^+$} \\
	
	\multicolumn{2}{l}{$\tilde{\eta}_0(t) < \KAPPAenne := \lnucl<\infty$, $\forall t \in \mathbb{R}^+$} \\

	\hline
			
	\end{tabular}
	\caption{Properties of factors $\tilde{\textbf{p}}(t) = 
	\{ \changegreen{\tilde{\alpha}_0}(t), \TDFdiffusion(t), \TDFpolymeriz(t), \tilde{\eta}_0(t) \}$ of rate functions \eqref{eqn:aggr_rate}-\eqref{eqn:nucl_rate}. The table provides the sign of factors $\tilde{\textbf{p}}$, whenever they admit zero as a possible value and the upper bounds on their time-evolution. We remark that the coefficients $\laggr,\ldiff$ in \eqref{eqn:aggr_rate}-\eqref{eqn:growth_rate} depend on the parameters $a,b$ (see \autoref{tab:PBE_lambdas_def}), however, all the summarised here properties hold for any finite value of $a,b\in\mathbb{R}$ and $\bar{\Psi}>0$.}
	\label{tab:factors_charact}
	\end{table} 

\renewcommand{\thesection}{B}
\section{LPMF PBM: Solutions in $B_0 := \{ (v,t) \in [0,\infty)^2: \, v < v_0, \, v_0 > 0 \}$}	
\label{sec:BoundaryBand_Sol}

\begin{theorem}
If $a \in \mathbb{R}$, $0 < b < 1$, $v_0 > 0$ and $m(v,t)$, $w(v,t)$ are the solutions to \eqref{eqn:PBE_latex}-\eqref{eqn:Sigma_m,w}, then $m(v,t) = w(v,t) = 0$, $\forall v \in [0,v_0)$, $\forall t \in \mathbb{R}^+$.
\label{eqn:boundary_band_sol_PBE}
\end{theorem}
	
\begin{proof} 
We provide the proof for $m(v,t)$ only, but the same argument applies to $w(v,t)$. 

Let us define $M^x_v(t) := \int_0^v \! u^x \, m(u,t) \, du$ for all $v,t \in \mathbb{R}^+$ and $x \in \mathbb{R}$. Then, for $v \in (0,v_0)$, $a \in \mathbb{R}$ and $0 < b < 1$, the evolution equation of $M^0_v(t)$ reads:

	\begin{equation}
	\begin{cases}
	\frac{d}{dt} M^0_v(t)
	& =
	- g(v,t) m(v,t)
	- \changegreen{\mu} \, M^0_v(t)
	- \changegreen{\tilde{\alpha}_0}(t) M^a_v(t) M^0_{\infty}(t)
	- \changegreen{\tilde{\alpha}_0}(t) M^0_v(t) M^a_{\infty}(t)
	+ \\
	& \quad
	+ \frac{\changegreen{\tilde{\alpha}_0}(t)}{2}
	\, \int_0^v \! m(u,t) 
	\, \int_0^{v-u} [ y^a + u^a ] \, m(y,t) \, dy
	\, du,
	\quad \forall v \in (0,v_0),	
	\, \forall t \in \mathbb{R}^+, \\
	M^0_v(0) & = 0, \quad \forall v \in (0,v_0),	
	\end{cases}
	\label{eqn:B1}
	\end{equation}	
	
\noindent \changeorange{with $\mu=\lphtr>0$ (see \autoref{tab:PBE_lambdas_def})} and $g$ given by \eqref{eqn:growth_rate}, such that $g(v,t) \ge 0$, $\forall v,t \in \mathbb{R}^+$ (see \autoref{tab:factors_charact}). Since $\changegreen{\tilde{\alpha}_0}(t)$ and $m(v,t)$ are non-negative for all $v,t \in \mathbb{R}^+$, as shown in \autoref{sec:time_dep_factors_rates} and \cite{RUSCONI2019106944}, we have

	\begin{equation}
	\changegreen{\tilde{\alpha}_0}(t) M^a_v(t) M^0_{\infty}(t)
	+
	\changegreen{\tilde{\alpha}_0}(t) M^0_v(t) M^a_{\infty}(t)	
	\ge 
	2 \changegreen{\tilde{\alpha}_0}(t) M^a_v(t) M^0_v(t),
	\label{eqn:B2}
	\end{equation}
	
\noindent and

	\begin{equation}
	\frac{\changegreen{\tilde{\alpha}_0}(t)}{2}
	\, \int_0^v \! m(u,t) 
	\, \int_0^{v-u} [ y^a + u^a ] \, m(y,t) \, dy
	\, du
	\le 
	\changegreen{\tilde{\alpha}_0}(t) M^a_v(t) M^0_v(t). 
	\label{eqn:B3}
	\end{equation}

\noindent Then from  \eqref{eqn:B1}-\eqref{eqn:B3} and $0 \le M^a_v(t) M^0_v(t) \le \infty$, it follows $dM^0_v(t)/dt \le 0$ and $M^0_v(t) \le M^0_v(0) = 0$, $\forall v \in (0,v_0)$, $\forall t \in \mathbb{R}^+$. Since $M^0_v(t)$ must be non-negative, we conclude that $M^0_v(t) = 0$, $\forall v \in (0,v_0)$, $\forall t \in \mathbb{R}^+$. 

Finally, one has:

	\begin{equation}
	m(v,t) = \frac{d}{dv} \int_0^v \! m(u,t) \, du 
	=
	\frac{d}{dv} M^0_v(t) = 0,
	\quad
	\forall v \in (0,v_0),
	\, \forall t \in \mathbb{R}^+,
	\end{equation}

\noindent and \autoref{eqn:boundary_band_sol_PBE} follows (rem. $m(0,t)=0$, $\forall t \in \mathbb{R}^+$). 
\end{proof}

\renewcommand{\thesection}{C}
\section{LPMF PBM: Finiteness of Distributions Moments}	
\label{sec:finite_moments}

\begin{theorem}
Given $a \in \mathbb{R}$, $0 < b < 1$ and $v_0 > 0$, the moments $M^0(t) := \int_0^{\infty} \! m(v,t) \, dv$ and $W^0(t) := \int_0^{\infty} \! w(v,t) \, dv$ of solutions to \eqref{eqn:PBE_latex}-\eqref{eqn:Sigma_m,w} are finite for all times $t \in \mathbb{R}^+$ and $M^0(t) \le \changegreen{ \KAPPAenne \, t}$, $W^0(t) \le \changegreen{\frac{ \KAPPAenne \, \lphtr }{2} \, t^2}$, $\forall t \in \mathbb{R}^+$. Here, $\KAPPAenne \in (0,\infty)$ is the upper bound of the factor $\tilde{\eta}_0(t)$ provided in \autoref{tab:factors_charact} and $\lphtr \in (0,\infty)$ is defined in \autoref{tab:PBE_lambdas_def}. 
\label{prop:finite_M,W0_PBE_latex}		
\end{theorem}

\begin{proof}
When $a \in \mathbb{R}$, $0 < b < 1$ and $v_0 > 0$, the evolution equations for the moments $M^0(t)$ and $W^0(t)$, $\forall t \in \mathbb{R}^+$, are given by
	
	\begin{equation}
	\begin{cases}
	\frac{dM^0(t)}{dt} 
	& =
	- \lim_{v \to \infty} g^0_m(v,t) 
	+ \tilde{\eta}_0(t) 
	- \changegreen{\mu} \, M^0(t) 
	- \changegreen{\tilde{\alpha}_0}(t)
	\left[ 2 M^a(t) M^0(t) - M^a(t) M^0(t) \right], \\
	\frac{dW^0(t)}{dt}
	& =
	- \lim_{v \to \infty} g^0_w(v,t) 
	+ \changegreen{\mu} \, M^0(t) 
	- \changegreen{\tilde{\alpha}_0}(t)
	\left[ 2 W^a(t) W^0(t) - W^a(t) W^0(t) \right], \\
	M^0(0) & = W^0(0) = 0,
	\end{cases}
	\label{eqn:ODE_M,W0_PBE_latex}
	\end{equation}

\noindent with

	\begin{equation}
	g^x_y(v,t) := 
	( \TDFdiffusion(t) v^{x+b} 
	+ \TDFpolymeriz(t) v^{x+1} )
	\, y(v,t) \ge 0,
	\quad y=m,w,
	\quad \forall x \in \mathbb{R},
	\quad \forall v,t \in \mathbb{R}^+.
	\label{eqn:flux_grow_mom_PBElatex}
	\end{equation}

\noindent For any $t \in \mathbb{R}^+$ such that $0 \le M^a(t) M^0(t) \le \infty$ and $0 \le W^a(t) W^0(t) \le \infty$, we can assure that $2 M^a(t) M^0(t) \ge M^a(t) M^0(t)$ and $2 W^a(t) W^0(t) \ge W^a(t) W^0(t)$. \changeorange{Since $\tilde{\alpha}_0(t), m(v,t)$ and $w(v,t)$ are non-negative for all $v,t \in \mathbb{R}^+$, as shown in \autoref{sec:time_dep_factors_rates} and \cite{RUSCONI2019106944}}, it follows

	\begin{equation}
	\frac{dM^0(t)}{dt} \le \KAPPAenne,
	\quad
	\frac{dW^0(t)}{dt} \le 
	\changegreen{\lphtr} \, M^0(t),
	\quad \forall t \in \mathbb{R}^+,
	\label{eqn:bounds_der_MW0}
	\end{equation}

\noindent with $\KAPPAenne \in (0,\infty)$ being the upper bound of the factor $\tilde{\eta}_0(t)$ provided in \autoref{tab:factors_charact} and \changeorange{$\mu=\lphtr \in (0,\infty)$ \eqref{eqn:phase_tr_rate}. Integrating \eqref{eqn:bounds_der_MW0} over $[0,t]$ and using the initial conditions in \eqref{eqn:ODE_M,W0_PBE_latex},} we get the bounds of $M^0(t)$, $W^0(t)$ stated in \autoref{prop:finite_M,W0_PBE_latex} and this concludes the proof. 
\end{proof}	

\changeorange{
\begin{theorem}
Given $a \le 0$, $0<b<1$, $v_0 > 0$ and any finite $x,y \in \mathbb{R}$ such that $x \le y$, there exist strictly positive quantities $K^x_m,K^x_w<\infty$, dependent on the chosen $x$, but independent of time $t$, such that the moments $M^x(t) := \int_0^{\infty} \! v^x \, m(v,t) \, dv$ and $W^x(t) := \int_0^{\infty} \! v^x \, w(v,t) \, dv$ of solutions to \eqref{eqn:PBE_latex}-\eqref{eqn:Sigma_m,w} satisfy $M^x(t) \le K^x_m + M^y(t)$, $W^x(t) \le K^x_w + W^y(t)$, $\forall t \in [0,T]$, with any fixed $T<\infty$. 
\label{prop:rel_order_mom_latex}		
\end{theorem}

\begin{proof}
	We obtain an upper bound on $M^x(t)$ only, as similar estimates can be made for $W^x(t)$. 

	\autoref{eqn:boundary_band_sol_PBE} implies that $ M^x(t) = \int_{v_0}^{\infty} \! v^x \, m(v,t) \, dv$, $\forall x \in \mathbb{R}$, $\forall t \in \mathbb{R}^+$. Then, if $\bar{v} \in \mathbb{R}$ is chosen such that $\max\{v_0,1\} < \bar{v} < \infty$, it follows ($\forall x \in \mathbb{R}$)

\begin{equation}
M^x(t) = 
\int_{v_0}^{\bar{v}} \! v^x \, m(v,t) \, dv
+
\int_{\bar{v}}^{\infty} \! v^x \, m(v,t) \, dv,
\quad
\forall t \in \mathbb{R}^+,
\quad \mbox{with} \quad
\max\{v_0,1\} < \bar{v} < \infty.
\end{equation}	
	
	Fixing $\bar{v}$ as above, we have $\int_{v_0}^{\bar{v}} \! v^x \, m(v,t) \, dv \le c_x \, S_m(t)$, $\forall t \in \mathbb{R}^+$, where $c_x := \int_{v_0}^{\bar{v}} \! v^x \, dv$, $0 < c_x < \infty$, $\forall x \in \mathbb{R}$, and $S_m(t) := \sup_{v \in \mathbb{R}^+} |m(v,t)|$, $\forall t \in \mathbb{R}^+$. 
	
	We also notice that $v^x \le v^y$, $\forall x \le y \in \mathbb{R}$, whenever $v > 1$. Indeed, the function $f:\mathbb{R}\to\mathbb{R}^+$, $f(x):=v^x$, is increasing with respect to its argument $x \in \mathbb{R}$ if $v>1$. Then, provided $m(v,t)$ to be non-negative (as shown in \cite{RUSCONI2019106944}) and $\bar{v}>1$, one has 

\begin{equation}
\int_{\bar{v}}^{\infty} \! v^x \, m(v,t) \, dv
\le 
\int_{\bar{v}}^{\infty} \! v^y \, m(v,t) \, dv
\le
M^y(t),
\quad
\forall x \le y \in \mathbb{R},
\quad
\forall t \in \mathbb{R}^+.
\end{equation}

	By using the derived above inequalities, one obtains the statement of \autoref{prop:rel_order_mom_latex}:

\begin{equation}
M^x(t) 
\le c_x \, S_m(t) + M^y(t)
\le K^x_m + M^y(t),
\quad
\forall x \le y \in \mathbb{R},
\quad
\forall t \in [0,T],
\end{equation}

\noindent where $T<\infty$ and, thus, the estimate \eqref{est:Sm} for $S_m(t)$ \eqref{eqn:Sup_norm_def} can be used (together with $c_x$ given above) to obtain the quantity $0 < K^x_m < \infty$ dependent on $x$, i.e. on $c_x$, but independent of time $t$.
\end{proof}	
}
		
\begin{theorem}
Given $a \le 0$, $0<b<1$, $v_0 > 0$, the moments $M^1(t) := \int_0^{\infty} \! v \, m(v,t) \, dv$ and $W^1(t) := \int_0^{\infty} \! v \, w(v,t) \, dv$ of solutions to \eqref{eqn:PBE_latex}-\eqref{eqn:Sigma_m,w} are finite for all times $t \in \mathbb{R}^+$.	
\label{prop:finite_M,W1_PBE_latex}		
\end{theorem}

\begin{proof}

If $a \le 0$, $0<b<1$, $v_0 > 0$, the evolution equations for $M^1(t)$ and $W^1(t)$ read, $\forall t \in \mathbb{R}^+$,

	\begin{equation}
	\begin{cases}
	\frac{dM^1(t)}{dt}
	& =
	- \, \lim_{v \to \infty} \, g^1_m(v,t) 
	\, + \,	
	( \TDFpolymeriz(t) - \mu ) \, M^1(t) 
	\, + \,
	\TDFdiffusion(t) \, M^b(t) 
	\, + \, 
	\tilde{\eta}_0(t) \, v_0 \, -	 
	\\
	& \quad - \,
	\tilde{\alpha}_0(t) \, 
	\left[ 
	\, M^{1+a}(t) \, M^0(t) 
	\, + \,
	M^1(t) \, M^a(t) \,
	\, - \,
	\, M^{1+a}(t) \, M^0(t) 
	\, - \,
	M^1(t) \, M^a(t) 
	\, \right],
	\\
	\frac{dW^1(t)}{dt}
	& =
	- \, \lim_{v \to \infty} \, g^1_w(v,t) 
	\, + \,	
	\TDFpolymeriz(t) \, W^1(t) 
	\, + \,
	\TDFdiffusion(t) \, W^b(t) 
	\, + \, 
	\mu \, M^1(t) \, -	 
	\\
	& \quad - \,
	\tilde{\alpha}_0(t) \, 
	\left[ 
	\, W^{1+a}(t) \, W^0(t) 
	\, + \,
	W^1(t) \, W^a(t) \,
	\, - \,
	\, W^{1+a}(t) \, W^0(t) 
	\, - \,
	W^1(t) \, W^a(t) 
	\, \right],
	\\
	M^1(0) & = W^1(0) = 0,
	\end{cases}
	\label{eqn:ODE_M,Wk_int_PBE_latex}
	\end{equation}

\noindent with $g^x_m,g^x_w$ given as \eqref{eqn:flux_grow_mom_PBElatex}, $M^x(t) := \int_0^{\infty} \! v^x \, m(v,t) \, dv$ and $W^x(t) := \int_0^{\infty} \! v^x \, w(v,t) \, dv$, $\forall x \in \mathbb{R}$.  

	Using the non-negativity of $g^1_m,g^1_w$ \eqref{eqn:flux_grow_mom_PBElatex}, $\mu$ \eqref{eqn:phase_tr_rate}, $m(v,t)$ and $w(v,t)$ \cite{RUSCONI2019106944}, together with the bounds provided in \autoref{tab:factors_charact}, one obtains $\forall t \in \mathbb{R}^+$

	\begin{equation}
	\begin{cases}
	\frac{dM^1(t)}{dt}
	& \le
	\KAPPApi \, M^1(t) 
	\, + \,
	\KAPPAdi \, M^b(t) 
	\, + \, 
	\KAPPAenne \, v_0 
	\, + \, 
	\KAPPAalpha \, 
	\left[ 
	\, M^{1+a}(t) \, M^0(t) 
	\, + \,
	M^1(t) \, M^a(t) \,
	\right],
	\\
	\frac{dW^1(t)}{dt}
	& \le
	\KAPPApi \, W^1(t) 
	\, + \,
	\KAPPAdi \, W^b(t) 
	\, + \, 
	\mu \, M^1(t) 
	\, + \,	 
	\KAPPAalpha \, 
	\left[ 
	\, W^{1+a}(t) \, W^0(t) 
	\, + \,
	W^1(t) \, W^a(t) 
	\, \right],
	\\
	M^1(0) & = W^1(0) = 0.
	\end{cases}
	\end{equation}
	
	When time $t$ belongs to the finite interval $[0,T]$, with any fixed $T<\infty$, \autoref{prop:rel_order_mom_latex} allows us writing (note that $a\le0$, $0<b<1$ and $v_0>0$)
		
	\begin{equation}
	\begin{cases}
	\frac{dM^1(t)}{dt}
	& \le
	\KAPPApi \, M^1(t) 
	\, + \,
	\KAPPAdi \, ( K^b_m + M^1(t) ) 
	\, + \, 
	\KAPPAenne \, v_0 \, +
	\\
	& \quad + 
	\, \KAPPAalpha  
	\left[ 
	\, M^0(t) \, ( K^{1+a}_m + M^1(t) ) 
	\, + \,
	( K^a_m + M^0(t) ) \, M^1(t)
	\, \right],
	\quad \forall t \in [0,T],
	\\
	\frac{dW^1(t)}{dt}
	& \le
	\KAPPApi \, W^1(t) 
	\, + \,
	\KAPPAdi \, ( K^b_w + W^1(t) ) 
	\, + \, 
	\mu \, M^1(t) \, + 
	\\
	& \quad +  
	\, \KAPPAalpha \, 
	\left[ 
	\, W^0(t) \,  ( K^{1+a}_w + W^1(t) ) 
	\, + \,
	( K^a_w + W^0(t) ) \, W^1(t) 
	\, \right],
	\quad \forall t \in [0,T],
	\\
	M^1(0) & = W^1(0) = 0,
	\end{cases}
	\end{equation}	
	
\noindent with $0 < K^x_m,K^x_w < \infty$, $\forall x \in \mathbb{R}$, independent of time $t$. Moreover, \autoref{prop:finite_M,W0_PBE_latex} holds for $M^0(t)$ and $W^0(t)$ (i.e. $M^0(t) \le \KAPPAenne \, t$, $W^0(t) \le \frac{ \KAPPAenne \, \lphtr }{2} \, t^2$, $\forall t \in \mathbb{R}^+$) providing

	\begin{equation}
	\begin{cases}
	\frac{dM^1(t)}{dt}
	& \le
	\KAPPApi \, M^1(t) 
	\, + \,
	\KAPPAdi \, ( K^b_m + M^1(t) ) 
	\, + \, 
	\KAPPAenne \, v_0 \, +
	\\
	& \quad + 
	\, \KAPPAalpha  
	\left[ 
	\, \KAPPAenne \, t \, ( K^{1+a}_m + M^1(t) ) 
	\, + \,
	( K^a_m + \KAPPAenne t ) \, M^1(t)
	\, \right],
	\quad \forall t \in [0,T],
	\\
	\frac{dW^1(t)}{dt}
	& \le
	\KAPPApi \, W^1(t) 
	\, + \,
	\KAPPAdi \, ( K^b_w + W^1(t) ) 
	\, + \, 
	\mu \, M^1(t) \, + 
	\\
	& \quad +  
	\, \KAPPAalpha \, 
	\left[ 
	\, \frac{ \KAPPAenne \, \lphtr }{2} \, t^2 
	\, ( K^{1+a}_w + W^1(t) ) 
	\, + \,
	( K^a_w + \frac{ \KAPPAenne \, \lphtr }{2} \, t^2 )
	\, W^1(t) 
	\, \right],
	\quad \forall t \in [0,T],
	\\
	M^1(0) & = W^1(0) = 0.
	\end{cases}
	\label{eqn:ineq_der_MW1_latex}
	\end{equation}
 
	In the case of $M^1(t)$, one obtains by integrating \eqref{eqn:ineq_der_MW1_latex} over $[0,t]$, with $t \in [0,T]$, $T<\infty$:

	\begin{equation}
	M^1(t) \le h(t) + \int_0^t \! k(s) \, M^1(s) \, ds,
	\quad
	\forall t \in [0,T],
	\quad T<\infty,
	\end{equation}

\noindent with $h(t) := k_1 \, t + k_2 \, t^2$, $k(t) := k_3 + k_4 \, t$ and $0 < k_i < \infty$ independent of time $t$, $\forall i=1,\dots,4$. Indeed, such quantities $k_i$, $\forall i=1,\dots,4$, depend only on the \changered{time-independent} constants in \eqref{eqn:ineq_der_MW1_latex}. Since $h(t)$ and $k(t)$ are continuous, non-negative and non-decreasing functions on $[0,T]$, a Gronwall's inequality (Theorem A in \cite{DingAhmad2018}) leads to

	\begin{equation}
	M^1(t) 
	\le 
	h(t) \, e^{\int_0^t \! k(\xi) \, d\xi}
	\le 
	( k_1 T + k_2 T^2 )
	\, e^{ k_3 T + k_4 T^2 }
	< \infty,
	\quad
	\forall t \in [0,T],
	\quad
	T < \infty.
	\label{est:M1}
	\end{equation}
	
\noindent As $T$ can be arbitrarily chosen in $(0,\infty)$, \eqref{est:M1} implies $M^1(t)<\infty$, $\forall t \in \mathbb{R}^+$. 

	To prove finiteness of $W^1(t)$, $\forall t \in \mathbb{R}^+$, we first notice that \eqref{est:M1} along with $0<\mu=\lphtr<\infty$ (\autoref{tab:PBE_lambdas_def}) confirm the finiteness of the term $\mu \, M^1(t)$ in the second inequality in \eqref{eqn:ineq_der_MW1_latex}. Then, it is possible to invoke a Gronwall's inequality for $W^1(t)$ similar to the discussed for $M^1(t)$ and, thus, to conclude the proof.
\end{proof}
	
\renewcommand{\thesection}{D}
\section{LPMF PBM: Boundedness of Solutions}	
\label{sec:Bounded_PBE_Sol}
		
\begin{theorem}
Given $a \le 0$, $0<b<1$ and $v_0 > 0$, the solutions $m(v,t)$ and $w(v,t)$ to \eqref{eqn:PBE_latex}-\eqref{eqn:Sigma_m,w} are bounded for $v \in \mathbb{R}^+$ and $t \in [0,T]$, with any fixed $T < \infty$.
\label{prop:bounded_sol_PBElatex}		
\end{theorem}

\begin{proof}
We provide a proof for $m(v,t)$ only, as similar arguments apply to $w(v,t)$. We start by denoting
	
	\begin{equation}
	\tilde{g}(v,t) 
	:=
	\begin{cases}
	g(v,t), & \mathrm{for} \quad v \ge v_0, \, t \in \mathbb{R}^+, 
	\\
	g(v_0,t), & \mathrm{for} \quad v < v_0, \, t \in \mathbb{R}^+, 
	\end{cases}
	\label{eqn:tilde_g_def}
	\end{equation}

\noindent and

	\begin{equation}
	\changegreen{\tilde{\alpha}}(v,u,t) 
	:=
	\begin{cases}
	\changegreen{\alpha}(v,u,t), 
	& \mathrm{if} \quad v,u \ge v_0, \, t \in \mathbb{R}^+, 
	\\
	\changegreen{\alpha}(v_0,v_0,t),
	& \mathrm{otherwise}, 
	\end{cases}
	\label{eqn:tilde_a_def}
	\end{equation}

\noindent where $g(v,t) = \TDFdiffusion(t) \, v^b + \TDFpolymeriz(t) \, v$ \eqref{eqn:growth_rate}, $v_0>0$, $\changegreen{\alpha}(v,u,t) = \changegreen{\tilde{\alpha}_0}(t) [ v^a + u^a ]$ \eqref{eqn:aggr_rate} and $a \le 0$. We consider now \eqref{eqn:PBE_latex} but with $\tilde{g}$ and $\changegreen{\tilde{\alpha}}$ instead of $g$ and $\changegreen{\alpha}$ respectively:

	\begin{equation}
	\partial_t m(v,t)
	\, + \,
	\tilde{g}(v,t) \, \partial_v m(v,t)
	\, + \,
	\partial_v \tilde{g}(v,t) \, m(v,t)
	\, + \, 
	\changegreen{\mu} \, m(v,t) 
	= 
	\tilde{\eta}_0(t) \, \delta(v-v_0) 
	\, + \,
	\mathcal{I}_m(v,t).
	\label{eq:PBE_latex_tilde_g}
	\end{equation}
	
\noindent Here

	\begin{align}
	\mathcal{I}_m(v,t)
	:=
	& \, \frac{1}{2} \, \int_{-\infty}^{\infty}
	\! \changegreen{\tilde{\alpha}}(v-u,u,t) 
	\, m(v-u,t) \, m(u,t) \, du
	\, - 
	\nonumber \\
	& - \, m(v,t) \,
	\int_{-\infty}^{\infty}
	\! \changegreen{\tilde{\alpha}}(v,u,t) \, m(u,t) \, du
	\label{eqn:Im_def}
	\end{align}

\noindent and $m(v,t)$ has been set to $0$ for all $v<0$ and $t \in \mathbb{R}^+$. We shall continue reasoning just for \eqref{eq:PBE_latex_tilde_g} and at the end, taking into account that \autoref{eqn:boundary_band_sol_PBE} still holds for this equation (as well as for \eqref{eqn:PBE_latex}) and $\tilde{g} \equiv g$, $\changegreen{\tilde{\alpha} \equiv \alpha}$ for $v,u \ge v_0$, our derivation will also hold for \eqref{eqn:PBE_latex}.
	
	We denote by $\varphi(t,v_*)$ the solution at time $t$, with initial data $v_* \in \mathbb{R}$, of the ODE:

	\begin{equation}
	\begin{cases}
	\frac{d}{dt}\varphi(t,v_*) 
	& = \tilde{g}(\varphi(t,v_*),t),
	\quad \forall t \in \mathbb{R}^+,
	\\
	\varphi(0,v_*) & = v_*.
	\end{cases}
	\label{eq:transportODE}
	\end{equation} 

\noindent The ODE provides a well-defined solution and a global flow $v = \varphi(t,v_*)$ and its inverse (in the second variable) $v_* = \varphi_*(t,v)$ as long as $\tilde{g}$ is Lipschitz (has bounded derivatives) in its variable $v$ (see \cite{BCDbook}, pp. $129-130$). Indeed,

	\begin{equation}
	\partial_v \tilde g(v,t)
	=
	\begin{cases}
	\TDFdiffusion(t) \, b \, v^{b-1} \, + \, \TDFpolymeriz(t),
	& \mathrm{for} \quad v \ge v_0, 
	\\
	0 & 
	\mathrm{for} \quad v < v_0,
	\end{cases}
	\label{eqn:partial_v_tilde_g}
	\end{equation}
	
\noindent (in the sense of distributions, hence we can ignore what happens at $v = v_0$) and thus, taking into account that $0<b<1$ and $\TDFdiffusion,\TDFpolymeriz$ are bounded functions of time (see \autoref{tab:factors_charact}), one can conclude that $\partial_v \tilde g$ is bounded as well.

\noindent We remark that $\varphi(t,v_*)$ is a strictly increasing function of its second argument $v_*$ and this property is used in \autoref{sec:Approx_Model_form}. Indeed, from \eqref{eq:transportODE} it follows:

	\begin{equation}
	\frac{d}{dt} \left( 
	\frac{\partial \varphi}{\partial v_*} \right)
	=
	\frac{\partial}{\partial v_*}
	\left( \frac{d\varphi}{dt} \right)
	=
	\frac{\partial \tilde{g}}{\partial \varphi}
	\frac{\partial \varphi}{\partial v_*},
	\quad
	\frac{\partial \varphi(0,v_*)}{\partial v_*} = 1,
	\end{equation}
	
\noindent which gives $\partial_{v_*} \varphi(t,v_*) = e^{\int_0^t \partial_v \tilde{g}(\varphi(s,v_*),s) ds} > 0$, meaning that $\varphi(t,v_*)$ monotonically increases with respect to $v_*$. 
	
	After such an analysis of the ODE \eqref{eq:transportODE}, we consider equation \eqref{eq:PBE_latex_tilde_g} aiming to show the boundedness of solution $m$. 
	
	Using \eqref{eq:transportODE} and the chain rule, we have:

	\begin{equation}
	\frac{d}{dt} m(\varphi(t,v_*),t) 
	=
	\left[
	\partial_t m(v,t) \, + \, \tilde{g}(v,t) \, \partial_v m(v,t)
	\right]_{v=\varphi(t,v_*)}.
	\label{eqn:chain_rule_ODE_m}
	\end{equation}

\noindent The Right Hand Side of \eqref{eqn:chain_rule_ODE_m} provides the first two terms in \eqref{eq:PBE_latex_tilde_g} evaluated at $v=\varphi(t,v_*)$. Hence, \eqref{eq:PBE_latex_tilde_g} can be rewritten (when evaluated at $v=\varphi(t,v_*)$) as:

	\begin{align}
	& \frac{d}{dt} m(\varphi(t,v_*),t)
	\, + \,
	( \, \partial_v \tilde{g}(\varphi(t,v_*),t) 
	\, + \, 
	\changegreen{\mu} \, ) 
	\, m(\varphi(t,v_*),t)
	= 
	\nonumber \\
	& = \, \tilde{\eta}_0(t) \, \delta(\varphi(t,v_*)-v_0)
	\, + \, 
	\mathcal{I}_m(\varphi(t,v_*),t).
	\label{eq:mPsi0}
	\end{align}

\noindent We denote:

	\begin{equation}
	\changegreen{\mathcal{F}}(t,v_*) :=
	\int_0^t 
	\! ( \, \partial_v \tilde{g}(\varphi(s,v_*),s) 
	\, + \, 
	\changegreen{\mu} \, ) \, ds,
	\end{equation}
	
\noindent and rewrite \eqref{eq:mPsi0} as

	\begin{equation}
	\frac{d}{dt} 
    \left( m(\varphi(t,v_*),t) 
    \, e^{\changegreen{\mathcal{F}}(t,v_*)} 
    \right) =
	\tilde{\eta}_0(t) 
	\, e^{\changegreen{\mathcal{F}}(t,v_*)} 
	\, \delta(\varphi(t,v_*)-v_0)
	\, + \,
 	e^{\changegreen{\mathcal{F}}(t,v_*)} 
 	\, \mathcal{I}_m(\varphi(t,v_*),t).
	\label{eq:eFm}
	\end{equation}
	
\noindent Integrating of \eqref{eq:eFm} over $[0,t]$ gives:

	\begin{align}
    m(\varphi(t,v_*),t) 
    \, e^{\changegreen{\mathcal{F}}(t,v_*)}
    -
	m(v_*,0) 
	=
	& \int_0^t \, \tilde{\eta}_0(s) 
	\, e^{\changegreen{\mathcal{F}}(s,v_*)} 
	\, \delta(\varphi(s,v_*)-v_0) \, ds
	\, +
	\nonumber \\
	& + \,
	\int_0^t \, e^{\changegreen{\mathcal{F}}(s,v_*)} 
	\, \mathcal{I}_m(\varphi(s,v_*),s) \, ds.
  	\label{eqn:int_delta_to_solve}
	\end{align}

	Since $0 < \TDFpolymeriz(t) \, v_0 \le \tilde{g}(v,t) < \infty$ for any finite $v \in \mathbb{R}$ and $t \in \mathbb{R}^+$ (see \autoref{tab:factors_charact}), the solution $v = \varphi(t,v_*)$ to \eqref{eq:transportODE} is a continuous (differentiable with finite derivative) and strictly increasing function of time $t$. Then, its inverse (in the time variable) $t = \varphi^{-1}(v,v_*)$ is well defined, with derivative

	\begin{equation}
	\frac{d}{dv} \varphi^{-1}(v,v_*) 
	=
	\frac{1}{\frac{d\varphi}{dt}(\varphi^{-1}(v,v_*),v_*)}
	=
	\frac{1}{
	\tilde{g}(
	\varphi( \varphi^{-1}(v,v_*), v_* ),
	\varphi^{-1}(v,v_*)
	)}
	=
	\frac{1}{ \tilde{g}(v, \varphi^{-1}(v,v_*) )}.	
	\end{equation}

\noindent As a consequence, it follows 

	\begin{align}
	& \int_0^t \! \tilde{\eta}_0(s) 
	\, e^{\changegreen{\mathcal{F}}(s,v_*)} 
	\, \delta(\varphi(s,v_*)-v_0) \, ds
	=
	\nonumber \\
	& =
	\int_{\varphi(0,v_*) - v_0}^{\varphi(t,v_*) - v_0} 
	\! \tilde{\eta}_0(\varphi^{-1}(x+v_0,v_*))
	\, e^{\changegreen{\mathcal{F}}(\varphi^{-1}(x+v_0,v_*),v_*)} 
	\, \frac{\delta(x) \, dx}{\tilde{g}(x+v_0,\varphi^{-1}(x+v_0,v_*))} =
	\nonumber \\
	& =
	H(v_0-\varphi(0,v_*)) \, H(\varphi(t,v_*)-v_0)
	\, \frac{ \tilde{\eta}_0(\varphi^{-1}(v_0,v_*)) }
	{ \tilde{g}(v_0,\varphi^{-1}(v_0,v_*)) }
	\, e^{\changegreen{\mathcal{F}}(\varphi^{-1}(v_0,v_*),v_*)},	
	\end{align}

\noindent where the change of variables $x = \varphi(s,v_*) - v_0$ has been applied and $H(y) := (y+|y|)/(2y)$ is the Heaviside step function. 

	Setting $\varphi(t,v_*)=v$ we have $v_*=\varphi_*(t,v)$, hence, \eqref{eqn:int_delta_to_solve} gives

	\begin{align}
	m(v,t) =
	& \, 
	m(\varphi_*(t,v),0) 
	\, e^{-\changegreen{\mathcal{F}}(t,\varphi_*(t,v))} 
	\, + \, 
	H(v_0-\varphi_*(t,v)) \, H(v-v_0)	
	\nonumber \\
	& \,
	\frac{ \tilde{\eta}_0(\varphi^{-1}(v_0,\varphi_*(t,v))) }
	{ \tilde{g}(v_0,\varphi^{-1}(v_0,\varphi_*(t,v))) }
	\, e^{-\changegreen{\mathcal{F}}(t,\varphi_*(t,v))} 
	\, e^{\changegreen{\mathcal{F}}(\varphi^{-1}(v_0,\varphi_*(t,v)),\varphi_*(t,v))}
	\, + 
	\nonumber \\
	& \,+ \,
	e^{-\changegreen{\mathcal{F}}(t,\varphi_*(t,v))} 
	\int_0^t 
	\, e^{\changegreen{\mathcal{F}}(s,\varphi_*(t,v))} 
	\, \mathcal{I}_m	(\varphi(s,\varphi_*(t,v)),s) \, ds,
 	\label{eqn:sol_m_PBE_latex}
	\end{align}
	
\noindent where $\varphi^{-1}(v_0,\varphi_*(t,v))$ corresponds to the time at which the flow $\varphi$ reaches the value $v_0$, starting from $\varphi_*(t,v)$ (at time $0$) and getting to $v$ at time $t$. Since $\varphi$ is a strictly increasing function of time, we can ensure that

	\begin{equation}
	0 \le \varphi^{-1}(v_0,\varphi_*(t,v)) \le t
	\quad \Leftrightarrow \quad
	\varphi_*(t,v) \le v_0 \le v.
	\label{eqn:time_phi_reach_v0}
	\end{equation}		

\noindent With help of \eqref{eqn:partial_v_tilde_g}, \autoref{tab:factors_charact} and $0 < b < 1$ one obtains
	
	\begin{equation}
	0 \le \partial_v \tilde{g}(v,t) \le \kappa_0, 
	\quad
	0  < \kappa_0 := \KAPPAdi \, b \, v_0^{b-1} + \KAPPApi < \infty,
	\quad
	\forall v \in \mathbb{R}, \forall t \in \mathbb{R}^+.
	\label{eqn:bounds_partial_tilde_g} 
	\end{equation}
	
\noindent Moreover, as $\mu = \lphtr \in (0,\infty)$, it follows 

	\begin{align}
	& 0 \le \changegreen{\mathcal{F}}(t,v_*) \le (\kappa_0 
	+ \changegreen{\lphtr} ) \, t,
	\quad \forall v_* \in \mathbb{R}, \, \forall t \in \mathbb{R}^+
	\quad \mbox{and thus,}
	\nonumber \\
	& \changegreen{\mathcal{F}}
	(\varphi^{-1}(v_0,\varphi_*(t,v)),\varphi_*(t,v))
	\le (\kappa_0 + \changegreen{\lphtr} ) \, t,
	\quad \mbox{for} \quad \varphi_*(t,v) \le v_0 \le v.	
	\label{eqn:bounds_mathcalF}
	\end{align}

\noindent Given $\changegreen{\tilde{\alpha}}$ \eqref{eqn:tilde_a_def}, we also have (see \autoref{tab:factors_charact})

	\begin{equation}
	0 < \changegreen{\tilde{\alpha}}(v,u,t) \le a_0,
	\quad
	0  < a_0 := 2 \, \changegreen{\KAPPAalpha} \, v_0^a < \infty,
	\quad
	\forall v,u \in \mathbb{R}, \, \forall t \in \mathbb{R}^+,	 
	\label{eqn:bounds_tilde_alpha}
	\end{equation}
	
\noindent and, as $m(v,t)$ is non-negative for any $v,t \in \mathbb{R}^+$ (\cite{RUSCONI2019106944}) and \autoref{prop:finite_M,W0_PBE_latex} holds for $M^0(t)$,

	\begin{equation}
	\mathcal{I}_m(v,t) 
	\le 
	\frac{a_0}{2} \, M^0(t) \, S_m(t)
	\le 
	\frac{a_0}{2} 
	\, \changegreen{\KAPPAenne \, t} \, S_m(t),
	\quad
	\forall v \in \mathbb{R}, \, \forall t \in \mathbb{R}^+,
	\end{equation}
	
\noindent where 

	\begin{equation}
	S_m(t) := \sup_{v \in \mathbb{R}^+} |m(v,t)|.
	\label{eqn:Sup_norm_def}
	\end{equation}
	
\noindent Finally, if time $t \in [0,T]$, $T<\infty$, one gets (according to estimations from \autoref{sec:time_dep_factors_rates})

	\begin{equation}
	\tilde{g}(v_0,t) \ge \TDFpolymeriz(t) \, v_0 
	\ge 
	\kappa_\varrho(T) 
	:=
	\frac{ \lpoly \, v_0 \, \Psi(T) }
	{ (\bar{\Psi}+1) ( V_{\mathrm{pol2}}(T) +
	\changegreen{\lpolone})} > 0,
	\quad \forall t \in [0,T].
	\label{eqn:tilde_g_lower_bound}
	\end{equation}

	The inequalities \eqref{eqn:time_phi_reach_v0}-\eqref{eqn:tilde_g_lower_bound} applied to \eqref{eqn:sol_m_PBE_latex} give the following estimation for the supremum norm $S_m(t)$ \eqref{eqn:Sup_norm_def} of solution $m(v,t)$ to \eqref{eqn:PBE_latex}-\eqref{eqn:Sigma_m,w} (rem. \changeorange{$S_m(t)=\sup_{v \in \mathbb{R}^+} m(v,t)$, $\forall t \in \mathbb{R}^+$, as $m$ is non-negative \cite{RUSCONI2019106944} and} $S_m(0)=0$ \eqref{eqn:PBE_latex}):		

	\begin{equation}
	S_m(t) \le 
	h(t) + \int_0^t \! k(\xi) \, S_m(\xi) \, d\xi,
	\quad \forall t \in [0,T],
	\quad T < \infty,
	\end{equation}

\noindent with

	\begin{equation}
	h(t) := 
	\frac{\KAPPAenne}{\kappa_\varrho(T)}
	\, e^{(\kappa_0+\changegreen{\lphtr}) t}
	\quad \mbox{and} \quad
	k(t) := 
	\frac{a_0 \, \KAPPAenne}{2}
	\, \changegreen{t} 
	\, e^{(\kappa_0+\changegreen{\lphtr}) t}
	\end{equation}
	
\noindent continuous, non-negative and non-decreasing functions on $[0,T]$. Then, a Gronwall's inequality (Theorem A in \cite{DingAhmad2018}) leads to

	\begin{equation}
	S_m(t) 
	\le 
	h(t) \, e^{\int_0^t \! k(\xi) \, d\xi}
	\le 
	\frac{\KAPPAenne 
	\, e^{(\kappa_0+\changegreen{\lphtr}) T}}
	{\kappa_\varrho(T)}
	\, e^{ \frac{a_0 \, \KAPPAenne}{2} 
	\, \changegreen{T^2} 
	\, e^{(\kappa_0+\changegreen{\lphtr}) T} }
	< \infty,
	\quad
	\forall t \in [0,T],
	\label{est:Sm}
	\end{equation}
	
\noindent and the consequent boundedness of $m(v,t)$ for $v \in \mathbb{R}^+$ and $t \in [0,T]$, with any fixed $T < \infty$.
\end{proof}

\renewcommand{\thesection}{E}
\section{Justification of Discarding Integral Terms in the LPMF PBM}
\label{sec:approx_arg}

	In the context of equation \eqref{eq:meps}, let us introduce 
	
	\begin{equation}
	S_{\me}(t) := \sup_{v \in \mathbb{R}^+} |{\me}(v,t)|,
	\quad
	\forall t \in \mathbb{R}^+.	
	\label{eqn:Sup_norme_def}
	\end{equation}
	
\noindent Then using \eqref{est:Sm} one obtains
	 		 
	\begin{equation}\label{est:unifmeps}
	S_{\me}(t)
	\le 
	\frac{\KAPPAenne 
	\, e^{(\kappa_0+\changegreen{\lphtr}) T}}
	{\kappa_\varrho(T)}
	\, e^{ \frac{\epsilon \, a_0 \, \KAPPAenne}{2} 
	\, \changegreen{T^2} 
	\, e^{(\kappa_0+\changegreen{\lphtr}) T} },
	\quad
	\forall t\in [0,T].
	\end{equation}
 
\noindent To this end we considered the space of square-integrable functions on the interval $[0,V]\subset \mathbb{R}$ defined as
 
$$
 L^2(0,V):=\{f:[0,V]\to\mathbb{R}; \int_0^V f^2(s)\,ds<\infty\},
$$ endowed with the norm
 
\begin{equation}\label{def:L2norm}
\|f\|_{L^2(0,V)}:=\left( \int_0^V f^2(s)\,ds\right)^{\frac 12}.
\end{equation}
 
\noindent 
We remark that if $S_g<\infty$ for some $g:\mathbb{R}\to \mathbb{R}$ then for any $V>0$ the inequality $\|g\|_{L^2(0,V)}\le  V^{\frac 12} S_g$ holds. 
We return to this useful property later but now we 
need to define the space $H^{-1}(0,V)$, a space larger than $L^2$ that is roughly speaking the space of functions in $L^2$ together with their distributional derivatives. More precisely,
 
$$
H^{-1}(0,V):=\{f:(0,V)\to\mathbb{R}; \textrm{ there exist } f_0,f_1\in L^2(0,V)\textrm{ such that }f=f_0+f'_1\},
$$ i.e. any function in $H^{-1}(0,V)$ can be written as the sum of a function in $L^2$ and the distributional derivative of a function in $L^2$ (with $f_0$ and $f_1$ in general different from each other; see for instance \cite{Evans98} sec 5.9.1). It should be noted that the decomposition into an $L^2$ part and the distributional derivative of an $L^2$ function might not be unique (i.e. we might have $f_0+f_1'=g_0+g_1'$ but $f_0\not=g_0$ and $f_1\not=g_1$). This is taken into account when defining the $H^{-1}$ norm as follows  (see for instance \cite{Evans98} sec 5.9.1)

$$
\|f\|_{H^{-1}(0,V)}:=\inf \bigg\{\left(\int_0^V f_0^2(s)+f_1^2(s)\,ds\right)^{\frac 12}; f=f_0+f_1'\bigg\}
$$ and the infimum is taken over all possible representations of $f$  of the form  $f_0+f_1'$. 

Also, we need to introduce spaces mixed in $v$ and $t$, i.e. 

$$
L^2(0,T; H^{-1}(0,V)):=\{ f: (0,V)\times (0,T) \to\mathbb{R};\int_0^T \|f(\cdot,s)\|_{H^{-1}(0,V)}^2\,ds<\infty\}
$$

\noindent with the norm

$$
\|f\|_{L^2(0,T; H^{-1}(0,V))}:=\left(\int_0^T \|f(\cdot,s)\|_{H^{-1}(0,V)}^2\,ds\right)^{\frac 12}
$$

\noindent and the space

$$
L^\infty(0,T; L^2(0,V)):=\{ f: (0,V)\times (0,T)\to\mathbb{R};\sup_{s\in [0,T]}\|f(\cdot,s)\|_{L^2(0,V)}<\infty\}
$$

\noindent \changeorange{with the norm}

$$
\changeorange{\|f\|_{L^\infty(0,T; L^2(0,V))}:=\sup_{s\in [0,T]}\|f(\cdot,s)\|_{L^2(0,V)}.}
$$

	Now we are in the position to recall the Aubin-Lions lemma (see for instance \cite{roubivcek2013}, Sec. $7.3$) which, adapted to our specific situation,  says that for a class of functions $(f_\eps)_\eps:(0,V)\times (0,T) \to\mathbb{R}$, if we have simultaneously

\begin{equation}\label{cond:AubinLions}
\|f_\eps\|_{L^\infty(0,T; L^2(0,V))}\le C_1, 
\quad
\|\partial_t f_\eps\|_{L^2(0,T; H^{-1}(0,V))}\le C_2,
\quad
\forall 0<\eps<\eps_0,
\end{equation} 
 for some $\eps_0>0$ and $C_1,C_2$ independent of $\eps$, then there exists a sequence $(\eps_k)_{k\in\N}$ with $\eps_k\to 0$ as $k\to \infty$ and a function $f_0\in C([0,T];L^2(0,V))$ such that

\begin{equation}\label{lim:mepsm0_appx}
\sup_{t\in [0,T]} \int_0^V 
\! (f_{\eps_k}(v,t)-f_0(v,t)) \, \varphi(v,t) \, dv \to 0,
\textrm{ as } k\to\infty,
\end{equation} for any function $\varphi:(0,V)\times [0,T]\to \mathbb{R}$ which is at least differentiable in both variables and for each $t\in [0,T]$ is zero on intervals of type $(0,\delta_t)\cup (V-\delta_t,V)$.

Multiplying the equation \eqref{eq:meps} by $\varphi$, with $\varphi$ as above, and integrating by parts to put the $\partial_t$ and $\partial_v$ derivatives onto the test function yield a weak form of the equation \eqref{eq:meps} in which we can pass to the limit using relation \eqref{lim:mepsm0_appx} and obtaining the weak form of equation \eqref{eq:meps} with $\eps=0$. 

The remaining step is to check if the conditions \eqref{cond:AubinLions} are met in our case. The first one is already achieved thanks to estimates \eqref{est:unifmeps} and the remark following \eqref{def:L2norm}. For the second one we inspect the equation \eqref{eq:meps} and note that all the terms on the right-hand side are suitably bounded due to the bounds mentioned above.

\changered{
\renewcommand{\thesection}{F}
\section{Glossary}
\label{sec:gloss}

\glsaddall
\printglossaries
}

\addcontentsline{toc}{section}{References}

\bibliography{refs.bib}

\end{document}